\documentclass[onecolumn,useamsfonts]{pasj02}
\usepackage{natbib}
\usepackage{aas2pasj}   
 
\usepackage{float}
\usepackage{amsmath}
\usepackage{tikz}
\usetikzlibrary{positioning, arrows.meta}
 
\usepackage{xr}
 
\newif\ifXRFROZEN \XRFROZENtrue
\ifXRFROZEN
  \input{xref_frozen_S1_pasj}
\else
\fi

\usepackage{xcolor}

\def\amm{NH$_3$}
 
\jyear{2026}
\Received{}
\Accepted{}

\graphicspath{{./}{figures/}}

\begin{document}
 
\title{B-fields In STar-forming Region Observations (BISTRO): \\ Dual-Wavelength Constraints on Size-Dependent Grain Alignment in Star-Forming Regions}
 
\author{%
Ray S.~\textsc{Furuya},\altaffilmark{1,2}\altemailmark\orcid{0000-0002-0786-7307} \email{rsfuruya@gmail.com}
David \textsc{Berry},\altaffilmark{3}\orcid{0000-0001-6524-2447}
Pierre \textsc{Bastien},\altaffilmark{4}\orcid{0000-0002-0794-3859}
Tetsuo \textsc{Hasegawa},\altaffilmark{5}\orcid{0000-0003-1853-0184}
Masafumi \textsc{Matsumura},\altaffilmark{6}\orcid{0000-0002-6906-0103}
Thiem \textsc{Hoang},\altaffilmark{7,8}\orcid{0000-0003-2017-0982}
Lapo \textsc{Fanciullo},\altaffilmark{9}\orcid{0000-0001-9930-9240}
Kazunari \textsc{Iwasaki},\altaffilmark{10}\orcid{0000-0002-9892-1881}
Simon \textsc{Coud\'{e}},\altaffilmark{5}\orcid{0000-0002-0859-0805}
Takashi \textsc{Onaka},\altaffilmark{11}\orcid{0000-0002-8234-6747}
Pham Ngoc \textsc{Diep},\altaffilmark{12}\orcid{0000-0002-2808-0888}
Sarah \textsc{Sadavoy},\altaffilmark{13}\orcid{0000-0001-7474-6874}
Hua-bai \textsc{Li},\altaffilmark{14}\orcid{0000-0003-2641-9240}
Koji \textsc{Kawabata},\altaffilmark{15,16,17}\orcid{0000-0001-6099-9539}
Doris \textsc{Arzoumanian},\altaffilmark{18,19}\orcid{0000-0002-1959-7201}
Yasuo \textsc{Doi},\altaffilmark{20}\orcid{0000-0001-8746-6548}
Jia-Wei \textsc{Wang},\altaffilmark{3}\orcid{0000-0002-6668-974X}
Tao-Chung \textsc{Ching},\altaffilmark{21}\orcid{0000-0001-8516-2532}
Janik \textsc{Karoly},\altaffilmark{22}\orcid{0000-0001-5996-3600}
Jihye \textsc{Hwang},\altaffilmark{18,19}\orcid{0000-0001-7866-2686}
Glenn J. \textsc{White},\altaffilmark{23}\orcid{0000-0002-7126-691X}
A-Ran \textsc{Lyo},\altaffilmark{7}
Kohji \textsc{Tomisaka},\altaffilmark{24}\orcid{0000-0003-2726-0892}
Le Ngoc \textsc{Tram},\altaffilmark{25}\orcid{0000-0002-6488-8227}
Fr\'{e}d\'{e}rick \textsc{Poidevin},\altaffilmark{26,27}\orcid{0000-0002-5391-5568}
Valentin J.~M.\ \textsc{Le Gouellec},\altaffilmark{28,29}\orcid{0000-0002-5714-799X}
Hiroko \textsc{Shinnaga},\altaffilmark{30}\orcid{0000-0001-9407-6775}
Masato I.~N. \textsc{Kobayashi},\altaffilmark{31}\orcid{0000-0003-3990-1204}
Junhao \textsc{Liu},\altaffilmark{32}\orcid{0000-0002-4774-2998}
Mehrnoosh \textsc{Tahani},\altaffilmark{33}\orcid{0000-0001-8749-1436}
Nguyen Bich \textsc{Ngoc},\altaffilmark{12}\orcid{0000-0002-5913-5554}
Jennifer \textsc{Hatchell},\altaffilmark{34}\orcid{0000-0002-4870-2760}
Archana \textsc{Soam},\altaffilmark{35}\orcid{0000-0002-6386-2906}
Hsi-Wei \textsc{Yen},\altaffilmark{21}\orcid{0000-0003-1412-893X}
Chakali \textsc{Eswaraiah},\altaffilmark{36,37,38}\orcid{0000-0003-4761-6139}
Vera \textsc{K\"{o}nyves},\altaffilmark{39}\orcid{0000-0002-3746-1498}
Jintai \textsc{Wu},\altaffilmark{40}
Jeong-Eun \textsc{Lee},\altaffilmark{41}\orcid{0000-0003-3119-2087}
Yusuke \textsc{Tsukamoto},\altaffilmark{30}\orcid{0000-0001-6273-805X}
Ekta \textsc{Sharma},\altaffilmark{36}\orcid{0000-0002-4541-0607}
Tie \textsc{Liu},\altaffilmark{42}\orcid{0000-0002-5286-2564}
Qilao \textsc{Gu},\altaffilmark{43}
Hiroyuki \textsc{Nakanishi},\altaffilmark{30}
Tsuyoshi \textsc{Inoue},\altaffilmark{44}\orcid{0000-0002-7935-8771}
Fumitaka \textsc{Nakamura},\altaffilmark{24,45}\orcid{0000-0001-5431-2294}
Rachel \textsc{Friesen},\altaffilmark{46}\orcid{0000-0001-7594-8128}
Jean-Fran\c{c}ois \textsc{Robitaille},\altaffilmark{47}\orcid{0000-0001-5079-8573}
Yoshiaki \textsc{Misugi},\altaffilmark{48}\orcid{0000-0002-4093-6925}
Raiga \textsc{Kashiwagi},\altaffilmark{48}\orcid{0000-0002-1461-3866}
Per \textsc{Friberg},\altaffilmark{3}\orcid{0000-0002-8010-8454}
Sarah \textsc{Graves},\altaffilmark{3}
Mark \textsc{Rawlings},\altaffilmark{3}\orcid{0000-0002-6529-202X}
Harriet \textsc{Parsons},\altaffilmark{3}\orcid{0000-0002-6327-3423}
David \textsc{Eden},\altaffilmark{49,50}\orcid{0000-0002-5881-3229}
Huei-Ru Vivien \textsc{Chen},\altaffilmark{51}\orcid{0000-0002-9774-1846}
Najimudeen \textsc{Bijas},\altaffilmark{52}\orcid{0000-0002-6398-7530}
Motohide \textsc{Tamura},\altaffilmark{5,11,53}\orcid{0000-0002-6510-0681}
Tyler L.\ \textsc{Bourke},\altaffilmark{54,55}\orcid{0000-0001-7491-0048}
Nagayoshi \textsc{Ohashi},\altaffilmark{21}\orcid{0000-0003-0998-5064}
Woojin \textsc{Kwon},\altaffilmark{56,57}\orcid{0000-0003-4022-4132}
Jongsoo \textsc{Kim},\altaffilmark{7,8}\orcid{0000-0002-1229-0426}
Shih-Ping \textsc{Lai},\altaffilmark{58,21}\orcid{0000-0001-5522-486X}
Keping \textsc{Qiu},\altaffilmark{40,59}\orcid{0000-0002-5093-5088}
Di \textsc{Li},\altaffilmark{36,60}\orcid{0000-0003-3010-7661}
Shu-ichiro \textsc{Inutsuka},\altaffilmark{61}\orcid{0000-0003-4366-6518}
Kate \textsc{Pattle},\altaffilmark{22}\orcid{0000-0002-8557-3582}
Derek \textsc{Ward-Thompson},\altaffilmark{39}\orcid{0000-0003-1140-2761}
and the BISTRO Collaboration}
\altaffiltext{1}{Institute of Liberal Arts and Sciences, Tokushima University, Minami Jousanjima-Cho 1-1, Tokushima, Tokushima 770-8502, Japan}
\altaffiltext{2}{Visiting Associate Professor, Center for Computational Astrophysics (CfCA), National Astronomical Observatory of Japan (FY~2023), Osawa 2-21-1, Mitaka, Tokyo 181-8588, Japan}
\altaffiltext{3}{East Asian Observatory, 660 N.\ A`oh\={o}k\={u} Place, University Park, Hilo, HI 96720, USA}
\altaffiltext{4}{Centre de recherche en astrophysique du Qu\'ebec \& D\'epartement de physique, Universit\'e de Montr\'eal, C.P.~6128, Succursale Centre-ville, Montr\'eal, QC H3C~3J7, Canada}
\altaffiltext{5}{National Astronomical Observatory of Japan, National Institutes of Natural Sciences, Osawa, Mitaka, Tokyo 181-8588, Japan}
\altaffiltext{6}{Faculty of Education \& Center for Educational Development and Support, Kagawa University, Saiwai-cho 1-1, Takamatsu, Kagawa 760-8522, Japan}
\altaffiltext{7}{Korea Astronomy and Space Science Institute, 776 Daedeokdae-ro, Yuseong-gu, Daejeon 34055, Republic of Korea}
\altaffiltext{8}{University of Science and Technology, Korea, 217 Gajeong-ro, Yuseong-gu, Daejeon 34113, Republic of Korea}
\altaffiltext{9}{Tamkang University, No.\ 151, Yingzhuan Rd., Tamsui Dist., New Taipei City 251301, Taiwan}
\altaffiltext{10}{Center for Computational Astrophysics, National Astronomical Observatory of Japan, Mitaka, Tokyo 181-8588, Japan}
\altaffiltext{11}{Department of Astronomy, Graduate School of Science, The University of Tokyo, 7-3-1 Hongo, Bunkyo-ku, Tokyo 113-0033, Japan}
\altaffiltext{12}{Vietnam National Space Center, Vietnam Academy of Science and Technology, 18 Hoang Quoc Viet, Hanoi, Vietnam}
\altaffiltext{13}{Department for Physics, Engineering Physics and Astrophysics, Queen's University, Kingston, ON K7L 3N6, Canada}
\altaffiltext{14}{Department of Physics, The Chinese University of Hong Kong, Shatin, N.T., Hong Kong}
\altaffiltext{15}{Hiroshima Astrophysical Science Center, Hiroshima University, Kagamiyama 1-3-1, Higashi-Hiroshima, Hiroshima 739-8526, Japan}
\altaffiltext{16}{Department of Physics, Hiroshima University, Kagamiyama 1-3-1, Higashi-Hiroshima, Hiroshima 739-8526, Japan}
\altaffiltext{17}{Core Research for Energetic Universe (CORE-U), Hiroshima University, Kagamiyama 1-3-1, Higashi-Hiroshima, Hiroshima 739-8526, Japan}
\altaffiltext{18}{Institute for Advanced Study, Kyushu University, Japan}
\altaffiltext{19}{Department of Earth and Planetary Sciences, Faculty of Science, Kyushu University, Nishi-ku, Fukuoka 819-0395, Japan}
\altaffiltext{20}{Department of Earth Science and Astronomy, Graduate School of Arts and Sciences, The University of Tokyo, 3-8-1 Komaba, Meguro, Tokyo 153-8902, Japan}
\altaffiltext{21}{Academia Sinica Institute of Astronomy and Astrophysics, No.1, Sec.\ 4., Roosevelt Road, Taipei 10617, Taiwan}
\altaffiltext{22}{Department of Physics and Astronomy, University College London, WC1E 6BT London, UK}
\altaffiltext{23}{School for Physical Sciences, The Open University, Walton Hall, Milton Keynes MK7 6AA, United Kingdom, and RALSpace, The Rutherford Appleton Laboratory, Chilton, Didcot, OX11 0QX, United Kingdom}
\altaffiltext{24}{Division of Theoretical Astronomy, National Astronomical Observatory of Japan, Mitaka, Tokyo 181-8588, Japan}
\altaffiltext{25}{Leiden Observatory, Leiden University, PO Box 9513, 2300 RA Leiden, The Netherlands}
\altaffiltext{26}{Instituto de Astrof\'isica de Canarias, E-38205 La Laguna, Tenerife, Canary Islands, Spain}
\altaffiltext{27}{Universidad de La Laguna, Dpto.\ Astrof\'isica, E-38206 La Laguna, Tenerife, Spain}
\altaffiltext{28}{SOFIA Science Center, Universities Space Research Association, NASA Ames Research Center, Moffett Field, California 94035, USA}
\altaffiltext{29}{Universit\'e Paris-Saclay, CNRS, CEA, Astrophysique, Instrumentation et Mod\'elisation de Paris-Saclay, 91191 Gif-sur-Yvette, France}
\altaffiltext{30}{Department of Physics and Astronomy, Graduate School of Science and Engineering, Kagoshima University, 1-21-35 Korimoto, Kagoshima, Kagoshima 890-0065, Japan}
\altaffiltext{31}{Physikalisches Institut, University of Cologne, Z\"{u}lpicher Str.\ 77, D-50937 K\"{o}ln, Germany}
\altaffiltext{32}{Division of ALMA, National Astronomical Observatory of Japan, Mitaka, Tokyo 181-8588, Japan}
\altaffiltext{33}{University of South Carolina, Columbia, SC 29208, USA}
\altaffiltext{34}{Physics and Astronomy, University of Exeter, Stocker Road, Exeter EX4 4QL, UK}
\altaffiltext{35}{Indian Institute of Astrophysics, II Block, Koramangala, Bengaluru 560034, India}
\altaffiltext{36}{CAS Key Laboratory of FAST, National Astronomical Observatories, Chinese Academy of Sciences, People's Republic of China}
\altaffiltext{37}{National Astronomical Observatories, Chinese Academy of Sciences, A20 Datun Road, Chaoyang District, Beijing 100012, People's Republic of China}
\altaffiltext{38}{Indian Institute of Science Education and Research (IISER) Tirupati, Rami Reddy Nagar, Karakambadi Road, Mangalam (P.O.), Tirupati 517507, India}
\altaffiltext{39}{Jeremiah Horrocks Institute, University of Central Lancashire, Preston PR1 2HE, UK}
\altaffiltext{40}{School of Astronomy and Space Science, Nanjing University, 163 Xianlin Avenue, Nanjing 210023, People's Republic of China}
\altaffiltext{41}{Astronomy Program, Department of Physics and Astronomy, Seoul National University, 1 Gwanak-ro, Gwanak-gu, Seoul 08826, Republic of Korea}
\altaffiltext{42}{Key Laboratory for Research in Galaxies and Cosmology, Shanghai Astronomical Observatory, Chinese Academy of Sciences, 80 Nandan Road, Shanghai 200030, People's Republic of China}
\altaffiltext{43}{Shanghai Astronomical Observatory, Chinese Academy of Sciences, 80 Nandan Road, Shanghai 200030, People's Republic of China}
\altaffiltext{44}{Department of Physics, Konan University, Okamoto 8-9-1, Higashinada-ku, Kobe 658-8501, Japan}
\altaffiltext{45}{SOKENDAI (The Graduate University for Advanced Studies), Hayama, Kanagawa 240-0193, Japan}
\altaffiltext{46}{National Radio Astronomy Observatory, 520 Edgemont Road, Charlottesville, VA 22903, USA}
\altaffiltext{47}{Univ.\ Grenoble Alpes, CNRS, IPAG, 38000 Grenoble, France}
\altaffiltext{48}{Division of Science, National Astronomical Observatory of Japan, 2-21-1 Osawa, Mitaka, Tokyo 181-8588, Japan}
\altaffiltext{49}{Department of Physics, University of Bath, Claverton Down, Bath, BA2 7AY, UK}
\altaffiltext{50}{Armagh Observatory and Planetarium, College Hill, Armagh, BT61 9DB, UK}
\altaffiltext{51}{Institute of Astronomy and Department of Physics, National Tsing Hua University, Hsinchu 300044, Taiwan}
\altaffiltext{52}{Jodrell Bank Centre for Astrophysics, School of Physics and Astronomy, The University of Manchester, Manchester M13 9PL, UK.}
\altaffiltext{53}{Astrobiology Center, National Institutes of Natural Sciences, 2-21-1 Osawa, Mitaka, Tokyo 181-8588, Japan}
\altaffiltext{54}{SKA Observatory, Jodrell Bank, Lower Withington, Macclesfield SK11 9FT, UK}
\altaffiltext{55}{Jodrell Bank Centre for Astrophysics, School of Physics and Astronomy, University of Manchester, Oxford Road, Manchester M13 9PL, UK}
\altaffiltext{56}{Department of Earth Science Education, Seoul National University, 1 Gwanak-ro, Gwanak-gu, Seoul 08826, Republic of Korea}
\altaffiltext{57}{SNU Astronomy Research Center, Seoul National University, 1 Gwanak-ro, Gwanak-gu, Seoul 08826, Republic of Korea}
\altaffiltext{58}{Institute of Astronomy and Department of Physics, National Tsing Hua University, Hsinchu 30013, Taiwan}
\altaffiltext{59}{Key Laboratory of Modern Astronomy and Astrophysics (Nanjing University), Ministry of Education, Nanjing 210023, People's Republic of China}
\altaffiltext{60}{University of Chinese Academy of Sciences, Beijing 100049, People's Republic of China}
\altaffiltext{61}{Department of Physics, Graduate School of Science, Nagoya University, Furo-cho, Chikusa-ku, Nagoya 464-8602, Japan}
 
 
\maketitle
 
\begin{abstract}
Dust polarization can trace magnetic fields through aligned grains.
We compared 450 and 850\,\micron\ POL-2 observations from the James Clerk Maxwell Telescope at $12\farcs6$ resolution.
We selected pixels with Stokes~$I$ signal-to-noise ratios above 50 in both bands.
The median absolute difference in polarization angle between bands is $6.7\arcdeg$ across 21 regions.
For total intensity $I\propto\nu^\alpha$ and polarized intensity $PI\propto\nu^\gamma$, 24--27 of 35 regions have $\gamma<\alpha$, allowing for $3\sigma$ limits on weak polarized intensities.
The median $\gamma-\alpha$ is $-0.51$, ranging from $-0.57$ to $-0.45$ under these limits.
The ratio of polarization fractions is $P_{450}/P_{850}=0.69\pm0.02_{\rm sys}\pm0.03_{\rm stat}$, with no detected dependence on the angle difference.
We suggest that the dependence of alignment on grain size offers the simplest explanation of these findings.
For optically thin components sharing a polarization angle and opacity law, temperature mixing reproduces the median $-0.51$ only if the warm dust is polarized less efficiently than the cold dust.
If the cold dust is at 10\,K or warmer and the warm dust at 40\,K or cooler, the median requires a ratio of warm to cold effective polarization efficiencies $r\lesssim0.18$.
We tested mixtures with opacity index $\beta=2$ in four fields against Herschel dust temperatures and Green Bank Ammonia Survey kinetic temperatures.
We assumed equal gas and dust temperatures, ammonia tracing cold gas, and model ranges of 3--20\,K for cold dust and 25--40\,K for warm dust carrying at most half the mass.
Mixtures matching both temperatures cannot reproduce $\gamma-\alpha$ within $1\sigma$ in three fields, including OMC\,2 tentatively, even for $r=0$.
Allowing warm dust at 10--100\,K and up to 95\% of the mass permits solutions only for $r\leq0.041$.
To explain these three fields by temperature mixing, we need wider model ranges and a very strong contrast in polarization efficiency.
The origin of this contrast remains undetermined.
\end{abstract}
 
\section{Introduction}
\label{sec:intro}
 
Magnetic fields ($\vec{B}$-fields) pervade the interstellar medium (ISM) and play a central role in the dynamics of molecular cloud formation, filament evolution, and star formation \citep{Hennebelle2019,Pattle2023PPVII}.
In star-forming regions, submillimeter polarimetry of thermal dust emission is the primary means of mapping $\vec{B}$-field morphology in the plane of the sky, complementing Zeeman measurements of the component along the line of sight \citep{Crutcher2012}.
Recent observations of nearby molecular clouds have established that field strengths in the plane of the sky lie between $\sim$1\,$\mu$G in diffuse envelopes and $\sim$10\,mG in the densest star-forming cores; in the OMC\,1 integral-shaped filament, Davis--Chandrasekhar--Fermi analysis yields $\sim$6\,mG \citep{Pattle2017,Pattle2023PPVII}.
The \textit{Planck} all-sky survey and column density maps from \textit{Herschel} provide the environmental context on larger scales for interpreting these results \citep{PlanckIntXIX2015,Andre2010}.
Single-dish submillimeter polarimeters such as JCMT/POL-2 bridge these cloud scales and the scales resolved by interferometers such as the Atacama Large Millimeter/submillimeter Array (ALMA), from several $\times 1000$\,au down to a few $\times 10$\,au \citep{Hull2019}.
For the present sample, the $12\farcs6$ POL-2 beam corresponds to $\sim$0.008--0.5\,pc at the adopted reference distances (section~\ref{sec:obs}; subsection~\ref{ss:Discussion_GammaAlpha}).

Submillimeter emission polarimetry probes magnetic fields in cold ($T \lesssim 100$\,K) gas through the alignment of non-spherical dust grains with their major axes perpendicular to the local $\vec{B}$-field direction \citep{DavisGreenstein1951}.
Thermal emission from these aligned grains is polarized, allowing us to infer the orientation of the $\vec{B}$-field in the plane of the sky.
These dust grains are a mixture of silicates and carbonaceous species comprising $\sim$1\% of the ISM by mass \citep{Draine2003}.
These grains emit optically thin radiation whose intensity traces the column density \citep{Hildebrand1983}.

On the James Clerk Maxwell Telescope (JCMT), the POL-2 polarimeter \citep{Friberg2016} and the Submillimetre Common-User Bolometer Array~2 (SCUBA-2) camera \citep{Holland2013} together provide submillimeter polarimetry over a wide field at two wavelengths (450/850\,$\mu$m).
The high dynamic range of this system enables polarization measurements across hydrogen column densities of $N_{\rm H} \sim 10^{20}$--$10^{24}$\,cm$^{-2}$, and its $\sim$3\arcmin\ field of view allows efficient mapping of extended structures.
The ongoing B-fields In STar-forming Region Observations (BISTRO) Large Program \citep{Ward-Thompson2017} has systematically mapped magnetic field morphology in various star-forming regions.
Observations at 450\,$\mu$m remain challenging because of atmospheric variability.
The companion technical paper (Paper~T) presents methods to establish and validate reliable 450 \, $\mu$m polarimetry and to quantify its systematic floors.

Optical and near-infrared extinction polarimetry differ conceptually from submillimeter emission polarimetry.
In extinction polarimetry, the polarization fraction $p_V$ and extinction $A_V$ both scale with the column density of dust along the line of sight.
Their ratio $p_V/A_V$ therefore cancels the dust column and isolates the intrinsic polarizing efficiency of the grain population, a quantity traditionally termed ``polarization efficiency'' \citep{Serkowski1975,Whittet2008}.
In emission polarimetry, in contrast, the polarized intensity $PI_\nu$ and the total intensity $I_\nu$ are both weighted by temperature through the Planck function $B_\nu(T)$.
Grains of different sizes reach distinct equilibrium temperatures, and their relative contributions vary with wavelength.
The ratio $P_\nu = PI_\nu/I_\nu$ therefore depends on the alignment efficiency, optical properties of the grains, and temperature structure along the line of sight.
Consequently, submillimeter emission polarimetry does not have a simple analog of $p_V/A_V$ that isolates alignment efficiency from other factors.

The most widely accepted theory for the alignment of dust grains is the Radiative Alignment Torque (RAT) mechanism \citep{Lazarian2007,LazarianHoang2007}, in which irregularly shaped dust grains are spun up by anisotropic radiation fields \citep{Andersson2015}.
Alignment efficiency should decrease as extinction increases because the RAT mechanism aligns grains through incident radiation \citep{Lazarian2007,Andersson2015}.
This trend appears as the \emph{polarization hole}: the polarization fraction $P$ decreases with increasing column density $N$.
The phenomenon was characterized in dark clouds by \citet{Goodman1995} and has been confirmed in near-infrared extinction polarimetry, as a decline in polarization efficiency with extinction \citep{Whittet2008}, and in emission polarimetry at submillimeter wavelengths \citep{Alves2014,Kwon2018,Soam2018,Coude2019}.

Both techniques integrate along the line of sight, and variations in $\vec{B}$-field orientation along that path can reduce the net polarization in either technique.
Extinction polarimetry samples a pencil beam toward each background star, whereas emission polarimetry additionally averages over the finite beam of the telescope.
Unresolved variations across this beam in the plane-of-sky $\vec{B}$-field orientation can further reduce the observed polarization \citep{Whittet2008}.

We argue that this simple picture of the polarization hole may not apply universally.
Near embedded protostars, the local radiation energy density can exceed that of the interstellar radiation field in the solar neighborhood by orders of magnitude \citep{Hoang2021}.
\citet{Whittet2008} and \citet{Coude2019} suggest that radiation from embedded young stellar objects may help sustain grain alignment at high extinction.
Such enhanced internal radiation fields may weaken or reverse the classical negative $P$--$N$ correlation.
Radiative torques can also disrupt large grains near protostars, reducing the polarization fraction and potentially counteracting this effect \citep{Hoang2021}.
With the survey at two wavelengths analyzed here, we test whether the $P$--$N$ correlation is positive or negative in actively star-forming regions and whether this behavior differs between wavelengths.

RAT theory predicts a strong increase in alignment efficiency with grain size, because larger grains experience greater radiative torques relative to gas damping \citep{LazarianHoang2007,HoangLazarian2016}.
This size dependence has direct implications for multiwavelength polarimetry.
If large grains align more efficiently than small grains, and if different grain sizes contribute preferentially at different wavelengths, then polarization properties should exhibit a systematic wavelength dependence.
 
In the models of \citet{Ysard2019}, silicate grains containing aromatic-rich amorphous carbon have a flatter far-infrared opacity spectrum than bare amorphous silicates, whereas in the models of dust evolution by \citet{Kohler2015}, the accretion of aliphatic-rich carbon mantles and coagulation into aggregates steepen it.
How grain composition and evolution change the spectral index of the far-infrared opacity is therefore not settled.
Dust models of the diffuse ISM also differ in whether silicate and carbonaceous material form separate grain populations \citep[e.g.,][]{Guillet2018} or a single composite material \citep{Hensley2023}.

If polarized emission is dominated by well-aligned large grains with a shallower effective emissivity spectrum, then the spectral index weighted by polarization should differ systematically from that weighted by mass.

Simple dust models predict that the submillimeter polarization fraction varies only weakly with wavelength \citep{Hildebrand1999}.
However, observations have revealed a complex polarization spectrum (the polarization fraction as a function of wavelength, $P(\lambda)$) with structure between 60\,$\mu$m and 1.3\,mm \citep{Vaillancourt2008}.
In contrast, flat submillimeter polarization spectra were found in Vela~C \citep{Gandilo2016} and in the Carina Nebula \citep{Shariff2019}.
We suggest that part of this diversity may arise from combining measurements obtained with different instruments, beam sizes, spatial filtering, and observing conditions.
Simultaneous measurements in two bands with matched beams, as analyzed here, reduce this ambiguity.
The THEMIS~2.0 dust model, designed for the high-latitude diffuse ISM, can produce a flat, slightly increasing, or slightly decreasing polarization fraction from the far-infrared to the submillimeter \citep{Ysard2024}.
Which of these trends the model produces depends on the mass ratio and alignment of its silicate and carbonaceous grain populations.

\citet{Fanciullo2022b} combined Stratospheric Observatory for Infrared Astronomy (SOFIA)/High-resolution Airborne Wideband Camera Plus (HAWC+) (154, 214\,$\mu$m) and JCMT/POL-2 (850\,$\mu$m) observations of NGC~2071.
They found that the polarization fraction decreases from 154 to 214\,$\mu$m, a ratio that their dust models with uniform alignment efficiency could \emph{not} reproduce.
They noted that a cold, poorly aligned dust component along the line of sight could lower this ratio, and that the dust models they used were developed for regions of moderate column density.
\citet{Michail2021} provided complementary constraints in Orion using HAWC+ polarimetry.
These studies focus on individual sources.
To test whether the observed behavior is universal or specific to each source, we analyze a survey of diverse environments.

We use three observational diagnostics to constrain how grain alignment depends on size without explicitly normalizing by dust column:
the ratio of polarization fractions $P_{450}/P_{850}$, which quantifies the wavelength dependence of the polarization fraction;
the difference in spectral indices $\gamma - \alpha$, which contrasts the wavelength dependence of polarized and total intensity;
and the $P$--$\mathcal{S}$ correlation between the polarization fraction $P$ and the dispersion function $\mathcal{S}$ of polarization angles, which constrains geometric depolarization from tangling of $\vec{B}$-fields.
These diagnostics serve a role similar to that of $p_V/A_V$ in extinction polarimetry.
These three quantities describe differences with wavelength and therefore depend less on the absolute column normalization than polarization fractions measured in a single band.

We pose three questions.
 
First, does submillimeter polarimetry trace a single $\vec{B}$-field geometry in the plane of the sky across wavelengths, or do different bands sample different magnetized structures?
For the 450/850\,$\mu$m comparison, we ask whether the two bands yield consistent polarization angles.
Similar field orientations at the two wavelengths would disfavor geometric effects that rotate the polarization angles between the bands as the origin of differences in polarization fraction.
 
Second, how does the spectral index of polarized emission compare with that of total emission across diverse star-forming environments?
At 450/850\,$\mu$m, this question contrasts the spectral index $\gamma$ of polarized intensity with the index $\alpha$ of total intensity.
The prediction that well-aligned large grains exhibit a shallower effective emissivity spectrum than the bulk grain population implies $\gamma < \alpha$.
For a given line of sight, $\gamma < \alpha$ is equivalent to $P_{450}/P_{850} < 1$ (equation~\ref{eq:Pratio_SpIdx}), that is, to a polarization fraction that is higher at 850 than at 450\,$\mu$m.
 
Third, is the wavelength dependence of the submillimeter polarization fraction set by intrinsic grain physics, or is it shaped by tangling of $\vec{B}$-fields along the line of sight?
At 450/850\,$\mu$m, we ask whether $P_{450}/P_{850}$ correlates with the difference in polarization angles between the bands, $|\chi_{450}-\chi_{850}|$, where $\chi_{\lambda}$ denotes the polarization angle measured in the plane of the sky at wavelength $\lambda$.
If $P_{450}/P_{850} < 1$ reflects intrinsic grain physics, such a correlation should not appear.
 
To address these three questions, we systematically analyze the 450 and 850\,$\mu$m data for the statistical subsample drawn from the broader survey of 69 sources presented in the companion census paper (Paper~S2).
This analysis requires data for the same positions in both bands, reduced consistently.
The 450\,$\mu$m sensitivity must allow comparisons of polarization fraction, spectral indices, and $\vec{B}$-field orientation at matched resolution.
Section~\ref{sec:obs} therefore defines the sample and data products needed to construct these diagnostics.

Section~\ref{sec:results} presents the three findings: consistency of $\vec{B}$-field angles (subsection~\ref{ss:bfield_consistency}; F1), ratios of polarization fractions (subsection~\ref{ss:P_ratio}; F3), and spectral-index analysis (subsection~\ref{ss:SpIdx}; F2).
Section~\ref{sec:discussion} develops the physical interpretation in terms of the dependence of grain alignment on size, the response to environment, and implications for RAT theory.
Subsections~\ref{ss:Discussion_Tmixing} and~\ref{ss:Discussion_DiscriminatingPower} then examine alternatives to this interpretation, including temperature mixing along the line of sight (appendix~\ref{app:Tmixing}).
Section~\ref{sec:summary} summarizes the three principal findings.
The method for determining distances is described in Paper~S2 appendix~\ref{S2-app:distance}; appendix~\ref{app:12CO} assesses $^{12}$CO($J$=3--2) contamination.

The three papers together address a broader question: what governs the wavelength dependence of submillimeter dust polarization in star-forming regions and how does it change with environment?
Paper~T establishes the calibration and noise characterization, demonstrates the correction of additive zero-level offsets in the 450\,$\mu$m Stokes~$Q$ and~$U$ intensities, and validates it against independent SOFIA/HAWC+ data at 154 and 214\,$\mu$m.
Paper~T also quantifies the floors of systematic uncertainty in the measurements of both bands used here.

The present paper (Paper~S1) identifies signatures of the wavelength dependence, and Paper~S2 extends the F1--F3 findings presented here to a unified F1--F6 program.
In Paper~S2, F4 establishes the robustness of F1--F3 for the population represented by the BISTRO {\sc PolSrc} sample.
F5 tests whether the wavelength dependence varies between environments.
The medians of $\gamma-\alpha$ and $R_p$ for each environment show no ordered sequence from low-mass star-forming (LMSF) through high-mass star-forming (HMSF) to Central Molecular Zone (CMZ) regions.
We do not detect strong differences between the classes, although the high-mass class has the lowest median.
F6 describes how associations between polarization and environment vary with wavelength within the Herschel Gould Belt Survey (HGBS) subsample of 9 fields.
 
\begin{figure*}[ht]
\centering
\includegraphics{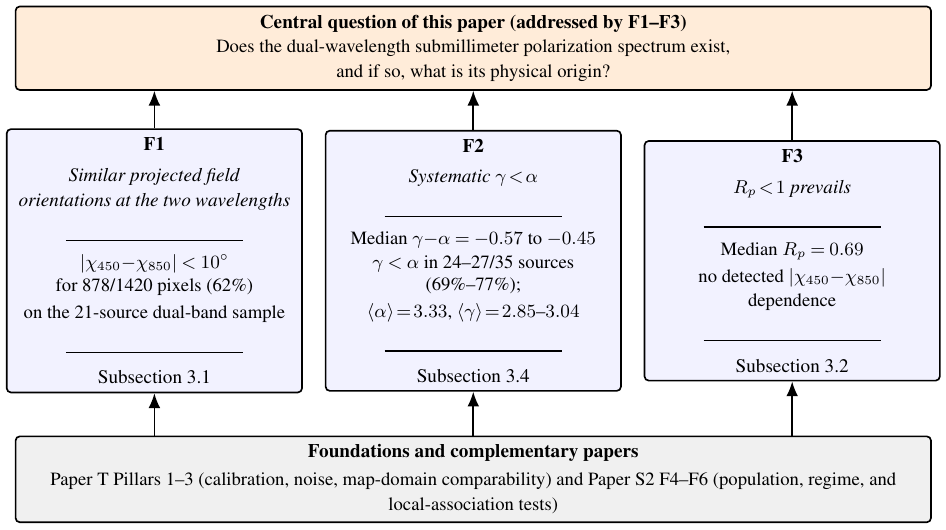}
\caption{
\textbf{Overview of the three findings in Paper~S1 (F1--F3 of the unified F1--F6 program).}
The central question of this paper---the existence and physical origin of the submillimeter polarization spectrum at two wavelengths---is addressed through three complementary observational findings.
Each finding uses the sample defined for its diagnostic in the text.
\textit{F1 (similar projected field orientations at the two wavelengths):} 878 of 1420 matched pixels (62\%) across 21 sources satisfy $|\chi_{450}-\chi_{850}|<10\degr$, with a median $|\chi_{450}-\chi_{850}|$ of $6.7\degr$.
This supports broadly consistent projected field orientations at the common resolution (subsection~\ref{ss:bfield_consistency}).
\textit{F2 (systematic} $\gamma<\alpha$): the spectral index $\gamma$ of polarized intensity is smaller than the index $\alpha$ of total intensity in 24--27 of 35 sources (median $\gamma-\alpha$ between $-0.57$ and $-0.45$).
We interpret this as a lower effective emissivity index $\beta$ for the grain population weighted by polarization (subsection~\ref{ss:SpIdx}).
\textit{F3 ($R_p<1$ prevails):} the ratio of polarization fractions $R_p \equiv P_{450}/P_{850} = 0.69$ (a polarization fraction typically higher at 850 than at 450\,$\mu$m) shows no detected dependence on the difference in polarization angles between the bands.
This disfavors geometric explanations that would produce such a dependence (subsection~\ref{ss:P_ratio}).
The three findings rest on the technical basis established in Paper~T (Pillars~1--3), and Paper~S2 extends them with tests across the population in F4--F6.
}
\label{fig:S1_findings_overview}
\end{figure*}

Figure~\ref{fig:S1_findings_overview} and table~\ref{tbl:S1_findings_index} summarize the three findings.
Figure~\ref{fig:S1_findings_overview} connects each question to its observational finding.
Table~\ref{tbl:S1_findings_index} identifies where we present each finding and the related sections in Papers~T and~S2: the Paper~T Pillar on which it relies, or the Paper~S2 finding that extends it.
 
\begin{longrotatetable}
\begin{deluxetable*}{lcllc}
\tablecaption{Index of the three findings in Paper~S1 and related material in Papers~T and~S2\label{tbl:S1_findings_index}}
\tablewidth{0pt}
\tablehead{
  \colhead{Topic} &
  \colhead{Finding} &
  \colhead{Location in this paper series} &
  \colhead{Key result / figure} &
  \colhead{Note}
}
\startdata
\multicolumn{5}{c}{\textit{Paper~S1 --- Findings F1--F3}} \\
\hline
$\vec{B}$-field angle agreement & F1 & Subsection~\ref{ss:bfield_consistency} & $\widetilde{|\Delta\chi|}$ within $10\degr$ in 10 of 13 regions ($N_{\rm samp}=1420$, 21 sources) & 1 \\
Ratio of polarization fractions $R_p$ & F3 & Subsection~\ref{ss:P_ratio} & sample median of $R_p = 0.69$, no detected $|\Delta\chi|$ dependence & 2 \\
Spectral indices $\alpha$, $\gamma$ & F2 & Subsection~\ref{ss:SpIdx} & median $\gamma-\alpha = -0.57$ to $-0.45$; $\gamma<\alpha$ in 24--27/35 sources (69\%--77\%) & 3 \\
\hline
\multicolumn{5}{c}{\textit{Paper~S1 --- Discussion synthesis}} \\
\hline
Synthesis of the three findings & --- & Section~\ref{sec:discussion} & size-dependent grain alignment (consistent with RAT) & 4 \\
\hline
\multicolumn{5}{c}{\textit{Material in Papers~T and~S2}} \\
\hline
Pillar~1 [flux conversion factor (FCF) + $C(T_{\rm d})$ correction] & --- & Paper~T~subsection~\ref{T-ss:flux} & provides $\alpha$, $\gamma$ used in F2 & 5 \\
Pillar~2 (Stokes\,$I$ + $Q,U$ noise model) & --- & Paper~T~subsections~\ref{T-ss:noiseI}--\ref{T-ss:noisemodel} & provides per-pixel detection masks used in F1, F3 & 6 \\
Pillar~3 (offset corr.\ + beam harmonization) & --- & Paper~T~subsections~\ref{T-ss:offset}--\ref{T-ss:beamval} & provides matched-resolution maps used throughout & 7 \\
F4: Population-level universality of F1--F3 & --- & Paper~S2 & {\sc PolSrc} sample confirms F1--F3 are sample-wide & 8 \\
F5: Regime-dependent wavelength response & --- & Paper~S2 & no ordered environmental sequence; no robust difference among classes & 9 \\
F6: Wavelength-dependent polarization--environment associations & --- & Paper~S2 & HGBS 9-field per-pixel $N_{\rm H}, T_d$ correlations & 10 \\
\enddata
\tablecomments{
The ``Finding'' column identifies F1--F3 in this paper and cross-references F4--F6 in Paper~S2.
The entry ``---'' denotes synthesis or material from Papers~T and~S2.
Topics follow their order in the text.
(1)~Angle comparison for 21 sources at $I/\sigma_I>30$ in both bands (brightest 10\% of the pixels of each source), establishing the adopted $10\arcdeg$ tolerance (subsection~\ref{ss:bfield_consistency}).
(2)~Comparison of the three estimators defined in Paper~S2 subsection~\ref{S2-ss:S2_P_ratio}.
We adopt the median of pixel ratios, $R_{\rm EW}$, for the results reported for individual sources (subsection~\ref{ss:definitions}).
(3)~Spectral indices for individual sources in the {\sc SpIdx} sample and cancellation of the common Planck shape factor $C(T_{\rm d})$ in $\gamma-\alpha$.
(4)~Interpretation of F2 and F3 in terms of alignment that depends on grain size and a shallower effective emissivity spectrum for the aligned grain population, subject to the geometric constraints from F1.
(5)~Extended-source FCF calibration and the $C(T_{\rm d})$ correction, $\beta=\alpha-C(T_{\rm d})$ (equation~\ref{eq:planck_correction}).
(6)~Two-component noise model and isotropy tests supporting the detection masks used in F1 and F3.
(7)~Three-step correction of the 450\,$\mu$m Stokes-$Q/U$ zero-level offset based on principal component analysis (PCA), and beam harmonization for the Stokes-$I,Q,U$ maps.
(8)~Population tests of F1--F3, with sample medians $R_{\rm EW}=0.69$ and $\gamma-\alpha$ between $-0.57$ and $-0.45$ and the comparison of angles between the bands in subsection~\ref{ss:bfield_consistency}.
(9)~LMSF/HMSF/CMZ comparisons: we do not detect an ordered sequence in $\gamma-\alpha$ or $R_p$, and the Kruskal--Wallis test does not robustly detect a difference among the groups.
(10)~Correlations with $N_{\rm H}$ and $T_d$ derived from HGBS in nine fields, testing how the wavelength dependence relates to local conditions.
}
\end{deluxetable*}
\end{longrotatetable}

\begin{longrotatetable}
\begin{deluxetable*}{cllllllc}
\tablecaption{Quantitative summary of findings F1--F3 in Paper~S1.\label{tbl:S1_F1F3_master}}
\tablewidth{0pt}
\tablehead{
  \colhead{Label} &
  \colhead{Observable} &
  \colhead{Sample} &
  \colhead{Result} &
  \colhead{Stat.\ unc.} &
  \colhead{Sys.\ floor} &
  \colhead{Section in this paper} &
  \colhead{Limitations}
}
\startdata
F1 &
\parbox[t]{2.8cm}{\raggedright $|\chi_{450}-\chi_{850}|$ inter-band agreement\strut} &
\parbox[t]{2.5cm}{\raggedright 21-source dual-band\\ ($N_{\rm samp}=1420$)\strut} &
\parbox[t]{3.5cm}{\raggedright $\widetilde{|\Delta\chi|}<10\arcdeg$ in 10 of 13 targets (overall $6.7\arcdeg$)\strut} &
\parbox[t]{2.8cm}{\raggedright per-pixel $\sigma_{\Delta\chi}\approx 3.5\arcdeg$\strut} &
\parbox[t]{3.0cm}{\raggedright $4$--$6\arcdeg$ floor\\ (Paper~T subsection~\ref{T-ss:beamval})\strut} &
Subsection~\ref{ss:bfield_consistency} &
\parbox[t]{3.0cm}{\raggedright {Mon~R2: sensitivity to the $Q/U$ zero-level offset correction (median $|\Delta\chi|$ of $12\arcdeg$ before and $4.0\arcdeg$ after it)}\strut} \\
F2 &
\parbox[t]{2.8cm}{\raggedright $\gamma-\alpha$ spectral-index difference\strut} &
\parbox[t]{2.5cm}{\raggedright 35-source {\sc SpIdx}\strut} &
\parbox[t]{3.5cm}{\raggedright sample median $\gamma-\alpha = -0.57$ to $-0.45$, 24--27/35 sources $\gamma<\alpha$\strut} &
\parbox[t]{2.8cm}{\raggedright {empirical SEM $0.10$ (35 sources)}\strut} &
\parbox[t]{3.0cm}{\raggedright $\lesssim 0.04$\\ (Paper~T table~\ref{T-tbl:systematics})\strut} &
Subsection~\ref{ss:SpIdx} &
\parbox[t]{3.0cm}{\raggedright Line-of-sight temperature mixing with a polarization-efficiency contrast (subsection~\ref{ss:Discussion_Tmixing})\strut} \\
F3 &
\parbox[t]{2.8cm}{\raggedright $R_p \equiv P_{450}/P_{850}$ ratio\strut} &
\parbox[t]{2.5cm}{\raggedright {30 {\sc SpIdx} sources with $R_{\rm EW}$\\ ({\sc DualBand}: 417 pixel pairs)}\strut} &
\parbox[t]{3.5cm}{\raggedright $R_{\rm EW} = 0.69$ sample median\strut} &
\parbox[t]{2.8cm}{\raggedright {empirical SEM $0.03$ (30 sources)}\strut} &
\parbox[t]{3.0cm}{\raggedright $\lesssim 0.02$\\ (Paper~T table~\ref{T-tbl:systematics})\strut} &
Subsection~\ref{ss:P_ratio} &
\parbox[t]{3.0cm}{\raggedright {OMC\,1 high-radiation regime; cancellation in $R_p$ is a limiting model}\strut} \\
\enddata
\tablecomments{
Quantitative summary of F1--F3, using the format of the F4--F6 summary in Paper~S2 table~\ref{S2-tbl:S2_F4F6_master}.
The F1 statistical entry is the typical formal uncertainty of an individual angle difference between the bands, not the standard error of a regional or sample median.
The F2 and F3 statistical entries are Maritz--Jarrett standard errors of the medians over 35 and 30 source entries, respectively; ``empirical SEM'' (standard error of the median) applies this estimator to the values for individual sources rather than to pixels within one source.
Interquartile ranges additionally describe the non-normal distributions of the values for individual sources.
The F2 range $-0.57$ to $-0.45$ spans the two extreme placements of the $3\sigma$ limits and is not a confidence interval.
The F2 and F3 systematic entries are the contributions of the terms quantified in Paper~T (calibration and residual 450\,$\mu$m Stokes-$Q/U$ zero-level offsets) to the uncertainties of the sample medians, taken through table~\ref{tbl:S1_dependency}.
They do not include the unevaluated dependence on beam matching and spatial filtering.
The F2 and F3 systematic entries also do not include a difference in the modulation efficiency of POL-2 between the two bands (Paper~T~subsection~\ref{T-ss:flux}).
Table~\ref{tbl:S1_findings_index} provides the broader index, including discussion and material from Papers~T and~S2.
}
\end{deluxetable*}
\end{longrotatetable}
 
Table~\ref{tbl:S1_F1F3_master} gives the quantitative summary of findings F1--F3.
The index of related material appears as table~\ref{tbl:S1_findings_index} above.
\section{Data and Methods}
\label{sec:obs}

\subsection{Data}
\label{ss:data}
 
We selected data from POL-2 observations available in the JCMT archive as of 2022 February.
We limited the analysis of 450\,$\mu$m data to observations taken under the highest-quality Weather Band (WB~1), corresponding to an atmospheric opacity $\tau_{225\,{\rm GHz}}^{\rm zenith} < 0.05$ at the zenith.
We also required a peak Stokes~$I$ signal-to-noise ratio $> 100$ at 450\,$\mu$m for polarization analysis in both bands.
The data in the two bands are co-spatial and co-temporal, because SCUBA-2 simultaneously acquires data at both 450 and 850\,$\mu$m in every observation.
The 450/850 calibration therefore largely cancels in the band ratios used here.
The resulting sample comprises 69 distinct star-forming regions (the {\sc PolSrc} sample).
Paper~S2 appendix~\ref{S2-app:S2_survey_fields} (table~\ref{S2-tab:surv}) lists the complete sample, with coordinates, distances, and references to archival POL-2 observations for each source; Paper~T table~\ref{T-tbl:T_sample_inventory} shows how this sample is nested within the survey.

These three selection criteria inherently bias the sample toward nearby star-forming regions with high surface brightness and strong continuum emission.
Our statistical results therefore describe the polarization properties of the brightest subregions within molecular clouds, predominantly dense cores and filaments actively forming stars.
They do not represent a volume-complete or flux-limited census of star-forming environments in the solar neighborhood.
More diffuse or weakly illuminated regions are observationally underrepresented, because their polarized emission at the shorter wavelength falls below the practical sensitivity floor of the present JCMT/POL-2 survey.
We do not test here whether the trends derived below extend to more diffuse material.

We processed all regions analyzed in this paper within the shared reduction and calibration framework of Paper~T, using a single calibration convention and a common procedure to match the beams and spatial frequencies.
Paper~T documents the reduction settings for each field and where the 450\,$\mu$m Stokes-$Q/U$ zero-level offset correction applies; the adopted maps include mosaics to which that correction was not applied (subsection~\ref{ss:bfield_consistency}).
We reduced both wavelengths with the \texttt{pol2map} pipeline within \texttt{Starlink}, following the validated three-step procedure for POL-2 data (\citealt{Parsons2021}; e.g., \citealt{Ward-Thompson2017,Pattle2021}).
The \texttt{pol2map} pipeline makes the maps with the iterative map maker \texttt{makemap}~\citep{Chapin2013}.
To ensure that the 450 and 850\,$\mu$m maps sample identical angular scales, we adopted \texttt{flt.filt\_edge\_largescale=300}, which retains structure smaller than $\sim$300\arcsec.
We also adopted \texttt{smooth450=yes}, which deconvolves the double-Gaussian 450\,$\mu$m beam and reconvolves to the 12\farcs6 effective 850\,$\mu$m beam.
The common range of angular scales sampled is thus 12\farcs6--300\arcsec.
We regridded all Stokes parameters onto a common coordinate system with 12\arcsec\ pixels and matched them to the 850\,$\mu$m beam resolution for the analysis of both bands.
We paired the 450 and 850\,$\mu$m pixels one to one, taking the closest pairs within 7\arcsec.
In the two Galactic Center fields GC~North and GC~South, the pixel grids of the two bands are offset from each other, and the paired pixels there are about 6\arcsec\ apart.
The use of AI tools in the analysis and in the preparation of the text is described in the Acknowledgments.

Paper~T (section~\ref{T-sec:techresults}) describes flux calibration, conversion to surface brightness, noise characterization, and the budget of systematic uncertainties.
It also presents the $Q/U$ zero-level offset correction and its independent validation with SOFIA/HAWC+ at 154 and 214\,$\mu$m.
In the present paper, the residual $Q/U$ zero-level offset enters only through the systematic floor $\delta X_{\rm sys}$ in the unified uncertainty notation of subsection~\ref{ss:uncertainty}.

We calculated the debiased polarization fractions with the modified asymptotic estimator \citep{Plaszczynski2014,Montier2015}, as described in Paper~T appendix~\ref{T-app:mas_debias}.
We assess potential $^{12}$CO($J$=3--2) contamination of the 850\,$\mu$m continuum band in appendix~\ref{app:12CO}.
The inferred contamination is comparable to or below the uncertainty in the absolute calibration of the continuum data and is therefore unlikely to affect the principal polarization results.

We assigned distances to all survey fields using a hierarchy of catalogs derived from Gaia and of literature values for well-studied sources.
Paper~S2 appendix~\ref{S2-app:distance} describes the procedure.
\label{ss:distance}%
We use these distances as reference scaling factors to convert angular measurements to physical units, not as precise distances, because the clouds studied here are spatially extended along the line of sight.
The primary analyses presented in this paper do not require a distance conversion.
The analyses use ratios or angular quantities ($P_{450}/P_{850}$, $\gamma - \alpha$, $|\chi_{450} - \chi_{850}|$, and rank correlation statistics), whose calculation does not contain a scaling factor that depends on distance.
An error in the adopted distance therefore does not propagate into these quantities.
 
The absence of a distance factor concerns only their calculation; it does not remove differences in physical resolution.
The same $12\farcs6$ beam subtends physical scales that vary by a factor of $\sim$60 across the sample.
Whether a given observable is sensitive to the difference in scale is a separate, empirical question.
This question is answered in Paper~S2, where a trend with distance is detected for $\beta$ and none is detected for $\gamma-\alpha$.
Table~\ref{tbl:S1_dependency} summarizes the technical inputs from Paper~T and the findings F4--F6 for the population studied in Paper~S2, which together provide the basis for the F1--F3 analyses of section~\ref{sec:results}.
\begin{longrotatetable}
\begin{deluxetable*}{lll}
\tablecaption{Inputs from Papers~T and~S2 (basis of the F1--F3 analyses of section~\ref{sec:results})\label{tbl:S1_dependency}}
\tablewidth{0pt}
\tablehead{
  \colhead{Input from Paper~T or~S2} &
  \colhead{Source} &
  \colhead{Used in this paper}
}
\startdata
Extended-source FCF + $C(T_{\rm d})$-corrected $\beta$ (Pillar~1) &
Paper~T subsection~\ref{T-ss:flux} &
Subsection~\ref{ss:SpIdx} ($\alpha, \gamma$ definitions for F2) \\
Stokes-$I$ + $Q,U$ analytic noise model (Pillar~2) &
Paper~T subsections~\ref{T-ss:noiseI}--\ref{T-ss:noisemodel} &
per-pixel detection masks for F1 (subsection~\ref{ss:bfield_consistency}) and F3 (subsection~\ref{ss:P_ratio}) \\
450\,$\mu$m Stokes-$Q/U$ zero-level offset correction + beam harmonization (Pillar~3) &
Paper~T subsections~\ref{T-ss:offset}--\ref{T-ss:beamval} &
matched-resolution dual-band maps used throughout F1, F2, F3 \\
Systematic-uncertainty floor (quantified terms) $\sigma(P_{450}/P_{850}) \lesssim 0.02$, $\sigma(\gamma{-}\alpha) \lesssim 0.04$ &
Paper~T subsection~\ref{T-sec:uncertainty}, table~\ref{T-tbl:systematics} &
$\delta X_{\rm sys}$ entries throughout section~\ref{sec:results} \\
F4--F6 population-level extension &
Paper~S2 section~\ref{S2-sec:S2_results} &
Section~\ref{sec:discussion} integrated synthesis \\
\enddata
\tablecomments{
The four inputs from Paper~T provide the maps at matched resolution and the uncertainty framework used in section~\ref{sec:results}.
Paper~S2 provides the population tests used to discuss grain alignment in section~\ref{sec:discussion}.
Paper~S2 table~\ref{S2-tbl:S2_dependency} lists the corresponding inputs to F4--F6.
}
\end{deluxetable*}
\end{longrotatetable}
 
\subsection{Definitions and notation}
\label{ss:definitions}
We define all quantities used below, including those shared with Papers~T and~S2.
\begin{description}
\item[$\chi_\lambda$] Polarization position angle, measured east
  of north in the range $0\degr$--$180\degr$; the inferred projected $\vec{B}$-field
  orientation is $\chi_\lambda + 90\degr$ (subsection~\ref{ss:bfield_consistency}).
\item[$\Delta\chi_{\rm obs} \equiv \chi_{450}-\chi_{850}$] Observed difference in polarization angle between the bands at each pixel; used to test the field geometry at the two wavelengths in this paper and with the same convention in Paper~S2 (subsection~\ref{ss:bfield_consistency}).
\item[$\sigma_{\Delta\chi}$] Formal statistical uncertainty on $\Delta\chi_{\rm obs}$ for each pixel, propagated from the noise model for Stokes~$Q,U$ in Paper~T~subsection~\ref{T-ss:noisemodel} (median over the survey: $\sigma_{\Delta\chi} \approx 3.5\degr$; subsection~\ref{ss:bfield_consistency}).
\item[$\widetilde{|\Delta\chi|}$] The tilde denotes a median and the vertical bars an absolute value taken before that median.
For $\widetilde{|\Delta\chi|}$ the median is taken over the matched pixels within the \emph{target} specified in the text or figure caption.
A target may be an individual source or a regional group of physically associated sources, as in figure~\ref{fig:PA450vsPA850}(b).
Paper~S2 retains this meaning for the measurements of angle differences in its figure~\ref{S2-fig:env_ladder}(c).
In its table~\ref{S2-tbl:env_4metric}, tildes on other quantities denote medians of the values for individual sources within each environmental class, as specified in that table's notes.
The technical residual $\widetilde{|\Delta\chi_{\rm tech}|}$ characterized in Paper~T is instead a median over the matched pixels of an individual dual-band source.
The quantity $\widetilde{|\Delta\chi|}$ is a robust summary of the raw observed angle differences only.
We subtract \emph{no} statistical or systematic uncertainty from it, so it should not be read as an intrinsic angle difference or one corrected for noise.
Comparisons of $\widetilde{|\Delta\chi|}$ with $\sigma_{\Delta\chi}$ and with the $10\degr$ tolerance scale are therefore comparisons between an uncorrected observed quantity and the two separate reference scales defined here and in subsection~\ref{ss:uncertainty}.
\item[$\Delta\chi_{\rm tech}$] Residual in polarization angle induced by matching the 450\,$\mu$m beam and spatial frequencies.
Paper~T~subsection~\ref{T-ss:beamval} characterizes this residual empirically, with medians for individual sources of $\widetilde{|\Delta\chi_{\rm tech}|} \sim 4\degr$--$6\degr$ and $29$ of the $30$ dual-band sources below $10\degr$ at $I/\sigma_I > 30$.
$\Delta\chi_{\rm tech}$ is a processing floor measured on the 30-source subsample used to validate beam matching in Paper~T.
It is separate from the observed agreement for each target in subsection~\ref{ss:bfield_consistency} ($\widetilde{|\Delta\chi|}<10\degr$ in 10 of the 13 targets analyzed here), from which nothing has been subtracted.
\item[$PI_\lambda$] Debiased polarized intensity at wavelength $\lambda$,
  $PI = \sqrt{Q^2+U^2}$ corrected for the positive bias caused by noise (Ricean bias) with the modified
  asymptotic estimator of \citet{Plaszczynski2014}, in mJy\,arcsec$^{-2}$ (subsections~\ref{ss:P_ratio} and \ref{ss:SpIdx}).
\item[$P_\lambda$] Polarization fraction, $P_\lambda \equiv PI_\lambda/I_\lambda$,
  quoted in percent (subsection~\ref{ss:P_ratio}).
Uppercase $P$ denotes the fraction throughout this series.
Polarized intensity is always written $PI$.
\item[$R_{\rm p}$] Ratio of polarization fractions, $R_{\rm p} \equiv P_{450}/P_{850}$, evaluated for each
  pixel; values for a source or a sample are obtained with the estimators below (subsection~\ref{ss:P_ratio}).
\end{description}
 
Throughout this series, we used three estimators to describe the polarization ratio of the population.
They differ only in the weighting of individual pixels:
\begin{description}
\item[$R_{\rm EW}$] Equal-weighted estimator: the median of the individual pixel
  ratios $P_{450,i}/P_{850,i}$ over pixels with $P\geq3\,\sigma_P$ in both bands that also satisfy the common
  Stokes-$I$ detection mask ($I_\lambda/\sigma_{I,\lambda} > 50$ and the brightest 5\% of the pixels of each source, with both criteria applied at both wavelengths).
  In defining the mask, we do not select pixels by the signal-to-noise ratio of polarized intensity, so that intrinsically
  low-polarization pixels are retained in the mask.
We refer below to the common Stokes-$I$ detection mask, without the floor on $P$, as the $R_{\rm EW}$ mask (equation~\ref{eq:R3}, subsection~\ref{ss:P_ratio}).
\item[$R_{\rm IWM}$] Intensity-weighted mean: the ratio of the intensity-weighted
  mean polarization fractions, $\langle P_{450}\rangle_I / \langle P_{850}\rangle_I$.
  This emphasizes regions of high column density that dominate the mass.
\item[$R_{\rm Corr}$] Correlation estimator: the slope of a weighted
  least-squares regression of $P_{450}$ on $P_{850}$ through the origin.
  This emphasizes bright, well-polarized pixels.
\end{description}
 
We compare these estimators across the full survey in Paper~S2.
Here we use $R_{\rm EW}$ unless otherwise stated.
 
\begin{description}
\item[$I_{450}/I_{850}$] Ratio of the Stokes-$I$ intensities at 450 and 850\,$\mu$m,
  evaluated for each pixel at the matched $12\farcs6$ resolution (subsection~\ref{ss:Iratio}).
\item[$\alpha$] Spectral index of total intensity,
  $\alpha \equiv \log(I_{450}/I_{850})/\log(\nu_{450}/\nu_{850})$ (equation~\ref{eq:spectral_indices}, subsection~\ref{ss:SpIdx}).
\item[$\gamma$] Spectral index of polarized intensity, defined identically with $PI$
  in place of $I$.
The observable of this paper is the difference
  $\gamma - \alpha$, in which the dependence on dust temperature cancels to first
  order (equation~\ref{eq:spectral_indices}, subsection~\ref{ss:SpIdx}).
\item[$\beta$] Dust opacity spectral index of modified-blackbody emission,
  $I_\nu \propto \nu^{\beta} B_\nu(T_{\rm d})$ (equation~\ref{eq:planck_correction}, subsection~\ref{ss:SpIdx}).
\item[$C(T_{\rm d})$] Shape factor of the Planck function between the two POL-2 bands at
  dust temperature $T_{\rm d}$, so that $\alpha = \beta + C(T_{\rm d})$ (equation~\ref{eq:planck_correction}).
\item[$\mathcal{S}$] Polarization-angle dispersion function, the root-mean-square
  angle difference between a pixel and its neighbors within a fixed kernel; it is
  non-negative by construction (equation~\ref{eq:S_def}, subsection~\ref{ss:S_dispersion}).
\item[$\mathcal{S}\times\mathcal{P}^{\rm frac}$] The product of $\mathcal{S}$ with
  the dimensionless polarization fraction $PI/I$, rather than its percentage value.
The product is constructed so that field tangling within the beam largely cancels
  between the two factors (subsection~\ref{ss:S_dispersion}; Paper~S2 subsection~\ref{S2-ss:S2_SP}).
We write the product with $\times$ throughout;
  it is an ordinary product, not a vector operation.
\end{description}

\subsection{Uncertainty notation and precision floors}
\label{ss:uncertainty}
Paper~T collects the systematic floors of the three dual-band observables in its equation~(\ref{T-eq:precision_floors}): $\sigma(P_{450}/P_{850}) \lesssim 0.02$ and $\sigma(\gamma - \alpha) \lesssim 0.04$ for the sample medians and $\widetilde{|\Delta\chi|} \sim 4\degr$--$6\degr$ for the difference in the angle between bands.
In sections~\ref{sec:results}--\ref{sec:summary}, we report every principal physical quantity in the form
\begin{equation*}
X \;=\; \bar{X} \;\pm\; \delta X_{\rm sys} \;\pm\; \delta X_{\rm stat},
\end{equation*}
where $\bar X$ is the central value, typically a median over pixels or sources, as specified for each result.
The systematic floor $\delta X_{\rm sys}$ comes from Paper~T~subsection~\ref{T-sec:uncertainty}.
Residual zero-level offsets in Stokes $Q/U$ at 450\,$\mu$m dominate the systematic floors for $P_{450}/P_{850}$ and $\gamma - \alpha$.
Flux calibration dominates the uncertainties in individual spectral indices (subsection~\ref{ss:SpIdx}).
The statistical term $\delta X_{\rm stat}$ is the standard error of the median in the relevant ensemble of pixels or sources.
We report the two terms separately because they describe different sources of uncertainty.
The term $\delta X_{\rm sys}$ measures how confidently the central value is anchored to the absolute calibration, while $\delta X_{\rm stat}$ measures how reproducibly we recover it within the sample.
When a closed-form $\delta X_{\rm stat}$ is not tabulated separately, typically for a single ensemble median, we apply the same convention for the standard error of the median (SEM) to the values for individual sources and label the result as empirical SEM.

We define every parent sample in this paper by conditions on the total intensity $I$ alone, after basic validity checks that exclude entries with non-finite or unphysical values.
We do not select any parent sample on the polarized intensity $PI$, the polarization fraction $P$, or the ratio of $P$ between the bands, because a selection on weak polarization, or on the sensitivity to it, would itself bias the quantities being measured.

To define the parent samples, we apply the stated Stokes-$I$ thresholds to the intensities and their per-pixel uncertainties in the adopted map catalogs.
In these criteria, $\sigma_I$ denotes the catalogued uncertainty for each pixel, wavelength, and map product.
Paper~T subsection~\ref{T-ss:noiseI} characterizes these uncertainties and separates the additive noise from the component that depends on intensity.
To reproduce our selection, the stated thresholds must be applied with these same catalog uncertainties.
Replacing them with an independently estimated root-mean-square (rms) noise level or a single additive noise floor can change which pixels pass a given threshold.
The blank-sky noise estimates quoted in the atlas captions of Paper~S2 are distinct from the per-pixel uncertainties used for selecting data.

Where the polarization signal enters an analysis, we treat its noise floor according to how the quantity uses the signal.
Quantities that depend directly on the value of $P$ include only pixels at or above the $3\sigma$ floor in the relevant bands, since $P$ below the floor is not treated as a detection.
A single-band correlation involving $P_\lambda$ requires $P_\lambda \geq 3\,\sigma_{P_\lambda}$, whereas $R_{\rm EW}$ and correlations involving $P_{450}/P_{850}$ require $P_{450} \geq 3\,\sigma_{P_{450}}$ and $P_{850} \geq 3\,\sigma_{P_{850}}$.

For the spectral index $\gamma$ of polarized intensity and the dispersion function $\mathcal{S}$, a pixel below the corresponding floor is kept as a $3\sigma$ limit.
A correlation that uses $\mathcal{S}_\lambda$ also requires the value of $\mathcal{S}_\lambda$ before debiasing to exceed $3\,\sigma_{\mathcal{S}}$ and excludes such limits.
The index $\gamma$ is an upper limit when $PI_{450} < 3\,\sigma_{PI_{450}}$ and a lower limit when $PI_{850} < 3\,\sigma_{PI_{850}}$.
We do not assign a value when both lie below the floor.
A value of $\mathcal{S}$ is an upper limit when $\mathcal{S} \leq 3\,\sigma_{\mathcal{S}}$.
Each limit is evaluated with the quantity below the floor set to $3\sigma$.
When the median for a source involves such limits, we quote the range between their two extreme placements.

Polarization angles are not subject to a floor.
An angle is determined by $Q$ and $U$, and we use its uncertainty directly.
The precision floor of the angle comparison, $4\arcdeg$--$6\arcdeg$ (Paper~T equation~\ref{T-eq:precision_floors}), is itself derived from pixels selected on $I$ alone.

At the total-intensity thresholds used here ($I/\sigma_I > 30$), the floors in $P$ and in $PI$ impose nearly the same conditions.
With $\sigma_P/P = [(\sigma_{PI}/PI)^2 + (\sigma_I/I)^2]^{1/2}$, $PI/\sigma_{PI} = 3$ corresponds to $P/\sigma_P \geq 2.98$.
 
In this paper, we introduce three uncertainty terms rather than inheriting them from Papers~T and~S2.
The first is the formal angle uncertainty for each pixel, $\sigma_{\Delta\chi} \approx 3.5\degr$, computed here from the Stokes $Q,U$ noise model of Paper~T (subsections~\ref{ss:definitions} and \ref{ss:bfield_consistency}).
The second is the floor from flux calibration, $\delta_{\rm sys} = 0.11$, on the median spectral indices $\langle\alpha\rangle$ and $\langle\gamma\rangle$ (subsection~\ref{ss:SpIdx}).
The third is the Maritz--Jarrett standard error of the median adopted for $\delta X_{\rm stat}$.
 
We use the $10\degr$ tolerance for the difference in angle between bands as an interpretive guideline.
It is an adopted \emph{physical tolerance scale}---a conservative budget for the systematic uncertainty in interpreting the angle difference (\emph{not} a measurement error).
Subsection~\ref{ss:bfield_consistency} lists the components of this budget.
Agreement within $10\degr$ means the difference between the bands is not larger than the systematics we can account for, and therefore is not attributed to an astrophysical cause.
The tolerance is a criterion, not an uncertainty or a \emph{hypothesis test}.
Exceeding it is not by itself a detection: we flag a source above the scale for individual examination, rather than classifying it.
We assess statistical significance separately, against $\sigma_{\Delta\chi}$.

\subsection{Analysis samples}
\label{ss:samples}
The three findings F1--F3 address different physical questions about the field orientation, the polarization spectrum, and its relation to local cloud conditions.
We use nested subsamples of the JCMT BISTRO 450/850\,\micron\ POL-2 survey, each appropriate for the measurements and statistical requirements of its test.
We distinguish the lines of sight used to compare polarization angles, the sources with reliable spectral indices, and the nearby fields with Herschel measurements of column density and dust temperature.
A single common subsample would either discard usable information or include pixels irrelevant to a particular test.
\emph{(i)} The full 69-field survey provides the basis for the finding that $R_{\rm p} < 1$ in roughly four-fifths of sources (subsection~\ref{ss:P_ratio}; the census of individual sources is given in Paper~S2 subsection~\ref{S2-ss:S2_P_ratio}).
\emph{(ii)} The {\sc SpIdx} sample (pixels in the $R_{\rm EW}$ mask with $N_{\rm samp} \geq 5$ for each source; LMSF / HMSF / CMZ $= 9/19/7$) underlies the F2 result $\gamma < \alpha$ in 69\,\%--77\,\% of sources (subsection~\ref{ss:SpIdx}, figure~\ref{fig:alpha_vs_gamma}).
\emph{(iii)} The sample of 21 sources used to compare angles in the two bands ($N_{\rm samp} = 1420$ pixels with valid $\chi_{450}$ and $\chi_{850}$ at the matched 12\farcs6 resolution) underlies the F1 test of angle consistency (subsection~\ref{ss:bfield_consistency}, figure~\ref{fig:PA450vsPA850}).
We use one map per region so that each pixel is counted once.
\emph{(iii')} The subset of 13 targets comprises individual sources and regional groups drawn from the 21-source sample, each with enough pixels to compute a reliable median $|\chi_{450}-\chi_{850}|$ for that target.
It underlies the inspection of individual targets in figure~\ref{fig:PA450vsPA850}(b).
\emph{(iv)} The HGBS subsample covers 9 regions and contains 417 pixels in the $R_{\rm EW}$ mask that overlap the \textit{Herschel} Gould Belt Survey \citep{Andre2010} maps of $N_{\rm H_2}$ and $T_{\rm d}$.
We use it to test for geometric depolarization through the relation between $P$ and $\mathcal{S}$ in individual pixels (subsection~\ref{ss:S_dispersion}, figure~\ref{fig:PvsS}) and to examine the $I_{450}/I_{850}$--$R_{\rm p}$ anticorrelation (subsection~\ref{ss:Iratio}, figure~\ref{fig:IratioRp}).
Each subsection restates the relevant selection at its point of use.
Paper~S2 lists the individual sources in the 35- and 21-source samples and gives the criteria for selecting sources for the spectral-index analysis (subsections~\ref{S2-ss:S2_P_ratio} and \ref{S2-ss:S2_SpIdx}).
These samples connect projected field geometry, the dust polarization spectrum, and the conditions in the interiors of nearby clouds.
Their different selection functions define the scope over which those constraints can be combined.
 
For the per-pixel tests in subsections~\ref{ss:Iratio} and \ref{ss:S_dispersion}, we use nine survey regions within the footprint of the \textit{Herschel} Gould Belt Survey \citep[HGBS;][]{Andre2010}: L1688~Ophiuchus~A, Taurus~B211/213, NGC~1333, Perseus~B1, Perseus~IC348, OMC\,1, OMC\,2, OMC\,3, and Serpens~Main.
The availability of public HGBS $N_{\rm H_2}$ and $T_{\rm d}$ maps at our POL-2 field positions determines these nine regions.
We do not select them by their physical properties.
The subsample matched to HGBS therefore inherits the selection function of the parent 69-field survey (subsection~\ref{ss:data}), restricted to the nearby clouds that the HGBS footprint covers.
 
This subsample provides co-spatial maps of column density $N_{\rm H_2}$ at $18\arcsec$ resolution and dust temperature $T_{\rm d}$ at $36\arcsec$ resolution.
The resolution of the column-density maps is therefore close to the $12\farcs6$ POL-2 beam, but the temperature maps derived from fits to the spectral energy distribution (SED) are a factor of about three coarser.
This difference in resolution applies to every statement below that uses $T_{\rm d}$: a single $T_{\rm d}$ pixel spans several independent polarization measurements.
The pooled set of pixels spans $\sim 10^{22}$--$10^{24}\,{\rm cm}^{-2}$ and $\sim 10$--$45\,{\rm K}$.
 
We restricted the analysis to pixels in the $R_{\rm EW}$ mask (the Method~3 mask of subsection~\ref{ss:P_ratio}) within these regions.
This yields a common set of 417 pixels used consistently in subsections~\ref{ss:Iratio} and \ref{ss:S_dispersion} (the {\sc DualBand} sample of Paper~T table~\ref{T-tbl:T_sample_inventory}).
The 12\arcsec\ grid is comparable to the matched 12\farcs6 beam, so the adjacent pixels are not strictly independent.
The quoted significances should therefore be read as upper bounds.
Both bands are pre-matched to a common 12\farcs6 resolution (subsection~\ref{ss:data}), so the 12\arcsec\ grid samples identical effective beams at both wavelengths.

\section{Results and Analysis}
\label{sec:results}
We present the results in the following order: $\vec{B}$-field angle consistency (F1, subsection~\ref{ss:bfield_consistency}); the polarization-fraction ratio (F3, subsection~\ref{ss:P_ratio}) and its per-pixel relation to the intensity ratio (subsection~\ref{ss:Iratio}); spectral indices (F2, subsection~\ref{ss:SpIdx}); and the angle-dispersion function $\mathcal{S}$ as a test of geometric depolarization (subsection~\ref{ss:S_dispersion}).
Section~\ref{sec:obs} gives the data, definitions, uncertainty notation, and samples on which the results rest.
 
\subsection{$\vec{B}$-field angle consistency}
\label{ss:bfield_consistency}
 
We first test whether the 450 and 850\,$\mu$m polarization angles indicate similar projected field orientations, a prerequisite for comparing the polarization properties of individual pixels in subsections~\ref{ss:P_ratio}--\ref{ss:S_dispersion}.
The difference between the 450 and 850\,$\mu$m polarization angles can reflect a change in the $\vec{B}$-field structure projected on the plane of the sky.
However, this difference can also arise from noise or map processing.
Interpreting that difference requires us to distinguish the observed angle difference from the processing residual and from the angular scale used to assess astrophysical significance.
Subsection~\ref{ss:definitions} defines the angle-related quantities used below, and subsection~\ref{ss:uncertainty} defines the $10\arcdeg$ tolerance.

All 450\,$\mu$m polarization data from single fields used here have undergone the 450\,$\mu$m Stokes-$Q/U$ zero-level offset correction of Paper~T~subsection~\ref{T-ss:offset}.
The same section validates this correction against SOFIA/HAWC+ at 154 and 214\,$\mu$m.
For regions in which Paper~T adopts a reduction specific to the field, we use that reduction.
For the other regions, we subtract the $Q/U$ offsets estimated by the three-step procedure.
The four mosaicked fields of the {\sc SpIdx} subsample that have no three-step fit --- PGCC~G203.21$-$11.20, Serpens~Main, G202.3$+$2.5, and the Rosette Molecular Cloud --- are used as reduced, without subtraction.

\begin{figure*}
\centering
\includegraphics[width=0.95\textwidth]{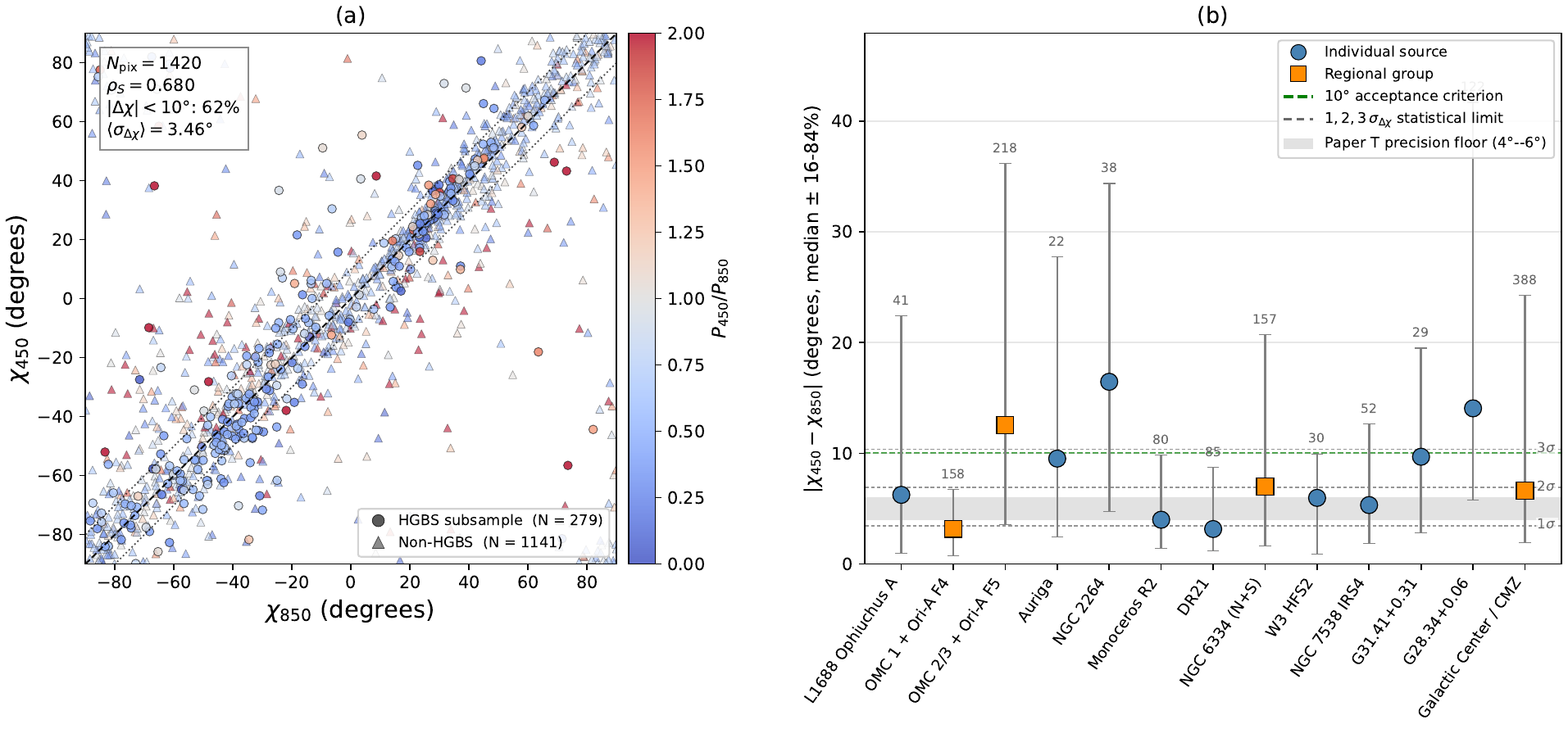}
\caption{
Comparison of 450 and 850\,\micron\ polarization angles at a common $12\farcs6$ resolution.
\textbf{(a)} Matched pixels with finite 450\,\micron\ angles across 21 sources, one map per region ($N_{\rm samp}=1420$), selected in Stokes $I$ only.
The angles $\chi_{450}$ are taken after the 450\,\micron\ Stokes-$Q/U$ zero-level offset correction of Paper~T, and the color scale uses $R_{\rm p}=P_{450}/P_{850}$ after the same correction.
Circles identify the HGBS subsample of nine fields used in figures~\ref{fig:PvsS} and~\ref{fig:IratioRp}; triangles identify other pixels.
The \texttt{coolwarm} scale shows $R_{\rm p}<1$ in blue, $R_{\rm p}=1$ in white, and $R_{\rm p}>1$ in red.
Dotted lines mark $\pm10\arcdeg$.
The inset gives Spearman $\rho_S$, the fraction within this band, and the median of the formal errors for individual pixels, $\langle\sigma_{\Delta\chi}\rangle$.
\textbf{(b)} Median absolute angle differences for individual sources (blue circles) and regional groups (orange squares), ordered by distance.
Bars span the 16th--84th percentiles of their distributions of pixel values, and labels give $N_{\rm samp}$.
Groups combine OMC~1 with Orion~A Field~4; OMC~2/3 with Orion~A Field~5; the North and South fields of NGC~6334; and the Galactic Center/CMZ clouds.
The green dashed line marks the $10\arcdeg$ acceptance criterion, which is not an uncertainty.
Light gray shading marks the Paper~T precision floor for median angle differences, $\widetilde{|\Delta\chi|}\sim4\arcdeg$--$6\arcdeg$ (Paper~T equation~\ref{T-eq:precision_floors}), evaluated with the same estimator as the plotted points.
Gray dashed lines mark one, two, and three times the typical statistical error for an individual pixel, $\sigma_{\Delta\chi}\simeq3.5\arcdeg$ (Paper~T subsection~\ref{T-ss:noisemodel}).
These lines are reference scales, not confidence limits on the median of each target.
Ten out of the thirteen targets lie within the adopted tolerance.
The larger medians of the other three (OMC\,2/3 + Orion~A Field~5, NGC~2264, and G28.34+0.06) occur where polarized intensities are weak and the angle uncertainties for individual pixels are higher.
The same figure with $\chi_{450}$ before the correction is shown in appendix~\ref{app:F1_uncorrected}.
There the median absolute angle difference for Monoceros~R2 is $12\arcdeg$, against $4.0\arcdeg$ after the correction, illustrating the sensitivity of the angles to it (subsection~\ref{ss:Discussion_GammaAlpha}).
}
\label{fig:PA450vsPA850}
\end{figure*}

We test wavelength independence in figure~\ref{fig:PA450vsPA850}(a) by comparing the 450 and 850\,$\mu$m polarization angles for the pixels with finite 450\,$\mu$m angles after the $Q/U$ zero-level offset correction.
The maps have a common angular resolution of $12\farcs6$ and are sampled on a $12\arcsec$ grid.
The sample contains $N_{\rm samp} = 1420$ pixels across 21 sources, with one map per region so that each pixel is entered once.
This selection applies signal-to-noise thresholds only to Stokes~$I$ in each band, with no pre-selection on the angle difference itself.
The observed wavelength independence is therefore an empirical property of the joint $\chi_{450}$--$\chi_{850}$ distribution rather than a consequence of the detection mask in each band.
The same holds for the narrower $R_{\rm EW}$ mask used in subsections~\ref{ss:P_ratio}--\ref{ss:S_dispersion}, which likewise applies a Stokes-$I$ $S/N$ threshold with no pre-selection on the angles.

The data cluster tightly along the 1:1 line, with $|\chi_{450} - \chi_{850}| < 10\arcdeg$ for $62\,\%$ of the $12\arcsec$ samples ($878$ of $1420$; 16th--84th percentile of a bootstrap over regions, $57\,\%$--$67\,\%$).
The agreement is closest where the polarized intensity is well detected.
Of the pixels with $PI/\sigma_{PI} \geq 20$ in both bands, $89\,\%$ lie within $10\arcdeg$ (median $3.3\arcdeg$), compared to $28\,\%$ of those with $PI/\sigma_{PI} < 3$ in either band (median $24\arcdeg$).
The $12\arcsec$ grid samples the common $12\farcs6$ resolution about once per beam, but adjacent samples can remain correlated.
The quoted fraction is therefore a fraction of sampled pixels rather than of independent measurements.
The observed median angle difference summarizes the agreement of the projected fields.
It does not remove measurement or processing effects, and comparison with the $10\arcdeg$ tolerance is an interpretive guide rather than a significance test.

The $10\arcdeg$ tolerance reflects the characteristic magnitude of systematic effects in the analysis of the two bands.
These effects include residuals from beam matching (section~\ref{sec:obs}), consistent with the empirical floor of $\widetilde{|\Delta\chi|} \sim$ a few to several degrees associated with beam processing in Paper~T~subsection~\ref{T-ss:beamval}.
In that test, $29$ of the $30$ dual-band sources reach $\widetilde{|\Delta\chi|} < 10\arcdeg$ after harmonization at $I/\sigma_I > 30$, and the median for L1688 is $\widetilde{|\Delta\chi|} = 4\fdg3$.
Other contributions are the residual $Q/U$ zero-level offsets (Paper~T~subsection~\ref{T-ss:offset}) and differences in spatial filtering between the bands.
The independent comparison of the 450\,$\mu$m data with SOFIA/HAWC+ gives median angle differences of $13\arcdeg$--$29\arcdeg$ after the $Q/U$ zero-level offset correction, comparable to the $11\arcdeg$--$28\arcdeg$ between 850\,$\mu$m and SOFIA (Paper~T~subsection~\ref{T-ss:offset}).
These differences may reflect dust weighting that depends on wavelength, as well as processing.

Small changes in polarization angle may be measurable even when both bands still indicate a similar projected field orientation.
We therefore separate the precision of an individual angle measurement from the larger tolerance used to judge whether a difference between the bands warrants an astrophysical explanation.
The corresponding \emph{statistical} detection limit derived from the Stokes $Q,U$ noise model of Paper~T~subsection~\ref{T-ss:noisemodel} is much smaller: the median uncertainty on $\Delta\chi$ for individual pixels is $\sigma_{\Delta\chi} \approx 3.5\arcdeg$, with $3\sigma_{\Delta\chi} \approx 10\arcdeg$.
Pixel-to-pixel deviations within the $10\arcdeg$ physical tolerance can therefore be statistically significant even though they remain physically small compared to the full $180\arcdeg$ angular range of polarization.

We measured $\widetilde{|\Delta\chi|} = 6.7\arcdeg$, slightly above the $4\arcdeg$--$6\arcdeg$ processing floor in Paper~T (equation~\ref{T-eq:precision_floors}).
The estimator is the same in both cases, and the processing floor is also evaluated on pixels selected in Stokes $I$ alone.
The 16th--84th percentile interval of a bootstrap over regions is $5.8\arcdeg$--$8.1\arcdeg$.
The signed median is $\widetilde{\Delta\chi} = +1.3\arcdeg$, with an interval of $+0.8\arcdeg$ to $+1.8\arcdeg$.
At every level of polarized intensity, the spread exceeds that expected from the formal $Q,U$ uncertainties alone: the median $|\Delta\chi|/\sigma_{\Delta\chi}$ is $1.6$--$3.7$, compared to $0.67$ for pure noise.

An angle difference that is formally resolved against $\sigma_{\Delta\chi}$ does not necessarily imply that the two wavelengths trace substantially different projected fields.
The physical interpretation of the angle difference also depends on the processing residuals and on the spatial pattern of the difference.

Most targets individually show agreement between the bands in figure~\ref{fig:PA450vsPA850}(b).
The median $|\chi_{450}-\chi_{850}|$ is below the $10\arcdeg$ criterion in 10 of 13 targets, and the median of the 13 medians for individual targets is $6.6\arcdeg$.
The three targets above the criterion are OMC\,2/3 + Orion~A Field~5 ($12.6\arcdeg$), G28.34+0.06 ($14.1\arcdeg$), and NGC~2264 ($16.5\arcdeg$).
Their pixels have weak polarized intensity and larger uncertainties in the angle of each pixel (median $\sigma_{\Delta\chi}$ of $6\arcdeg$--$8\arcdeg$).
The agreement therefore generally holds where the angles are well constrained.
Before the 450\,$\mu$m Stokes-$Q/U$ zero-level offset correction (appendix~\ref{app:F1_uncorrected}), six targets exceed the criterion.
Among them, Monoceros~R2 has a median $|\chi_{450}-\chi_{850}| = 12\arcdeg$ (16th--84th percentile range $4.6\arcdeg$--$27\arcdeg$) with a median signed angle difference of $+8.1\arcdeg$; after the correction, its median $|\chi_{450}-\chi_{850}|$ is $4.0\arcdeg$.
 
Monoceros~R2 is therefore a useful diagnostic of how sensitive the angles in the two bands are to the $Q/U$ zero-level offset correction.
It also has a strongly negative difference between the spectral indices, $\gamma - \alpha = -1.04$ (subsection~\ref{ss:SpIdx}), in the same direction as the trend throughout the survey.
L1688 Ophiuchus~A has the most negative value in the sample used for the spectral indices, $\gamma-\alpha = -1.14$.
Mon~R2 was independently excluded from the comparison of polarization angles with SOFIA/HAWC+ because of the large measurement uncertainties (Paper~T~subsection~\ref{T-ss:offset}).
After correction, the difference in the angle between the bands in Monoceros~R2 is within the adopted tolerance.
We return to its spectral-index difference in subsection~\ref{ss:Discussion_GammaAlpha}.

We interpret the generally similar angles in the two bands as a constraint on changes in the net projected polarization direction at the adopted resolution.
This similarity does not establish that the two bands have identical weights along the line of sight or within the beam.
Wavelength-dependent depolarization can therefore remain even where the net angles are similar.
The angles disfavor geometric explanations that predict large rotations between the bands.
However, unresolved mixtures of emitting components remain a possible contribution to the wavelength dependence of the polarization fraction and of the spectral index of polarized intensity documented in the following subsections.

\citet{VaillancourtMatthews2012} compared 350\,$\mu$m polarimetry from the Caltech Submillimeter Observatory with 850\,$\mu$m polarimetry from the JCMT at about 20\arcsec\ resolution in 17 Galactic clouds.
These authors found angle differences centered near zero, but with $64\,\%$ of the positions with $P \geq 3\sigma_P$ at both wavelengths differing by more than $10\arcdeg$.
In our data, in which both bands were observed with POL-2, $62\,\%$ of the $12\arcsec$ samples agree within $10\arcdeg$.
Between single-dish and interferometric scales, \citet{Hull2014} found consistent $\vec{B}$-field orientations for a subset of star-forming cores.
With the comparison of angles between the bands as a prerequisite, subsection~\ref{ss:P_ratio} examines the wavelength dependence of the polarization fraction directly through the ratio $R_{\rm p}$ for individual pixels (subsection~\ref{ss:definitions}).
Subsection~\ref{ss:SpIdx} expresses $R_{\rm p}$ in terms of the pair of spectral indices $(\alpha, \gamma)$.

\subsection{Polarization-fraction ratio}
\label{ss:P_ratio}
 
We use the polarization-fraction ratio $R_{\rm p}$ (subsection~\ref{ss:definitions}) to characterize the wavelength dependence of dust polarization at matched positions and angular resolution.
$R_{\rm p}$ can be evaluated without a separate estimate of dust column density, because each polarization fraction $P_\nu \equiv PI_\nu/I_\nu$ is normalized by the corresponding total intensity.
A departure from unity, $R_{\rm p} \neq 1$, indicates that polarized and total emission have different spectral dependences between 450 and 850\,$\mu$m.
We examine size-dependent grain alignment as a physically motivated explanation of this wavelength dependence, using the comparison of polarization angles between the bands and the $P$--$\mathcal{S}$ relation (subsection~\ref{ss:S_dispersion}) to constrain geometric contributions.
We also assess how dust optical properties, temperature structure along the line of sight, and unresolved $\vec{B}$-field geometry may affect this interpretation (subsections~\ref{ss:Discussion_Tmixing} and~\ref{ss:Discussion_DiscriminatingPower}).
 
The predominance of $R_{\rm p} < 1$ is one of the three findings (F1--F3) that jointly constrain submillimeter grain alignment.
In NGC\,2071, which is not part of our 35-source sample, the polarization fraction decreases with wavelength from 154 to 214\,$\mu$m \citep{Fanciullo2022b} and from 450 to 850\,$\mu$m (\citealt{Lyo2021}; Paper~S2 subsection~\ref{S2-ss:S2_P_ratio}).
This decrease is in the opposite sense to the predominance of $R_{\rm p} < 1$ reported below.
Section~\ref{sec:obs} and Paper~T~subsection~\ref{T-ss:offset} document the 450\,$\mu$m Stokes-$Q/U$ zero-level offset correction and its SOFIA/HAWC+ cross-validation, on which the input maps are based.
We first define the Method~3 estimator and its common dual-band mask, then report the prevalence of $R_{\rm p} < 1$ throughout the survey and the median for the 35-source sample.
 
\begin{figure}[htbp]
\centering
\includegraphics[width=0.80\columnwidth]{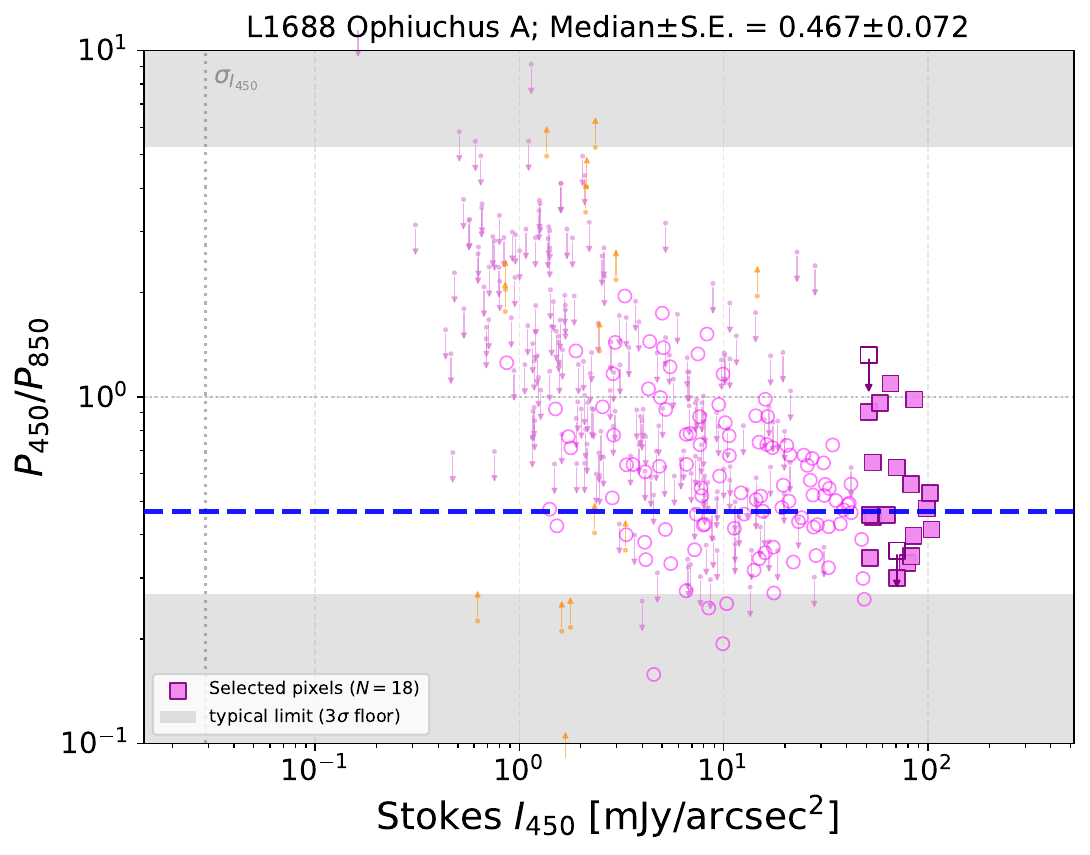}
\caption{
Method~3 selection for L1688 Ophiuchus~A: ratios $P_{450}/P_{850}$ of individual pixels versus Stokes $I_{450}$.
In the underlying maps, the pixel centers are spaced by $12\farcs0$, nearly one beam width ($12\farcs6$ FWHM).
Filled violet squares satisfy the Method~3 mask ($I/\sigma_I > 50$ and Stokes~$I$ above the 95th percentile of the source, with both criteria applied at both wavelengths), without a cut on the $S/N$ of polarized intensity.
Open magenta circles mark the 103 pixels outside the Method~3 mask whose $P_{450}$ and $P_{850}$ both exceed three times their uncertainties.
Light gray shading marks the typical measurement limits of the ratio.
These are the medians over the selected pixels of the ratios obtained by setting $P_{450}$ to its $3\sigma$ floor (below $R_{\rm p}=0.27$) or $P_{850}$ to its $3\sigma$ floor (above $R_{\rm p}=5.2$).
Pixels outside the mask whose $P_{450}$ ($P_{850}$) alone lies below $3\sigma$ are drawn at their $3\sigma$ upper (lower) limits with light magenta downward (orange upward) arrows (221 and 15 pixels, of which two and one lie outside the plotted range).
The 1241 pixels outside the mask that lie below $3\sigma$ at both wavelengths are not drawn.
Two of the 18 selected pixels have $P_{450}$ below $3\sigma$.
These two pixels are also drawn as open squares with downward arrows at their $3\sigma$ upper limits.
The horizontal dashed line is $R_{\rm EW}=0.47\pm0.07$, the median of the 16 selected pixels with polarization detections in both bands.
The median over all 18 selected pixels is identical, so the two pixels drawn as limits do not move it.
``S.E.'' in the figure denotes the standard error of the median $\delta_{\rm stat}$.
For this source, $R_{\rm IWM}=0.49$, $R_{\rm Corr}=0.47$, and $R_{\rm EW}=0.47$ agree within 0.02.
Paper~S2 subsection~\ref{S2-ss:S2_P_ratio} compares the estimators across sources.
}
\label{fig:Pratio_OphA_method3}
\end{figure}

We seek a representative polarization spectrum for each cloud without allowing its brightest core to dominate the result or excluding lines of sight with intrinsically weak polarization.
We adopt Method~3 as the canonical estimator for the $R_{\rm p}$ analysis of Paper~S1.
It is the equal-weighted median of polarization ratios for individual pixels over the subset of the common Stokes-$I$ mask with polarization detected in both bands (figure~\ref{fig:Pratio_OphA_method3}):
\begin{equation}
R_{\rm EW} \;\equiv\; \underset{i \,\in\, \mathcal{M}_{\rm det}}{\mathrm{median}}\!\left[\frac{P_{450,\,i}}{P_{850,\,i}}\right],
\label{eq:R3}
\end{equation}
where $\mathcal{M}$ is the set of pixels that satisfies the common dual-band Stokes-$I$ detection mask.
This mask requires $I_{450}/\sigma_{I,450} > 50$ and $I_{850}/\sigma_{I,850} > 50$, Stokes~$I$ above the 95th percentile of the source at both wavelengths, and Stokes~$I$ above an absolute floor at both wavelengths ($10\%$ of the mean intensity of the pixels at or above the 99th percentile).
For each source, we take the median over $\mathcal{M}_{\rm det}\subset\mathcal{M}$, the members with $P_{450}\geq3\,\sigma_{P_{450}}$ and $P_{850}\geq3\,\sigma_{P_{850}}$ (subsection~\ref{ss:uncertainty}).
This median therefore describes the detected ratios within the sample selected on intensity; it does not include the ratios below the floor.
We deliberately do not select pixels by the signal-to-noise ratio of polarized intensity in defining $\mathcal{M}$, so that intrinsically low-polarization pixels remain in the parent sample.
Selecting the parent mask on $PI/\sigma_{PI}$ would preferentially exclude the weak-$P$ tail of the pixel distribution from every analysis that uses this mask.
Defining $\mathcal{M}$ by Stokes~$I$ alone keeps the $P_{450}/P_{850}$ analysis internally consistent with the pixel selection used for the spectral-index analysis of subsection~\ref{ss:SpIdx} and for the correlations with environmental parameters that we examine for individual pixels later in this paper.
The $I/\sigma_I > 50$ threshold, together with the brightness cuts for each source, keeps the retained pixels at high signal-to-noise in both bands.
Adjacent pixels can remain correlated at the matched $12\farcs6$ resolution, so the number of pixels is not an effective number of independent measurements.
 
The estimator benefits from a generic \emph{common-mode cancellation}, because each term $P_{450,i}/P_{850,i}$ is formed from the same line of sight.
A quantity that acts as a common multiplicative factor on $P_{450}$ and $P_{850}$ cancels in the ratio for each pixel.
Variations between pixels in column density, dust temperature, or field tangling within the beam cancel only to the extent that they act in this way.
$R_{\rm EW}$ is therefore less sensitive to spatial intensity weighting than estimators weighted by intensity.
The common-mode cancellation does not apply to the residual additive 450\,$\mu$m Stokes-$Q/U$ zero-level offsets inherited from Paper~T~subsection~\ref{T-ss:offset}.
Paper~T evaluates their effect with Monte Carlo simulations (Paper~T~subsection~\ref{T-sec:uncertainty}).
For the terms quantified in Paper~T, the resulting systematic floor for the sample median, $\sigma_{\rm sys}(R_{\rm EW}) \lesssim 0.02$ (Paper~T~subsection~\ref{T-sec:precisionfloor}; Paper~T table~\ref{T-tbl:systematics}), lies an order of magnitude below the detected signal.
 
The alternative intensity-weighted estimator ($R_{\rm IWM}$) and the weighted regression over individual pixels ($R_{\rm Corr}$) yield sample medians that agree with the Method~3 value to within the quoted precision when taken over the sources with an $R_{\rm EW}$ entry.
Restricting Paper~S1 to Method~3 should not affect our conclusions for the sample as a whole.
Their complete derivation and tables for individual sources are presented in Paper~S2 (subsection~\ref{S2-ss:S2_P_ratio}).
In practice, Method~3 defines the common mask for selecting high-quality measurements in both bands throughout this paper.
We use this mask to estimate $R_{\rm p}$ across the survey (figure~\ref{fig:Pratio_OphA_method3}).
It also defines the {\sc DualBand} sample of 417 pixel pairs that overlap HGBS (subsection~\ref{ss:samples}), used in the tests of individual pixels in figures~\ref{fig:PvsS} and \ref{fig:IratioRp}.
With the additional requirement $N_{\rm samp}\geq 5$ for each source, it defines the {\sc SpIdx} sample (subsection~\ref{ss:SpIdx}, figure~\ref{fig:alpha_vs_gamma}).
The resulting median describes the typical wavelength dependence among the retained lines of sight through bright clouds.
Interpreting that dependence still requires separate constraints on magnetic geometry and dust properties.
 
The full 69-field census presented in Paper~S2 (subsection~\ref{S2-ss:S2_P_ratio}) shows that roughly four-fifths of sources have $R_{\rm p} < 1$.
The sample medians in this census are $R_{\rm p} = 0.68$--$0.78$ for the three methods.
This behavior holds across the sampled environments: nearby low-mass cores (L1688 Ophiuchus~A at 139\,pc), intermediate-distance high-mass regions (OMC\,1 at 399\,pc, DR\,21 at 1500\,pc), and distant Galactic Center clouds (Sagittarius~B2 at 8.3\,kpc).
The consistency of $R_{\rm p} < 1$ across roughly four-fifths of the sources, spanning nearly two orders of magnitude in heliocentric distance and more than two orders of magnitude in column density, motivates an interpretation in terms of intrinsic grain physics.
This consistency does not exclude an environmental or geometric origin.
In this survey, environment and distance are themselves strongly coupled (Spearman $\rho \simeq 0.83$, with no overlap in distance between the Central Molecular Zone and either disk class; Paper~S2).
Spanning distance is therefore also spanning environment, and the two cannot be separated here.

Across the {\sc SpIdx} sample (the 35-source spectral-index subsample; see table~\ref{tbl:S1_F1F3_master} for the samples used for F1--F3 in this paper), the median of the ratios estimated for individual sources yields
\begin{equation}
R_{\rm EW} \;=\; 0.69 \;\pm\; 0.02_{\rm sys} \;\pm\; 0.03_{\rm stat},
\label{eq:RGlobalSampleMedian}
\end{equation}
where $\delta R_{\rm sys} = 0.02$ is the systematic floor of Paper~T~subsection~\ref{T-sec:precisionfloor} and $\delta R_{\rm stat} = 0.03$ is the Maritz--Jarrett standard error of the median, evaluated on the unrounded values for individual sources (interquartile range 0.26).
The median is taken over the 30 of the 35 {\sc SpIdx} sources that carry an $R_{\rm EW}$ entry.
The five mosaicked fields NGC~1333, PGCC~G203.21$-$11.20, Serpens~Main, G202.3$+$2.5 and the Rosette Molecular Cloud have no entry in the table of ratios for individual sources in Paper~S2.
For the source with the highest signal-to-noise ratio, L1688 Ophiuchus~A, $R_{\rm EW} = 0.47$.
Several other sources with high signal-to-noise ratios (e.g., DR\,21 with $R_{\rm EW} = 0.51$ and NGC\,6334~North with $R_{\rm EW} = 0.46$) lie in the same range.
The CMZ stack has a higher value, $R_{\rm EW} = 0.82$, which is still below unity.
Paper~S2 (subsection~\ref{S2-ss:S2_P_ratio}) lists the $R_{\rm EW}$ values for individual sources and compares them with $R_{\rm IWM}$ and $R_{\rm Corr}$.
The sample medians are below unity for all three estimators.

\subsection{Intensity ratio and polarization ratio per pixel}
\label{ss:Iratio}
We use the {\sc DualBand} sample of 417 pixel pairs in the HGBS overlap region (subsection~\ref{ss:samples}) as the parent sample.
These pixels were selected with the Method~3 mask of subsection~\ref{ss:P_ratio}, without a cut on the signal-to-noise ratio of $PI$.
Each statistic uses the subset that meets its detection requirements (subsection~\ref{ss:uncertainty}).
Independent measurements of column density and dust temperature are available for these pixels.

The 450-to-850\,$\mu$m intensity ratio varies both within a cloud and between clouds.
This variation reflects differences in the dust emission sampled along each line of sight.
We ask where the apparent association of the intensity ratio with the polarization spectrum originates.
To describe the $\gamma - \alpha$ structure in individual pixels, we use the Spearman rank correlation between $I_{450}/I_{850}$ and $R_{\rm p} = P_{450}/P_{850}$.
Subsection~\ref{ss:SpIdx}, equation~\ref{eq:Pratio_SpIdx}, gives the conversion to spectral indices.
We evaluated this correlation for the pixels of the {\sc DualBand} sample (subsection~\ref{ss:samples}) with polarization detected in both bands.
As noted earlier, the pixel spacing of 12\arcsec\ is comparable to the full width at half maximum (FWHM) of the matched beam (12\farcs6).
We therefore treat the resulting $p$-values as nominal.
We detected a statistically significant negative correlation,
\begin{equation}
r_s(I_{450}/I_{850}, R_{\rm p}) = -0.49
\quad (N = 327, \; p = 1.9 \times 10^{-21}),
\label{eq:rS_IratioRp}
\end{equation}
and the anticorrelation remains $r_s = -0.35$ ($N = 265$, $p = 5.9 \times 10^{-9}$) when OMC\,1 is excluded, so the anticorrelation is not due to OMC\,1 alone.
 
The two quantities share $I_{450}/I_{850}$ because $R_{\rm p} = (PI_{450}/PI_{850})/(I_{450}/I_{850})$ by definition (subsection~\ref{ss:definitions}).
A negative rank correlation between them is therefore expected for algebraic reasons whenever $\gamma - \alpha$ is not constant across the sample.
We report this relation as another representation of the $\gamma - \alpha$ result for individual pixels rather than as independent evidence.
Pixels with higher $I_{450}/I_{850}$ have lower $R_{\rm p}$, while those with lower $I_{450}/I_{850}$ have $R_{\rm p}$ closer to unity.
 
The difference in polarization angle $|\chi_{450} - \chi_{850}|$ remains small ($\lesssim 20\arcdeg$) in the full range of $I_{450}/I_{850}$ and $R_{\rm p}$.
We did not detect a significant correlation between $R_{\rm p}$ and $|\Delta\chi|$.
The negative $P$--$\mathcal{S}$ associations in both bands (figure~\ref{fig:PvsS}) are consistent with depolarization accompanying variations in the projected field.
However, these associations do not demonstrate that the depolarization cancels in $R_{\rm p}$.
The absence of a detected correlation with $|\Delta\chi|$ indicates that geometric effects that rotate the angles are not the dominant driver of the observed wavelength dependence, although a subdominant contribution cannot be entirely excluded.
 
\begin{figure}[htbp]
\centering
\includegraphics[width=0.85\columnwidth]{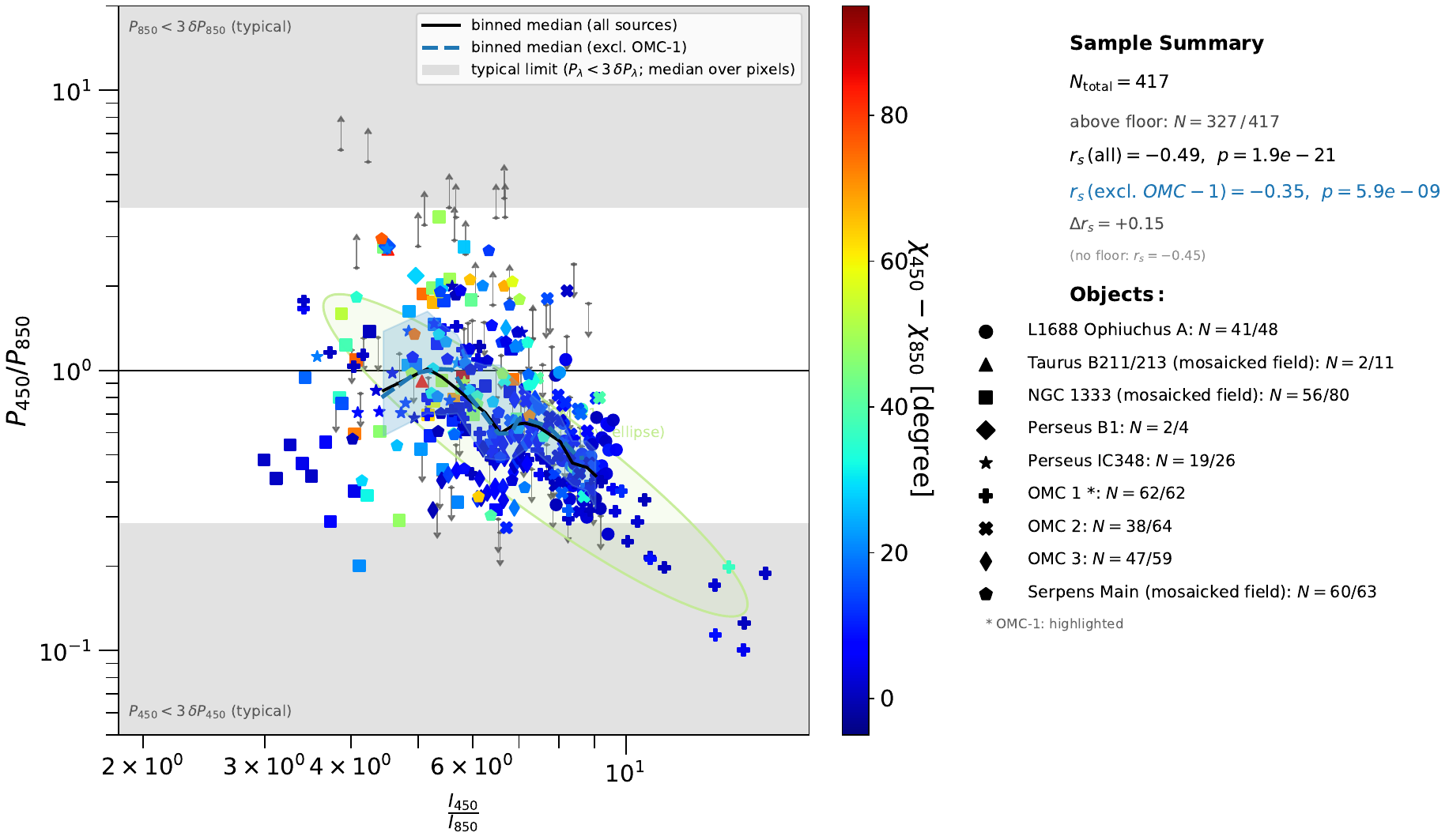}
\caption{
Anticorrelation between the intensity ratio $I_{450}/I_{850}$ and the polarization ratio $R_{\rm p} = P_{450}/P_{850}$ for individual pixels in the {\sc DualBand} sample (417 sampled pixel pairs in the HGBS overlap region).
Colors indicate the polarization-angle difference $|\chi_{450} - \chi_{850}|$.
The anticorrelation persists in subsamples that include OMC\,1 and those that exclude OMC\,1 ($r_s = -0.49$, $N = 327$ for the full subsample; $r_s = -0.35$, $N = 265$ excluding OMC\,1), so it is not driven by that single luminous source.
The colors show that the relation does not arise from a systematic variation of $|\Delta\chi|$ across the diagram.
Light gray shading marks the typical limits of $R_{\rm p}$ set by the detection floor $P_i \geq 3\,\delta P_i$ (Paper~T~subsection~\ref{T-ss:noisemodel}).
The symbol $\delta P_i$ denotes the uncertainty in the polarization fraction of pixel $i$.
These limits are the medians over pixels of $3\,\delta P_{450,i}/P_{850,i}$ (below $R_{\rm p} = 0.28$) and $P_{450,i}/(3\,\delta P_{850,i})$ (above $R_{\rm p} = 3.8$).
Pixels with $P_{450}$ below its floor are plotted at $3\,\delta P_{450}/P_{850}$ as upper limits of $R_{\rm p}$ (downward arrows, 56).
Pixels with $P_{850}$ below its floor are plotted at $P_{450}/(3\,\delta P_{850})$ as lower limits (upward arrows, 16).
The 18 pixels below both floors are not plotted, and limits are excluded from the Spearman statistics.
The light-blue band shows the interquartile range of the binned median.
Through equation~(\ref{eq:Pratio_SpIdx}), this anticorrelation directly expresses, for individual pixels, the $\gamma - \alpha$ spectral-index structure developed in subsection~\ref{ss:SpIdx}.
}
\label{fig:IratioRp}
\end{figure}

Figure~\ref{fig:IratioRp} shows this test for individual pixels: $R_{\rm p}$ versus $I_{450}/I_{850}$, with colors indicating $|\Delta\chi|$.
It also shows the typical detection limits and represents pixels below them as $3\sigma$ limits.
The anticorrelation in the pooled sample persists when OMC \,1 is excluded.
However, the median of the correlations within individual clouds is near zero, so the pooled relation does not establish a common local response across clouds.
It summarizes the same observables at the two wavelengths rather than providing an independent constraint on grain alignment.
 
\subsection{Spectral Indices of Total and Polarized Intensity}
\label{ss:SpIdx}
 
\begin{figure}
\centering
\includegraphics[width=0.85\columnwidth]{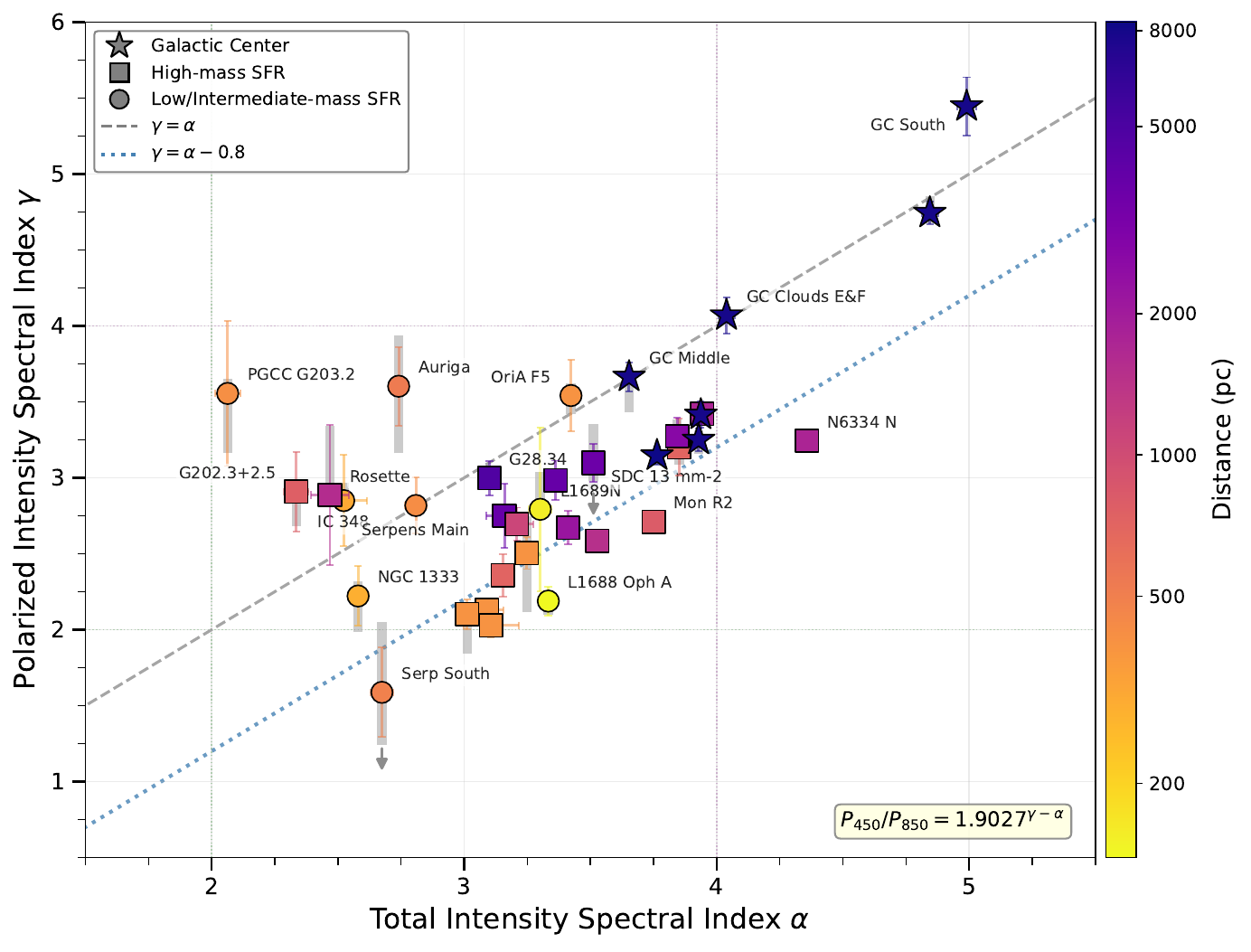}
\caption{Correlation between the Stokes~$I$ spectral index $\langle\alpha\rangle$ and the polarized-intensity spectral index $\langle\gamma\rangle$ for the 35 sources in the statistical sample ($N_{\rm samp} \geq 5$).
Symbols show the measured medians for individual sources, and error bars represent the standard error of each median.
The light-gray vertical bars show the range of each source's median $\gamma$ when measurements of polarized intensity below three times their uncertainty are treated as $3\sigma$ limits.
A nondetection at 450\,$\mu$m gives an upper limit on $\gamma$, whereas one at 850\,$\mu$m gives a lower limit.
For every source, the measured median lies within the corresponding range.
The gray bars and arrows indicate this range, not a correction to the measured median.
For Serpens~South and SDC~13 mm-2, the upper end of the gray bar gives an upper limit on the median $\gamma$, and the downward arrow indicates that the range extends without a lower bound.
The downward lengths of these two bars and arrows are chosen for display only and do not indicate lower bounds.
Under the two extreme placements of the limits, 24--27 of 35 sources lie below the $\gamma = \alpha$ line, with median $\gamma - \alpha$ between $-0.57$ and $-0.45$ ($P_{450}/P_{850} = 0.69$--$0.75$).
The medians of the values for individual sources are 3.33 for $\alpha$ and 2.85--3.04 for $\gamma$, with the latter range reflecting the same two extreme placements.
The dashed line shows $\gamma = \alpha$, and the dotted line shows $\gamma = \alpha - 0.8$ ($P_{450}/P_{850} \approx 0.6$).
For matched measurements, the $\gamma<\alpha$ side of the equality line corresponds to $P_{450}<P_{850}$, or a higher polarization fraction at the longer wavelength.
Symbols distinguish environment categories (stars: CMZ; squares: high-mass; circles: low/intermediate-mass).
Paper~S2 (subsections~\ref{S2-ss:S2_SpIdx} and \ref{S2-ss:S2_EnvStrat}) presents the breakdown by environment and the five sources with $\gamma > \alpha$ at the $1\sigma$ level.
}
\label{fig:alpha_vs_gamma}
\end{figure}

The analysis of the polarization-fraction ratio (subsection~\ref{ss:P_ratio}) established that $P_{450}/P_{850} < 1$ is the dominant behavior in our sample.
The spectral indices complement the analysis of $R_{\rm p}$.
Whereas $R_{\rm p}$ combines the two wavelengths into a ratio in which the dust column largely cancels, $\alpha$ and $\gamma$ describe the frequency dependence of the total and polarized intensity separately.
We can therefore examine their individual implications for dust emissivity and grain alignment.
We summarize here the framework and the principal F2 results from the {\sc SpIdx} sample.
We first connect the indices to $R_{\rm p}$ through equation~(\ref{eq:Pratio_SpIdx}) and to the $I_{450}/I_{850}$--$R_{\rm p}$ anticorrelation for individual pixels in subsection~\ref{ss:Iratio} (figure~\ref{fig:IratioRp}), the counterpart of $\gamma - \alpha$.
We then report the F2 finding, $\gamma < \alpha$ in $69\,\%$--$77\,\%$ of the {\sc SpIdx} sample (figure~\ref{fig:alpha_vs_gamma}), and examine its robustness.
We close by discussing biases associated with $S/N$, which are assessed in Paper~T (sub-subsection~\ref{T-ss:spidx_snrbias}).

Paper~S2 (subsection~\ref{S2-ss:S2_SpIdx}) gives the full table of spectral indices for individual sources, examines the biases, and illustrates the spatially resolved analysis with L1688 Ophiuchus~A.
It also presents the results separately for the LMSF, HMSF, and CMZ environments.

We define the spectral indices for total and polarized intensity as
\begin{equation}
\alpha \equiv \frac{\log(I_{450}/I_{850})}{\log(\nu_{450}/\nu_{850})}
\quad \text{and} \quad
\gamma \equiv \frac{\log(PI_{450}/PI_{850})}{\log(\nu_{450}/\nu_{850})},
\label{eq:spectral_indices}
\end{equation}
where $\nu_{450} = 665.0$\,GHz and $\nu_{850} = 349.5$\,GHz are the mid-frequencies of the SCUBA-2/POL-2 passbands.
Throughout this subsection, we apply the Stokes-$I$ criteria of the Method~3 mask of subsection~\ref{ss:P_ratio}: $I/\sigma_I > 50$ and the absolute brightness floor at both wavelengths.
We do not apply a cut on the signal-to-noise ratio of $PI$.
However, for the spectral-index analysis, we keep the brightest 10\% rather than 5\% of the pixels in each source (Paper~S2 sub-subsection~\ref{S2-sss:SpIdx_Global}).
The statistical sample is the {\sc SpIdx} sample of 35 sources (subsection~\ref{ss:samples}).

The polarization-fraction ratio compares how steeply the polarized and total dust emission rise with frequency between 850 and 450\,$\mu$m.
Writing the polarization-fraction ratio of subsection~\ref{ss:P_ratio} in terms of the spectral indices $\alpha$ and $\gamma$ connects the observations to the framework for grain physics adopted throughout the paper:
\begin{equation}
\frac{P_{450}}{P_{850}} = \left(\frac{\nu_{450}}{\nu_{850}}\right)^{\gamma - \alpha} = 1.90^{\gamma - \alpha}.
\label{eq:Pratio_SpIdx}
\end{equation}
For the same pair of measurements, $\gamma<\alpha$ means that polarized intensity rises less steeply with frequency than total intensity.
Equivalently, $P_{450}<P_{850}$: the polarization fraction is higher at the longer wavelength.
The same polarization fractions in the two bands correspond to $\gamma=\alpha$.
We cross-check equation~(\ref{eq:Pratio_SpIdx}) against the spatially resolved measurements of $\alpha$ and $\gamma$ described below.
 
For emission described by a modified blackbody, $I_\nu \propto \nu^{\beta} B_\nu(T_{\rm d})$ with dust opacity spectral index $\beta$, the spectral index of total intensity separates into a term for the opacity and one for the shape of the Planck function,
\begin{equation}
\alpha = \beta + C(T_{\rm d}),
\qquad
C(T_{\rm d}) \equiv
\frac{\log\left[B_{\nu_{450}}(T_{\rm d})/B_{\nu_{850}}(T_{\rm d})\right]}
     {\log(\nu_{450}/\nu_{850})},
\label{eq:planck_correction}
\end{equation}
where $C(T_{\rm d})$ is the shape factor of the Planck function between the two POL-2 bands ($C \approx 1.01$ at $T_{\rm d} = 15$\,K and $1.45$ at $25$\,K).
This relation is also given in Paper~S2 equation~(\ref{S2-eq:planck_correction}) and Paper~T equation~(\ref{T-eq:planck_correction}).
The observed slope of the intensity spectrum, $\alpha$, is thus the opacity slope, $\beta$, plus the slope of the Planck function, $C(T_{\rm d})$.
The slope of the Planck function approaches 2 in the Rayleigh--Jeans limit and falls well below 2 at the dust temperatures of molecular clouds.
We refer to obtaining $\beta$ by subtracting $C(T_{\rm d})$ from $\alpha$ as the $C(T_{\rm d})$ correction.
Paper~T~subsection~\ref{T-ss:flux} establishes the extended-source calibration used for $\alpha$, $\beta$, and $\gamma$.
 
Our primary observable, $\gamma - \alpha$, is independent of the $C(T_{\rm d})$ correction adopted for a common dust temperature.
The principal conclusions of this paper therefore do not depend on the particular value assumed for that temperature.

Paper~S2 (subsection~\ref{S2-ss:S2_SpIdx}) reports the dust temperatures for individual sources and comparisons of the resulting $\beta$ values among environments.
Uncertainty in dust temperature broadens $\beta$ for an individual source by $\sigma_{\beta, T} \approx 0.35$ (Paper~S2 sub-subsection~\ref{S2-sss:SpIdxFramework}).
This uncertainty cancels exactly from $\gamma - \alpha$ provided the dust within a given beam is isothermal.
If the components along the line of sight differ \emph{both} in temperature and in polarization efficiency, $\alpha$ and $\gamma$ sample different moments of the temperature distribution, and the cancellation no longer holds.
Differences in polarization angle between components at different temperatures can also affect our conclusions (subsection~\ref{ss:Discussion_Tmixing} and appendix~\ref{app:Tmixing}).
 
The ratio $R_{\rm p}$ and the difference $\gamma - \alpha$ therefore describe the same polarization measurement in the two bands at each line of sight.
Interpreting this measurement in terms of dust opacity or grain alignment requires additional information about how the material along each line of sight is weighted in thermal and polarized emission.

The F2 analysis uses the statistical sample of 35 sources with $N_{\rm samp} \geq 5$, selected by the criteria in Paper~S2 (subsection~\ref{S2-ss:S2_SpIdx}).
Across this sample, the medians of the Stokes~$I$ and polarized-intensity indices measured for individual sources are
\begin{equation}
\langle\alpha\rangle \;=\; 3.33 \;\pm\; 0.11_{\rm sys} \;\pm\; 0.12_{\rm stat}
\quad (\text{range } 2.06\text{--}4.99),
\label{eq:alphaGlobal}
\end{equation}
\begin{equation}
\langle\gamma\rangle \;=\; 2.91 \;\pm\; 0.11_{\rm sys} \;\pm\; 0.13_{\rm stat}
\quad (\text{range } 1.59\text{--}5.45),
\label{eq:gammaGlobal}
\end{equation}
respectively.
The medians of $\alpha$ and $\gamma$ for individual sources are the primary observables used throughout this paper.
 
For the individual indices, the flux calibration determines $\delta_{\rm sys}$.
The calibration tolerances of 17\% and 7\% in the two bands (Paper~T~subsection~\ref{T-sec:uncertainty}) contain two components \citep{Mairs2021}.
The scatter of the calibrator flux conversion factors (FCFs) varies independently between observations.
This scatter contributes $\simeq 0.26$ to $\alpha$ or $\gamma$ for a single observation.
The contribution decreases as $1/\sqrt{N}$ over the $N$ observations coadded into a field, and averages further over the 35 sources, to $\lesssim 0.03$ in the median.

The 5\% uncertainty of the Uranus flux model, together with the POL-2 attenuation factors, is common to all data.
It shifts every source by the same amount and therefore does not average down.
It gives an uncertainty of $0.11$ when the two bands are treated as independent and thus determines $\delta_{\rm sys} = 0.11$ for the medians.
Both calibration components enter $\alpha$ and $\gamma$ in common and cancel in their difference, whose floor is $\sigma(\gamma - \alpha) \lesssim 0.04$.
However, the residual 450 \, $\mu$m Stokes-$Q/U$ zero-level offset acts on $\gamma$ alone.
The statistical term $\delta_{\rm stat}$ is the Maritz--Jarrett standard error of the median of the 35 values for individual sources (``empirical SEM'').
 
The median of the spectral-index differences for individual sources is
\begin{equation}
\gamma - \alpha \;=\; -0.51 \;\pm\; 0.04_{\rm sys} \;\pm\; 0.10_{\rm stat}
\quad (\text{range } -1.14 \text{ to } +1.49),
\label{eq:gammaMinusAlphaGlobal}
\end{equation}
where $\delta_{\rm stat}=0.10$ is the Maritz--Jarrett standard error of the median of the 35 $\gamma - \alpha$ values for individual sources.
The values in equations~(\ref{eq:alphaGlobal})--(\ref{eq:gammaMinusAlphaGlobal}) are the medians of the measured values for individual sources.
We also treat pixels with polarized intensity below three times its uncertainty as $3\sigma$ limits on $\gamma$.
At the two extreme placements of these limits, the median in equation~(\ref{eq:gammaGlobal}) spans $2.85$--$3.04$, and the median in equation~(\ref{eq:gammaMinusAlphaGlobal}) spans $-0.57$ to $-0.45$.
Equation~(\ref{eq:Pratio_SpIdx}) then gives $P_{450}/P_{850} = 1.90^{-0.51} = 0.72 \pm 0.02_{\rm sys} \pm 0.05_{\rm stat}$, or $0.69$--$0.75$ at the two extremes of the $3\sigma$ limits.
The predicted ratio agrees with the directly measured Method~3 sample median, $R_{\rm EW} = 0.69 \pm 0.02_{\rm sys} \pm 0.03_{\rm stat}$ (subsection~\ref{ss:P_ratio}), within the combined systematic and statistical uncertainties of the two approaches.
The principal F2 result is that 24--27 of 35 sources (69\,\%--77\,\%) lie below the $\gamma = \alpha$ line, i.e.\ $\gamma < \alpha$.
This range spans the two extreme placements of the $3\sigma$ limits where the polarized intensity lies below its noise floor.
This result is shown in figure~\ref{fig:alpha_vs_gamma} and tested in appendix~\ref{app:S1_secondary}.
 
At the measured medians, a minority of sources (ten of 35; eight to eleven with the $3\sigma$ limits at their two extremes) lie above the $\gamma = \alpha$ line in figure~\ref{fig:alpha_vs_gamma} ($\gamma > \alpha$, corresponding to $P_{450}/P_{850} > 1$).
The error bars for five of them (Orion~A Field~5, Serpens~Main, the Rosette Molecular Cloud, GC~Clouds~E and~F, and GC~Middle) reach $\gamma = \alpha$.
The remaining five lie above the line by more than their individual statistical errors.
The combined systematic floor of $\sigma(\gamma - \alpha) \lesssim 0.04$ from Paper~T~subsection~\ref{T-sec:precisionfloor} is small compared with the typical statistical uncertainty in $\gamma - \alpha$ for individual sources ($\delta_{\rm stat} \sim 0.13$).
This statistical uncertainty arises mainly from the uncertainty
in $\gamma$.
Including the systematic floor does not change which of these
ten sources lie above $\gamma = \alpha$ by more than $1\sigma$.
Paper~S2 tabulates these five sources as $\gamma > \alpha$ at $1\sigma$ and discusses their environments.
The seven CMZ sources have a median $\gamma-\alpha$ of $-0.10$, above the sample median, but the present sample does not establish a difference between disk and CMZ clouds (Paper~S2 subsection~\ref{S2-ss:S2_EnvStrat}).
We discuss the sources above the $\gamma=\alpha$ line as exceptions to the dominant trend in subsection~\ref{ss:Discussion_GammaAlpha}.

\begin{figure*}
\centering
\includegraphics[width=0.95\textwidth]{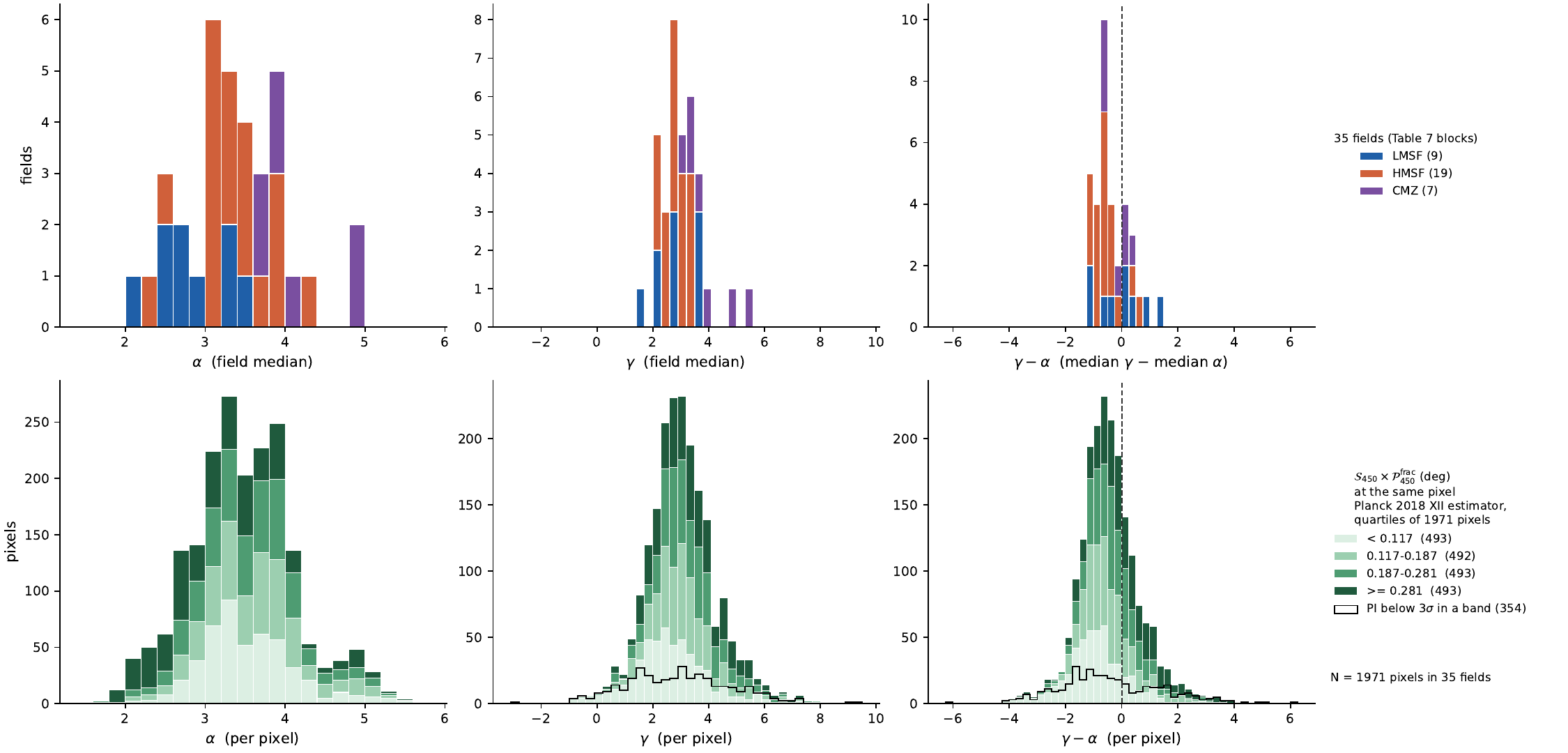}
\caption{Distributions of the spectral indices of the {\sc SpIdx} sample.
Top row: one entry for each source (35 sources), with the definitions of Paper~S2 table~\ref{S2-tbl:SpIdx_Global}.
$\alpha$ and $\gamma$ are the medians for each source, and $\gamma-\alpha$ is the median of $\gamma$ minus the median of $\alpha$.
Bars are stacked by environment (LMSF 9, HMSF 19, CMZ 7).
Bottom row: one entry for each pixel (1971 pixels), stacked by the quartiles of $\mathcal{S}_{450}\times\mathcal{P}^{\rm frac}_{450}$ measured at the same pixel (bin edges in degrees in the legend).
The shading is not a classification of physical state.
The black outline marks the 354 pixels whose polarized intensity lies below $3\sigma$ in at least one band.
For these pixels, $\gamma$ is an upper or lower limit.
The $\gamma-\alpha$ for individual pixels and the shading share $P_{450}$ algebraically, so the two covary in part by construction.
The two rows summarize the measurements differently: the top row uses a difference of medians for each source, whereas the bottom row shows the distribution of differences for individual pixels.
The G28.34+0.06 mosaic contributes 121 pixels (6.1\,\%) and the five largest fields 25\,\%.
The OMC~1 and Orion~A Field~4 grids are offset by $3\farcs0$, so that 70 pairs of 12\arcsec\ pixels overlap; both fields are plotted.
Dashed lines mark $\gamma-\alpha=0$.
}
\label{fig:spidx_hist}
\end{figure*}

Figure~\ref{fig:spidx_hist} shows the distributions underlying these medians.
The top row contains the values for the 35 sources, stacked by environment; the bottom row contains the 1971 pixels of the {\sc SpIdx} selection.
The median of $\gamma-\alpha$ over individual pixels is $-0.51$, equal to the median of the values for individual sources in equation~(\ref{eq:gammaMinusAlphaGlobal}).
Pixels with polarized intensity below three times its uncertainty in at least one band, for which $\gamma$ is a limit, dominate the wings.
They account for 102 of the 152 pixels whose $\gamma-\alpha$ differs from the median by two or more, compared with 18\,\% of all pixels.
In this pixel sample, $\gamma-\alpha$ covaries with the proxy for grain alignment efficiency, $\mathcal{S}_{450}\times\mathcal{P}^{\rm frac}_{450}$ (Spearman $\rho=+0.36$ over the 1971 pixels).
The covariation is partly algebraic because both quantities contain $P_{450}$.
We discuss its physical interpretation in Paper~S2.

The relevant observational question is whether residual errors in the 450\,$\mu$m measurements could mimic the lower polarization fraction relative to 850\,$\mu$m and the negative difference between the spectral indices of polarized and total intensity.
We adopt the systematic floors $\sigma(P_{450}/P_{850}) \lesssim 0.02$ and $\sigma(\gamma - \alpha) \lesssim 0.04$ from Paper~T~subsection~\ref{T-sec:precisionfloor} (Paper~T table~\ref{T-tbl:systematics}) as $\delta_{\rm sys}$ in equation~(\ref{eq:gammaMinusAlphaGlobal}) and the propagated $P_{450}/P_{850}$ above.
The residual zero-level offsets in Stokes $Q/U$ at 450\,$\mu$m dominate both floors.
Neither floor includes a difference in the modulation efficiency of POL-2 between the two bands (sub-subsection~\ref{sss:class2_instrumental}).
These floors are an order of magnitude below the detected signals $P_{450}/P_{850} \approx 0.46$--$1.28$ and $|\gamma - \alpha| \approx 0.51$.
The sample median of $\gamma-\alpha$ is also resolved against its statistical uncertainty, the Maritz--Jarrett standard error of 0.10 over the 35 source entries (interquartile range 0.78).
No plausible revision of the $Q/U$ zero-level offset correction can reverse the sign of the effect, $\gamma < \alpha$ in 69\,\%--77\,\% of sources, or remove its central magnitude, $\gamma - \alpha$ between $-0.57$ and $-0.45$.

Paper~S2 (subsection~\ref{S2-ss:S2_SpIdx}) compares the spectral-index method with the estimators of the polarization-fraction ratio in subsection~\ref{ss:P_ratio} for individual sources.
They agree within the combined systematic and statistical uncertainties for L1688 Ophiuchus~A and for the sample medians.
The difference measured across the sample is the observational result.
To attribute it to dust physics, we must show, on the same selected lines of sight, that residual $Q/U$ zero-level offsets and errors in the beam response cannot account for it.
Multiplicative calibration terms that cancel in a ratio must be distinguished from the residual offsets and the errors in the beam response.

We examine whether the inferred spectral indices change systematically as the precision of the polarized measurements improves.
Noise can most strongly distort the ratio of polarized intensities where the intrinsic polarization is weak.
Polarized intensities are intrinsically weaker than total intensities, so the spectral index $\gamma$ is more susceptible to biases associated with $S/N$ than the Stokes~$I$ spectral index $\alpha$.
At low $S/N$, the modified-asymptotic debiasing applied to $PI$ does not completely remove the residual Ricean bias.
Fluctuations in $PI_{450}$ and $PI_{850}$ can therefore propagate as a systematic downward bias in $\gamma$.
 
We do not apply a cut on $S/N(\gamma)$ to the F2 statistical sample (Paper~S2 subsection~\ref{S2-ss:S2_SpIdx}).
Paper~T (sub-subsection~\ref{T-ss:spidx_snrbias}) uses running medians to assess how each index depends on its own formal $S/N$.
A flat running median alone does not establish an unbiased estimate because $S/N(\gamma) = |\gamma|/\sigma_\gamma$ contains $\gamma$ itself.
We therefore test the result directly by varying the selection (Paper~T sub-subsection~\ref{T-ss:spidx_snrbias_sens}).
Applying a common threshold to all sources and varying it over $4 \le S/N(\gamma) \le 8$ retains at least one pixel in 28--34 of the 35 sources.
The threshold moves the sample median of $\gamma-\alpha$ from $-0.51$ to between $-0.40$ and $-0.26$ and gives $\gamma<\alpha$ in 65\,\%--79\,\% of the retained sources, compared with 71\,\% without a threshold.
The sign of the F2 offset is therefore preserved at every threshold tested, whereas its amplitude depends on the pixel selection.
A stable estimate of the spectral index over the precisely measured part of a source helps identify the range over which the polarization spectrum can be interpreted reliably.
The $S/N(\gamma)$ threshold test complements the noise model.
However, the adopted selection still limits the conclusions to the cloud lines of sight that remain measurable at 450\,$\mu$m.
The result is a shallower spectrum of polarized emission than of total dust emission in most of the sampled sources, with a higher polarization fraction at 850 than at 450\,$\mu$m.

\subsection{Polarization-angle dispersion function $\mathcal{S}$ and $\mathcal{S}\times\mathcal{P}^{\rm frac}$}
\label{ss:S_dispersion}
 
We use $\mathcal{S}$ to test for geometric depolarization in individual pixels.
A pixel with small $\mathcal{S}$ lies in a region where the projected field is ordered on the 12\arcsec\ scale of the pixel grid, whereas a pixel with large $\mathcal{S}$ lies where the angle changes between neighboring positions.
If such changes continue below the beam scale, averaging within the beam reduces the polarization fraction regardless of grain physics.
We probe geometric depolarization with the $P$--$\mathcal{S}$ relation in each band, as described below.
 
We quantify the variation of the $\vec{B}$-field orientation between neighboring positions in the plane of the sky with the polarization-angle dispersion function of \citet{Planck2020-XII} (see also \citealt{Hildebrand2009}).
Following Paper~S2 (sub-subsection~\ref{S2-sss:S_definition}), we calculate it by
\begin{equation}
\mathcal{S}(\mathbf{r}) \;=\; \sqrt{\frac{1}{N}\sum_{i=1}^{N}\left[\chi(\mathbf{r}+\boldsymbol{\delta}_i)-\chi(\mathbf{r})\right]^{2}},
\label{eq:S_def}
\end{equation}
Here $\chi(\mathbf{r})$ is the polarization angle at pixel $\mathbf{r}$.
The sum runs over the $N$ valid pixels $\mathbf{r}+\boldsymbol{\delta}_i$ among the eight nearest neighbors on the 12\arcsec\ grid (a lag of one pixel), with angle differences wrapped to $(-90\arcdeg,+90\arcdeg)$.
We require $N \geq 5$ and use the value debiased for noise, $(\mathcal{S}^2-\sigma_{\mathcal{S}}^2)^{1/2}$, set to zero where $\mathcal{S}<\sigma_{\mathcal{S}}$.
We propagate $\sigma_{\mathcal{S}}$ from the angle uncertainties for individual pixels using equation~8 of \citet{Planck2020-XII}.
We obtain the angle uncertainties from the Stokes $Q,U$ noise model of Paper~T~subsection~\ref{T-ss:noisemodel}.
Thus $\mathcal{S}$ is non-negative, and a completely random angle field gives $\mathcal{S}\approx 52\arcdeg$ ($\pi/\sqrt{12}$).
We calculate $\mathcal{S}$ in the same way at 450 and 850\,$\mu$m, so differences between the bands should reflect field structure rather than choices of method.

A low polarization fraction can result from either inefficient grain alignment or cancellation within a disordered $\vec{B}$-field.
Combining the polarization fraction with the local angle dispersion provides a complementary way to examine these two contributions.
We use the product $\mathcal{S}\times\mathcal{P}^{\rm frac}$ as a proxy for grain alignment efficiency, following \citet{Planck2020-XII}.
\citet{Fissel2016} analyzed the joint dependence of the polarization fraction on $\mathcal{S}$ and column density, and $\mathcal{S}$ itself builds on the angular dispersion function of \citet{Hildebrand2009} and \citet{Houde2009}.
The product is constructed so that the effect of field tangling on the scale of the beam largely cancels.
A persistent variation may then reflect intrinsic grain alignment efficiency.

To interpret this proxy, we need a range in both $\mathcal{S}$ and $\mathcal{P}^{\rm frac}$ that only the full census of 69 sources provides, including environments that span the range of $\mathcal{S}$ across the sample.
We therefore present its distribution among sources in Paper~S2 subsection~\ref{S2-ss:S2_SP} and its dependence on environment in Paper~S2 subsection~\ref{S2-ss:S2_EnvStrat}.
In the present paper, we use the proxy only in the comparison with $\gamma-\alpha$ for individual pixels (subsection~\ref{ss:SpIdx}, figure~\ref{fig:spidx_hist}).
Variations of the product do not necessarily provide a unique measure of alignment efficiency.
Its interpretation must remain consistent with the measured $P$--$\mathcal{S}$ relation and with the unresolved structure along the line of sight.

We examine whether regions with greater angle dispersion also have lower polarization fractions, and whether that association appears at both POL-2 wavelengths.
Disorder in the $\vec{B}$-field can reduce the observed polarization when different orientations are averaged within the beam or along the line of sight.
We analyzed the correlation between the polarization fraction $P$ and the polarization-angle dispersion function $\mathcal{S}$ at each wavelength to probe geometric depolarization from averaging tangled fields within the beam.
This analysis does not test whether geometric depolarization drives the observed $R_{\rm p} < 1$.
 
\begin{figure}[htbp]
\centering
\includegraphics{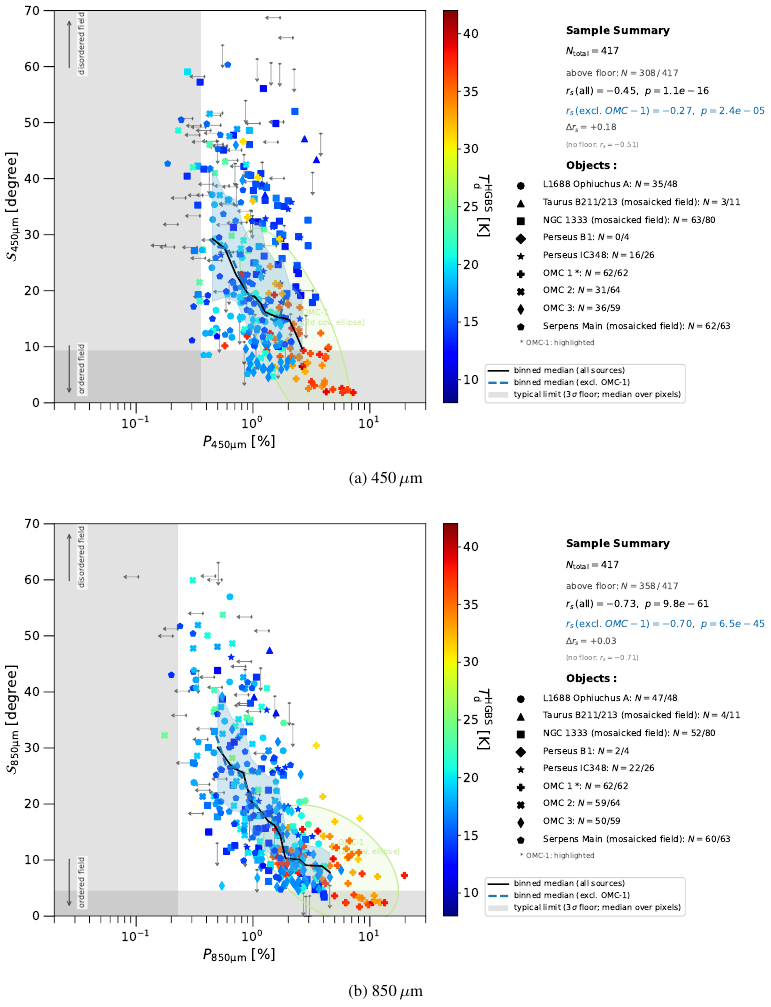}
\caption{
Polarization fraction versus angle dispersion $\mathcal{S}$ at \textbf{(a)} 450 and \textbf{(b)} 850\,$\mu$m for Method~3 pixels in nine HGBS regions, colored by HGBS dust temperature.
$\mathcal{S}$ is the debiased angle dispersion of equation~(\ref{eq:S_def}), computed as for the $\mathcal{S}$ tables of Paper~S2.
For pixels satisfying $P_i\geq3\,\delta P_i$ at the relevant band (Paper~T subsection~\ref{T-ss:noisemodel}) and with $\mathcal{S}$ detected at $3\sigma_{\mathcal{S}}$, the pooled Spearman coefficients are $r_s=-0.45$ ($N=308$, $p=1.1\times10^{-16}$) at 450\,$\mu$m and $r_s=-0.73$ ($N=358$, $p=9.8\times10^{-61}$) at 850\,$\mu$m.
Excluding OMC~1 gives $r_s=-0.27$ ($N=246$, $p=2.4\times10^{-5}$) and $-0.70$ ($N=296$, $p=6.5\times10^{-45}$), respectively, both statistically significant (subsection~\ref{ss:S_dispersion}).
OMC~1 samples the high-radiation regime with $T_d\gtrsim30$--40\,K.
Changing the polarization threshold to $2\sigma_P$ or $4\sigma_P$ alters $r_s$ by at most 0.05 at 450\,$\mu$m and 0.01 at 850\,$\mu$m.
Both panels draw from a parent sample of 417 pixel pairs.
Light gray shading marks the typical $3\sigma$ detection limits, obtained as medians of the floors for individual pixels: $P_{450} < 0.36\%$ and $P_{850} < 0.23\%$ (horizontal axes; $3\,\delta P_i$) and $\mathcal{S}_{450} < 9.4\arcdeg$ and $\mathcal{S}_{850} < 4.5\arcdeg$ (vertical axes; $3\sigma_{\mathcal{S}}$).
Each pixel is tested against its own floor, so individual limits scatter about these medians.
Pixels below their own floor are plotted at it as $3\sigma$ limits (small gray symbols with arrows): leftward arrows for $P_i < 3\,\delta P_i$ (55 at 450\,$\mu$m, 29 at 850\,$\mu$m) and downward arrows for undetected $\mathcal{S}$ (35 and 25).
Pixels below both floors (19 and 5) are not plotted.
Six of these limits at 450\,$\mu$m (five downward, one leftward) and four at 850\,$\mu$m (downward) lie above $\mathcal{S} = 70\arcdeg$, outside the plotted range.
The light-blue band shows the interquartile range of the binned median.
The labeled arrows at the left edge indicate ordered fields at low $\mathcal{S}$ and disordered fields at high $\mathcal{S}$.
These correlations describe associations with the resolved angle dispersion in the subsets selected by detections in each band.
The $p$-values calculated from these pixels are nominal, and the correlations alone do not establish cancellation of geometric depolarization in $R_{\rm p}$.
}
\label{fig:PvsS}
\end{figure}

Both subsamples are drawn from the {\sc DualBand} sample (417 sampled pixel pairs in the HGBS overlap region; subsection~\ref{ss:samples}).
For each correlation, we required $P_i \geq 3\,\delta P_i$ in the relevant band, following the detection mask for individual pixels in Paper~T~subsection~\ref{T-ss:noisemodel}.
We also required $\mathcal{S}$ before debiasing to exceed $3\sigma_{\mathcal{S}}$, following the detection criterion of the $\mathcal{S}$ tables in Paper~S2 (sub-subsection~\ref{S2-sss:S_definition}).
We detected statistically significant negative rank correlations at both wavelengths (figure~\ref{fig:PvsS}),
\begin{equation}
r_s(P_{450}, \mathcal{S}_{450}) = -0.45
\quad (N = 308, \; p = 1.1 \times 10^{-16}),
\label{eq:rS_P450}
\end{equation}
\begin{equation}
r_s(P_{850}, \mathcal{S}_{850}) = -0.73
\quad (N = 358, \; p = 9.8 \times 10^{-61}),
\label{eq:rS_P850}
\end{equation}
Both correlations remain statistically significant when we exclude OMC\,1, the warmest and most luminous single source in the HGBS subsample, from the sample above the detection floors.
The coefficients are then $r_s = -0.27$ ($N = 246$, $p = 2.4 \times 10^{-5}$) at 450\,$\mu$m and $r_s = -0.70$ ($N = 296$, $p = 6.5 \times 10^{-45}$) at 850\,$\mu$m.
The anticorrelation of equations~(\ref{eq:rS_P450})--(\ref{eq:rS_P850}) is therefore not produced by the high-radiation regime of OMC\,1 (characterized in our HGBS subsample by $T_{\rm d} \gtrsim 30$--$40$\,K and the highest $N_{\rm H_2}$).
At 850\,$\mu$m it is strong and nearly unchanged without OMC\,1, and at 450\,$\mu$m it is weaker without OMC\,1 but remains significant.
The anticorrelation also appears within individual clouds.
In each of the seven HGBS fields with at least five qualifying pixels, the coefficient is negative at both wavelengths (the medians of the coefficients for individual clouds are $-0.45$ at 450\,$\mu$m and $-0.55$ at 850\,$\mu$m).

The negative $P$--$\mathcal{S}$ associations in both bands are consistent with depolarization accompanying variations in the projected field across the HGBS subsample.
The $P$--$\mathcal{S}$ relation is a qualitative indicator of geometric depolarization rather than a quantitative measure of its magnitude.
The two correlations use different subsets selected by the detections in each band ($N = 308$ and 358), rather than the same pixels.
They describe how polarization and the measured angle dispersion vary across the selected lines of sight.
They do not directly quantify the geometric contribution to $R_{\rm p}$ or demonstrate its cancellation in the ratio.
The physical interpretation of the correlations depends on which cloud conditions contribute to the pooled relation.

$P$ and $\mathcal{S}$ are derived from the same Stokes $Q$ and $U$.
At low signal-to-noise ratio, the noise increases $\mathcal{S}$ while lowering $P$, which can by itself produce an anticorrelation.
As a check on this noise-induced anticorrelation, we raised the polarization threshold.
At 850\,$\mu$m, the coefficient stays between $-0.74$ and $-0.66$ for $P_i \geq 2$--$10\,\delta P_i$ (between $-0.70$ and $-0.66$ without OMC\,1), so the 850\,$\mu$m relation is robust to the choice of threshold.
At 450\,$\mu$m, the coefficient for the pooled sample stays between $-0.49$ and $-0.40$.
However, without OMC\,1, the anticorrelation weakens from $-0.34$ at $2\,\delta P_i$ to $-0.03$ at $10\,\delta P_i$ ($N = 92$).
With these pixels we cannot tell whether the weaker 450\,$\mu$m relation outside OMC\,1 reflects noise or changes in which pixels remain in the sample.

This check does not exclude every contribution related to noise.
Both $P$ and $\mathcal{S}$ share the noise in $Q$ and $U$.
The dispersion $\mathcal{S}$ also depends on the signal-to-noise ratios of the neighboring pixels, which a threshold at the central pixel does not control.
A higher threshold also changes the composition of the retained sample.

Unlike the $P$--$\mathcal{S}$ relation, the $T_{\rm d}$--$P$ and $N_{\rm H_2}$--$P$ correlations for individual pixels depend on OMC\,1 (appendix~\ref{app:tier2_env}, table~\ref{tbl:Tier2}; subsection~\ref{ss:Discussion_theory}).
A contribution of OMC\,1 that acts as a common multiplicative factor at both wavelengths would cancel in the ratio $R_{\rm p}$ between the bands.
Pooled pixels mix relations within clouds with differences between clouds, and adjacent pixels are not independent.
The coefficients are therefore descriptive, and their $p$-values are nominal.
 
\noindent
Taken together, the three principal findings of subsections~\ref{ss:bfield_consistency}--\ref{ss:S_dispersion} indicate that the polarized emission at 450 and 850\,$\mu$m shows similar projected field orientations yet a shallower effective spectrum than the bulk dust emission.
These findings are similar projected field orientations, within $10\arcdeg$ in 10 of 13 targets (F1, subsection~\ref{ss:bfield_consistency}); the prevalence of $R_{\rm p} < 1$, with the $P$--$\mathcal{S}$ probe of geometric depolarization (F3, subsections~\ref{ss:P_ratio} and \ref{ss:S_dispersion}); and $\gamma < \alpha$ in 69\,\%--77\,\% of the statistical sample (F2, subsection~\ref{ss:SpIdx}).
We interpret this combined evidence in section~\ref{sec:discussion}, where we also discuss the high-radiation regime at the warm end of the sample.
\section{Discussion}
\label{sec:discussion}
 
Our polarimetric survey at two wavelengths establishes three observational constraints on the physics of submillimeter polarization, which we summarize as Findings~1--3 in subsection~\ref{ss:FourFindings}.
We regard a wavelength-dependent alignment efficiency linked to the grain-size distribution as the most straightforward interpretation of these constraints (subsections~\ref{ss:Discussion_GammaAlpha}--\ref{ss:Discussion_theory}).
We cannot entirely exclude an origin in compositional variations or a contribution from multi-component temperature structure (subsections~\ref{ss:Discussion_Tmixing} and~\ref{ss:Discussion_DiscriminatingPower}).
Subsection~\ref{ss:Discussion_Tmixing} shows that, for a common polarization angle, temperature mixing reaches the observed median $\gamma-\alpha$ only if the warm dust is polarized less efficiently than the cold dust.

\subsection{Summary of the Three Findings}
\label{ss:FourFindings}
Section~\ref{sec:results} establishes three observational facts that constrain the origin of submillimeter polarization.
We summarize them in the order of the physical argument in subsections~\ref{ss:Discussion_GammaAlpha}--\ref{ss:Discussion_theory}, rather than as presented in section~\ref{sec:results}.
Finding~1 ($\vec{B}$-field angle consistency, from subsection~\ref{ss:bfield_consistency}) constrains the projected field orientations at the two wavelengths.
Finding~2 ($\gamma < \alpha$, from subsection~\ref{ss:SpIdx}) provides the principal diagnostic from the spectral indices.
Finding~3 ($R_{\rm p} < 1$ prevails, from subsections~\ref{ss:P_ratio} and \ref{ss:S_dispersion}) adds the ratio diagnostic, which shows no detected dependence on the angle difference between the bands.

\paragraph*{Finding 1: Similar Projected Field Orientations at the Two Wavelengths.}
The polarization angles at 450 and 850\,$\mu$m agree to within $|\chi_{450} - \chi_{850}| \lesssim 10\arcdeg$ in most of the sample (figure~\ref{fig:PA450vsPA850}).
This value is small compared to the angular range $180\arcdeg$ and lies within the combined systematic budget.
The median $|\Delta\chi|$ is $6.7\arcdeg$, slightly above the $4\arcdeg$--$6\arcdeg$ processing floor of Paper~T (equation~\ref{T-eq:precision_floors}), and the signed median is $+1.3\arcdeg$.
The Stokes $Q,U$ noise model gives a median per-pixel statistical uncertainty of $\sigma_{\Delta\chi} \approx 3.5\arcdeg$ (subsection~\ref{ss:bfield_consistency}; Paper~T~subsections~\ref{T-ss:noisemodel}--\ref{T-ss:offset}), about a third of the $10\arcdeg$ tolerance.
Ten of the thirteen regional medians lie within the adopted tolerance of $10\arcdeg$, and the larger medians occur in the other three regions where the polarized intensities are weak and the angle uncertainties for individual pixels are greater.
The Monoceros~R2 median is $12\arcdeg$ before and $4.0\arcdeg$ after the 450\,$\mu$m Stokes-$Q/U$ zero-level offset correction (subsection~\ref{ss:bfield_consistency}).
This agreement holds across the intensity ratios, polarization ratios, and environmental conditions in our sample.
We interpret this agreement as indicating that the net field direction in the plane of the sky changes little between the two wavelengths.
The wavelength dependence of the polarization fraction most likely reflects intrinsic grain properties rather than geometrically distinct field configurations, although individual regions may contain localized exceptions.

\paragraph*{Finding 2: Systematic $\gamma < \alpha$ Across Diverse Environments.}
The polarized-intensity spectral index $\gamma$ is smaller than the total-intensity spectral index $\alpha$ in 24--27 of the 35 sources in the {\sc SpIdx} sample (69\%--77\%; figure~\ref{fig:alpha_vs_gamma}).
The medians of the values for individual sources are $\langle\alpha\rangle = 3.33 \pm 0.11_{\rm sys} \pm 0.12_{\rm stat}$ and $\langle\gamma\rangle = 2.91 \pm 0.11_{\rm sys} \pm 0.13_{\rm stat}$ (subsection~\ref{ss:SpIdx}).
The statistical uncertainties are the standard errors (SEs) of the medians over the 35 values for individual sources; the interquartile ranges are 0.75 and 0.72.
The median difference over the sources is
\begin{equation*}
\gamma - \alpha \;=\; -0.51 \;\pm\; 0.04_{\rm sys} \;\pm\; 0.10_{\rm stat}.
\end{equation*}
With pixels below the noise floor in polarized intensity entered as $3\sigma$ limits, these medians span $\langle\gamma\rangle = 2.85$--$3.04$ and $\gamma - \alpha = -0.57$ to $-0.45$.
This systematic difference is observed across low-mass cores, intermediate-mass regions, high-mass star-forming regions, and Galactic Center clouds.
We infer that polarized emission has a shallower spectral dependence than the total intensity, consistent with a dominant contribution from the aligned grain population.

\paragraph*{Finding 3: $P_{450}/P_{850} < 1$ Prevails.}
The polarization-fraction ratio $P_{450}/P_{850}$ is less than unity for the majority of pixels and sources, with the sample median $P_{450}/P_{850} = 0.69 \pm 0.02_{\rm sys} \pm 0.03_{\rm stat}$ and values for individual sources spanning $0.46$--$1.28$ (subsection~\ref{ss:P_ratio}).

Comparing $R_{\rm p}$ with $|\Delta\chi|$ tests a particular signature of a geometric contribution to the polarization spectrum.
If integration along the line of sight through differently oriented fields, or averaging spatially distinct structures within the beam, produced the wavelength dependence of polarization, we would expect systematic variations in $\Delta\chi$ with $P_{450}/P_{850}$.
The negative $P$ -- $\mathcal{S}$ associations in each band (subsection~\ref{ss:S_dispersion}), present with or without OMC \,1, are consistent with depolarization accompanying variations in the projected field.
They do not measure the relative depolarization at the two wavelengths or demonstrate its cancellation in $R_{\rm p}$.
We do not detect a systematic dependence of $P_{450}/P_{850}$ on $|\Delta\chi|$ in this comparison (subsection~\ref{ss:Iratio}).
Together with Finding~1, this non-detection indicates that $P_{450}/P_{850} < 1$ most likely reflects intrinsic grain physics.
We consider size-dependent alignment efficiency to be the most natural interpretation.
The non-detection constrains geometric explanations that would produce such a dependence, but does not exclude depolarization by unresolved components with similar projected orientations.

In their respective samples, similar polarization angles constrain the orientation of the $\vec{B}$-field traced by the emission, whereas a lower 450-to-850\,$\mu$m polarization ratio restricts the relative spectral contributions of the polarized and total dust emission.
Combining these constraints narrows the possible origins of the polarization spectrum.

Finding~1 constrains changes in the net projected field direction between the bands and thus disfavors geometric effects that would rotate the polarization angles.
It does not establish identical weights along the line of sight at the two wavelengths.
In Finding~3, we do not detect a dependence of the polarization ratio $P_{450}/P_{850}$ on the angle difference between the bands.
This disfavors a geometric origin that would produce such a dependence and is consistent with intrinsic grain physics.
Finding~2 is most simply interpreted in terms of size-dependent alignment.
This picture is consistent with the findings if the grains contributing most strongly to polarization have a shallower effective emission spectrum.
Emission from the aligned grain population then appears to rise less steeply with frequency than that from the bulk dust, producing $\gamma < \alpha$.
The observations constrain that picture without uniquely identifying the grain population or excluding unresolved thermal, compositional, and geometric contributions.

We interpret these findings in terms of grain alignment in subsection~\ref{ss:Discussion_GammaAlpha} and RAT theory in subsection~\ref{ss:Discussion_theory}.
Subsection~\ref{ss:Discussion_Tmixing} examines temperature mixing along the line of sight.
We then discuss the discriminating power of the observations and the remaining degeneracies in subsection~\ref{ss:Discussion_DiscriminatingPower}.
Section~\ref{sec:summary} summarizes the principal results and benchmarks.

\subsection{Size-Dependent Grain Alignment: Origin of $\gamma < \alpha$}
\label{ss:Discussion_GammaAlpha}
 
The systematic finding that $\gamma < \alpha$ in 69\%--77\% of sources (24--27 out of 35; Finding~2; figure~\ref{fig:alpha_vs_gamma}) demonstrates that polarized emission has a shallower spectral dependence than total-intensity emission.
In this subsection, we argue that this result is most simply explained by size-dependent grain alignment, where larger grains align more efficiently than smaller grains.
We first show why a geometric origin is disfavored, and then develop an interpretation in terms of size-dependent alignment.
Subsection~\ref{ss:Discussion_DiscriminatingPower} discusses the remaining degeneracies in the properties of the dust.

A dependence of the polarization fraction on wavelength can arise if the two bands give unequal weights to magnetic components with different orientations.
The angle comparison and the polarization--dispersion relation probe observable consequences of this possibility, although neither resolves the structure within the beam or along the line of sight.
The similar polarization angles in the two bands disfavor line-of-sight field mixing that would rotate the net angle between the bands.
The absence of a detected $R_{\rm p}$--$|\Delta\chi|$ correlation does not show a signature of beam averaging that would produce such rotations.
Neither test excludes line-of-sight depolarization itself, including wavelength-dependent depolarization by components with similar projected orientations or by two nearly orthogonal components at different temperatures (appendix~\ref{app:Tmixing_r1}).
 
Our data also probe geometric depolarization from tangling of the $\vec{B}$-field through the angle dispersion $\mathcal{S}$.
With the detection mask for individual pixels from Paper~T~subsection~\ref{T-ss:noisemodel}, we found statistically significant negative rank correlations between the polarization fraction and $\mathcal{S}$ at both wavelengths (figure~\ref{fig:PvsS}).
The coefficients are $r_s = -0.45$ ($N = 308$, $p = 1.1 \times 10^{-16}$) at 450\,$\mu$m and $r_s = -0.73$ ($N = 358$, $p = 9.8 \times 10^{-61}$) at 850\,$\mu$m.
Both remain significant without OMC\,1, with $r_s = -0.27$ and $-0.70$, respectively (subsection~\ref{ss:S_dispersion}).
These associations are consistent with depolarization accompanying variations in the projected field in both bands, but they do not measure the relative depolarization at the two wavelengths.
We regard intrinsic grain properties, rather than primarily geometric artifacts, as the most natural explanation of Finding~2, on the basis of Findings~1 and~3 (subsection~\ref{ss:FourFindings}).
 
The negative rank correlation between $P$ and $\mathcal{S}$ has a distinct role among the three diagnostics introduced in section~\ref{sec:intro}: the $P$--$\mathcal{S}$ relation is a qualitative indicator of geometric depolarization.
The other two, $P_{450}/P_{850}$ and $\gamma - \alpha$, quantify the wavelength dependence directly.
Their signatures persist in subsamples that include OMC\,1 and those that exclude OMC\,1 (subsection~\ref{ss:Discussion_theory}).
If the depolarization indicated in both bands acts as the same multiplicative factor at both wavelengths, the factor cancels in $R_{\rm p}$.
Such an equal factor is a limiting model, not a result established by the observed $P$--$\mathcal{S}$ correlations.
We do not detect a signature in the polarization angles that requires a geometric origin for the predominantly higher polarization fraction at 850 than at 450 \, $\mu$m.
The measurements leave unresolved how strongly mixing along the line of sight and averaging over different physical scales contribute to individual sources.

We also consider how the survey selection function (section~\ref{sec:obs}) affects the frequency of $\gamma<\alpha$ reported in subsection~\ref{ss:SpIdx}.
The selection on Stokes~$I$ at 450\,$\mu$m preferentially samples bright interiors of clouds with active star formation and substantial local radiation fields.
The statistical sample is therefore weighted toward environments where RATs are expected to align grains efficiently according to their size.
The predominance of $\gamma<\alpha$ in this sample (69\,\%--77\,\%; 24--27 out of 35) demonstrates that the effect is common in the regions of high surface brightness currently accessible to JCMT polarimetry in both bands.
Our sample does not test whether $\gamma<\alpha$ is equally frequent in colder, more diffuse cloud envelopes.
The polarized emission at the shorter wavelength is much weaker there, so such a comparison will require substantially deeper observations.

Monoceros~R2 provides a useful test of how sensitive the angle comparison between the bands is to the 450\,$\mu$m Stokes-$Q/U$ zero-level offset correction.
Without the correction, its median absolute angle difference exceeds the $10\arcdeg$ physical tolerance.
Its median $|\chi_{450}-\chi_{850}|$ is $12\arcdeg$, with a median signed angle difference of $+8.1\arcdeg$ (appendix~\ref{app:F1_uncorrected}).
After the correction, the median $|\chi_{450}-\chi_{850}|$ falls to $4.0\arcdeg$, below the adopted tolerance (subsection~\ref{ss:bfield_consistency}).
We therefore do not attribute the angle difference before correction alone to grain physics.
Mon~R2 also lies well below $\gamma=\alpha$, with $\gamma-\alpha=-1.04$.
This value is strongly negative, although L1688 Ophiuchus~A has the most negative difference, $-1.14$.

We treat Monoceros~R2 as a test case for both the sensitivity of the F1 angle comparison to the $Q/U$ zero-level offset correction and the lower edge of the F2 ($\gamma<\alpha$) distribution.
Distinguishing radiation anisotropy, grain growth, and thermal stratification along the line of sight in an individual source requires modeling beyond the scope of this survey.
Paper~S2 table~\ref{S2-tbl:influential_source_policy} summarizes how Mon~R2 and the other influential sources, OMC\,1 and Sgr\,B2, are treated in the BISTRO {\sc PolSrc} census.

The observation $\gamma<\alpha$ means that polarized intensity rises less steeply with frequency than total intensity, so the polarization fraction is higher at 850 than at 450\,$\mu$m.
With a geometric origin disfavored by Findings~1 and~3, this difference in frequency dependence can arise through two complementary mechanisms in a multicomponent dust model.
 
First, if polarization is dominated by well-aligned large grains \citep{LazarianHoang2007,HoangLazarian2016,Andersson2015}, and if those grains exhibit a shallower effective emissivity spectrum than the smaller grains that dominate the bulk emission, then the polarization-weighted spectral index $\gamma$ will naturally be lower than the mass-weighted $\alpha$.
Whether the aligned grains have such a flatter spectrum depends on the dust model.
In the models of \citet{Ysard2019}, silicate grains containing aromatic-rich amorphous carbon have a far-infrared opacity spectral index of $\approx 1.35$, compared to $\approx 2.1$ for bare amorphous silicates; neither porosity nor ice mantles change the index.
In the dust-evolution models of \citet{Kohler2015}, accretion of aliphatic-rich carbon mantles and coagulation into aggregates raise the index from $\sim 1.5$ to $\sim 2$.
In the astrodust model of \citet{Hensley2023}, large grains are a single composite material.
In this model, the polarization fraction of the diffuse ISM changes only from 0.96 to 1.01 times its 850\,$\mu$m value between 250\,$\mu$m and 3\,mm.
If polarized emission preferentially traces grains with a flatter opacity spectrum than those that dominate the total intensity, $\gamma < \alpha$ follows.
 
Second, the intrinsic polarization efficiency may decrease with frequency.
In the RAT framework, larger grains are expected to align more efficiently in stronger radiation fields \citep{LazarianHoang2007,HoangLazarian2016}.
Both POL-2 bands lie deep in the Rayleigh regime for the grain sizes considered ($2\pi a/\lambda \lesssim 0.014$ at 450\,$\mu$m for grain radii $a \lesssim 1\,\mu$m), where the absorption cross-section per unit mass is independent of grain size at fixed composition \citep{Draine2006}.
The wavelength dependence therefore cannot arise because each wavelength directly selects a different grain size.
This dependence must enter through differences in temperature or in the opacity law between the aligned grains and the bulk dust.
This size-dependent alignment produces an effective wavelength dependence of the polarization efficiency.
This effective dependence reduces $\gamma$ relative to $\alpha$ even without requiring the same spectral behavior for the aligned and bulk-emitting dust components.

Multiwavelength polarimetry of NGC~2071 combined HAWC+ (154, 214\,$\mu$m) and POL-2 (850\,$\mu$m) data.
The polarization fraction decreases from 154 to 214\,$\mu$m (median $P_{214}/P_{154} = 0.48$--$0.63$), and the dust models with uniform alignment efficiency compared in that study could not reproduce this ratio \citep{Fanciullo2022b}.
The dust models \citet{Fanciullo2022b} compared to their data were developed for regions with moderate column density.
In contrast, grains in the dense regions sampled here may grow to $\sim 1\,\mu$m and become substantially colder \citep{Ysard2019}.
In addition, the 154 and 214\,$\mu$m bands trace warmer dust than the POL-2 bands do.
The present survey therefore probes a different part of the grain population, and a size-dependent signature here does not require one at the shorter wavelengths.
In the POL-2 bands, the polarization fraction is lower at 850 than at 450\,$\mu$m in NGC~2071 \citep{Lyo2021} and in L43 \citep{Scheiter2026}.
Our Method~1 ratios for these two sources, which are not part of the 35-source sample, are also above unity (Paper~S2 subsection~\ref{S2-ss:S2_P_ratio}).
In OMC\,1, by contrast, the polarization fraction increases with wavelength from the HAWC+ to the POL-2 bands in the dense gas and decreases in the warm regions, as mapped against dust temperature and H$_2$ column density in figure~2 of \citet{Tram2024}.
At some sky positions, the spectrum appears to be V-shaped, decreasing with wavelength from 54 to $\sim 300\,\mu$m and increasing at longer wavelengths \citep{Tram2024}.
For OMC\,1, the median pixel ratio in the Method~1 mask, $\tilde{R}_1=0.58\pm0.04$, likewise indicates a higher polarization fraction at 850 than at 450\,$\mu$m.
This direction, which also holds in most of our sources, is consistent with the rise of the V-shaped polarization spectrum at long wavelengths.
Earlier measurements of several molecular clouds, taken together, suggested such a spectrum with a minimum near 350\,$\mu$m \citep{Vaillancourt2008,Gandilo2016}.
However, our measurements in two bands alone do not establish the presence or wavelength of a minimum.
\citet{Gandilo2016} noted that the minimum had been seen directly only in OMC\,1.
They found a relatively flat spectrum without a pronounced minimum in Vela~C, so the V shape cannot be considered universal.

The direction of the change in the polarization fraction between the two bands therefore differs among regions.
In our 35-source sample, $\gamma < \alpha$ in 69\%--77\% of sources (24--27 out of 35) within the bright regions accessible to JCMT polarimetry in both bands.
The 12\farcs6 beam samples physical scales that vary by nearly two orders of magnitude at the adopted reference distances (section~\ref{sec:obs}).
It resolves substructure within dense cloud cores at $\sim 0.008$\,pc for nearby sources ($139$\,pc; L1688 Ophiuchus~A), but averages over regions on the scale of clumps at $\sim 0.5$\,pc in the CMZ ($8.3$\,kpc; Sagittarius~B2).
Despite this range in physical scale, the $R_{\rm p} < 1$ and $\gamma < \alpha$ behavior is recovered across the sampled environments.
This result is consistent with size-dependent alignment being a widespread, though not necessarily universal, physical characteristic of interstellar dust in star-forming regions.
 
We do not detect a monotonic dependence of $\gamma-\alpha$ on physical beam size over that range (Paper~S2: Spearman $\rho = 0.18$, $p = 0.30$), but this non-detection places only a weak constraint on such a dependence.
The fitted slope is $0.09 \pm 0.19$ per dex, with a $95\%$ interval of $-0.30$ to $+0.48$; over the $\sim$1.7 dex spanned here, this interval still allows a change comparable to the measured spectral-index difference itself.
The range of beam scales therefore remains a limitation of this comparison; the non-detection does not remove it.

A minority of sources exhibit $\gamma > \alpha$, including several objects in the Galactic Center environment.
These exceptions do not change the dominant trend, $\gamma < \alpha$, in the statistical sample, but indicate that the polarization spectrum is not strictly universal.
Paper~S2 (subsections~\ref{S2-ss:S2_SP} and \ref{S2-ss:S2_EnvStrat}) discusses these outliers in the context of their environments.

The systematic result $\gamma < \alpha$ holds across 69\%--77\% of sources (24--27 out of 35).
Combined with constraints on geometric effects (Findings~1, 3 and the $P$--$\mathcal{S}$ relation as an indicator of geometric depolarization), it indicates that submillimeter polarization is most likely dominated by well-aligned large grains whose polarized emission exhibits a shallower effective spectral dependence than the bulk dust emission.
This size-dependent alignment provides a unified explanation for $P_{450}/P_{850} < 1$ and offers a quantitative benchmark for theoretical dust models.

\subsection{Implications for Radiative Alignment Torque Theory}
\label{ss:Discussion_theory}
 
We use RAT theory as a physically motivated interpretation of the difference between the bands.
This theory provides the leading theoretical framework for understanding grain alignment in the interstellar medium \citep{LazarianHoang2007,HoangLazarian2016,Andersson2015}.
Radiative torques provide a physical reason for grain alignment to depend on grain size and on the local radiation field.

The observed polarization spectrum can test this picture only in combination with assumptions about which grains dominate the polarized emission and how their optical properties vary.
These observations do not uniquely determine the grain-size distribution or the function describing alignment efficiency.
Subsections~\ref{ss:Discussion_Tmixing} and~\ref{ss:Discussion_DiscriminatingPower} address the remaining degeneracies in dust properties and in the structure along the line of sight.
Several aspects of our results agree with standard RAT predictions, while others require the qualifications discussed below.
 
First, the systematic finding that $\gamma < \alpha$ (Finding~2) is consistent with size-dependent alignment efficiency, a core prediction of RAT theory.
\citet{HoangLazarian2016} showed that alignment efficiency increases strongly with grain size because larger grains experience greater radiative torques relative to gas damping.
Our observation that polarized emission has a lower spectral index than total intensity supports this prediction indirectly, through the dependence of the effective opacity slope on size or structure.
Second, the similar projected field orientations at the two wavelengths (Finding~1) are consistent with both wavelengths tracing the same population of aligned grains.
This similarity is expected if alignment is dominated by a single mechanism operating across grain sizes.
 
The anticorrelation between the intensity ratio and the polarization ratio for individual pixels (subsection~\ref{ss:Iratio}) expresses the same $\gamma - \alpha$ structure algebraically, rather than providing a separate test.
For completeness, the coefficients are $r_s(I_{450}/I_{850}, R_{\rm p}) = -0.49$ for the full HGBS subsample and $-0.35$ without OMC\,1 (figure~\ref{fig:IratioRp}).
The analysis of individual clouds in Paper~S2 (sub-subsection~\ref{S2-sss:S2_source_aware_correlation}) finds a median correlation near zero ($r_s = -0.03$, 4 of 7 fields with the same correlation sign as the full pooled sample).
It interprets the pooled relation as a mixture of differences between clouds and structure within OMC\,1.
The $I_{450}/I_{850}$--$R_{\rm p}$ relation persists in subsamples that include OMC\,1 and those that exclude OMC\,1.
By contrast, the $T_{\rm d}$--$P$ and $N_{\rm H_2}$--$P$ correlations in each band depend on OMC\,1 (see below).
This persistence is consistent with interpreting the signal of size-dependent grain alignment as a property of the bulk sample rather than of a single luminous source.
The measurements are compatible with radiative alignment acting on grains whose effective emission spectrum differs from that of the bulk dust.
They do not uniquely determine the alignment mechanism, and the pooled correlation between the ratios does not independently identify it.

OMC\,1, the warmest and most luminous source in the HGBS subsample, defines the high-radiation end of our {\sc DualBand} sample of 417 pixel pairs in the HGBS overlap region.
The warm, luminous interior of OMC\,1 extends the sample into conditions where radiation may affect grain alignment differently than in colder clouds.
We therefore need to distinguish its contribution from trends that characterize the other nearby regions.
The $T_{\rm d}$--$P$ and $N_{\rm H_2}$--$P$ correlations in each band \citep[historically the most-cited diagnostics of grain alignment in submillimeter polarimetric studies; see][]{Andersson2015,Pattle2023PPVII} are sensitive to whether OMC\,1 is included.
Excluding OMC\,1 reverses the sign of the $T_{\rm d}$--$P_{450}$ rank correlation, from $r_s = +0.13$ (full HGBS subsample, $N = 343$) to $r_s = -0.29$ ($N = 281$).
This exclusion also reduces the strong $T_{\rm d}$--$P_{850}$ correlation from $r_s = +0.36$ ($N = 383$) to $r_s = +0.03$ ($N = 321$).
The $N_{\rm H_2}$--$P$ anticorrelations likewise strengthen at 450\,$\mu$m ($r_s = -0.07$ to $-0.46$) or change sign at 850\,$\mu$m ($r_s = +0.21$ to $-0.13$) once OMC\,1 is excluded (full tabulation in appendix~\ref{app:tier2_env}, table~\ref{tbl:Tier2}).
 
The sensitivity of the correlations in each band to OMC\,1 is compatible with RAT alignment models in which stronger radiation lowers the minimum size of aligned grains when other local conditions are held fixed \citep{Hoang2021}.
However, radiative torque disruption (RATD) can also deplete large grains in strong radiation fields, so the net polarization response depends on both alignment and grain survival \citep{Hoang2021}.
The $T_{\rm d}$--$P$ and $N_{\rm H_2}$--$P$ relations in each band acquire contributions that are specific to OMC\,1 and absent in colder, less irradiated HGBS sources.
A contribution specific to OMC\,1 would cancel in the observables comparing the two wavelengths ($\gamma - \alpha$, $P_{450}/P_{850}$, and $r_s(I_{450}/I_{850}, R_{\rm p})$) if it acts as a common multiplicative factor at both wavelengths.
These observables have similar values in subsamples that include OMC\,1 and those that exclude OMC\,1.
We therefore interpret OMC\,1 as marking the high-radiation boundary of the physical conditions in which the size-dependent alignment picture holds, rather than as a counter-example to it.
Paper~S2 (subsections~\ref{S2-ss:S2_SpIdx} and \ref{S2-ss:S2_EnvStrat}) compares the polarization properties in each band across the LMSF, HMSF, and CMZ environments of the full {\sc PolSrc} sample.
OMC\,1 substantially changes several pooled environmental correlations.
We recommend that it be retained as a distinct observational regime in comparisons with alignment models.
The persistence of a pooled wavelength-ratio relation after its exclusion does not by itself identify a common local alignment response in the remaining clouds.
We defer comparison with previous studies in this area to Paper~S2.

Models that reproduce our observations would naturally include size-dependent alignment efficiencies, with larger grains better aligned, as implemented in current RAT theory.
They would further include environmental processing that modifies grain size distributions and optical properties, and saturation, in which the alignment efficiency plateaus at high radiation intensities.
Such models would need to reproduce the observables comparing the two wavelengths ($\gamma - \alpha$, $P_{450}/P_{850}$), which are common to the sample.
They would also need to reproduce the $T_{\rm d}$--$P$ and $N_{\rm H_2}$--$P$ behavior in each band that appears only in OMC\,1, the highest-radiation regime probed here (appendix~\ref{app:tier2_env}).
The robustness of $\gamma < \alpha$ across 69\%--77\% of sources (24--27 out of 35) provides a quantitative benchmark.
Dust models that cannot reproduce $\gamma - \alpha = -0.51 \pm 0.04_{\rm sys} \pm 0.10_{\rm stat}$ ($-0.57$ to $-0.45$ with the $3\sigma$ limits at their two extremes) and the corresponding $P_{450}/P_{850} = 0.69 \pm 0.02_{\rm sys} \pm 0.03_{\rm stat}$ (range among sources $0.46$--$1.28$) from first principles do not account for the observed polarization spectrum.
Reproducing $\gamma<\alpha$ with optically thin emission and a common polarization angle requires that the polarized emission arise preferentially from colder dust or from grains with a flatter effective emissivity spectrum (subsections~\ref{ss:Discussion_Tmixing} and~\ref{ss:Discussion_DiscriminatingPower}).
The flatter spectrum may arise from size-dependent alignment or from differences in composition between the aligned grains and the bulk dust (sub-subsection~\ref{sss:class3_dust}).
Subsection~\ref{ss:Discussion_Tmixing} first quantifies how much less efficiently the warm dust must be polarized than the cold dust for temperature mixing to reproduce the observed median $\gamma-\alpha$.
Subsection~\ref{ss:Discussion_DiscriminatingPower} brings together the remaining degeneracies in dust emissivity and temperature (Class~3), the geometric alternatives (Class~1), and the instrumental alternatives (Class~2) considered in this paper.
 
\subsection{Line-of-Sight Temperature Mixing}
\label{ss:Discussion_Tmixing}

To assess how temperature structure contributes to the observed wavelength dependence of polarization, we examined the combined emission from cold and warm dust along the same line of sight.
We extend a single-temperature description to two components representing cold dust in shielded regions and warmer dust heated by embedded stars or external radiation.
Both components contribute to polarized thermal emission.
We assume that the two components share a common polarization angle.
We denote the effective polarization efficiencies of the cold and warm components by $p_{\rm c}$ and $p_{\rm w}$, respectively.
Each efficiency is the ratio of polarized to total emission for that component.

\begin{figure}[htbp]
\centering
\includegraphics[width=0.6\textwidth]{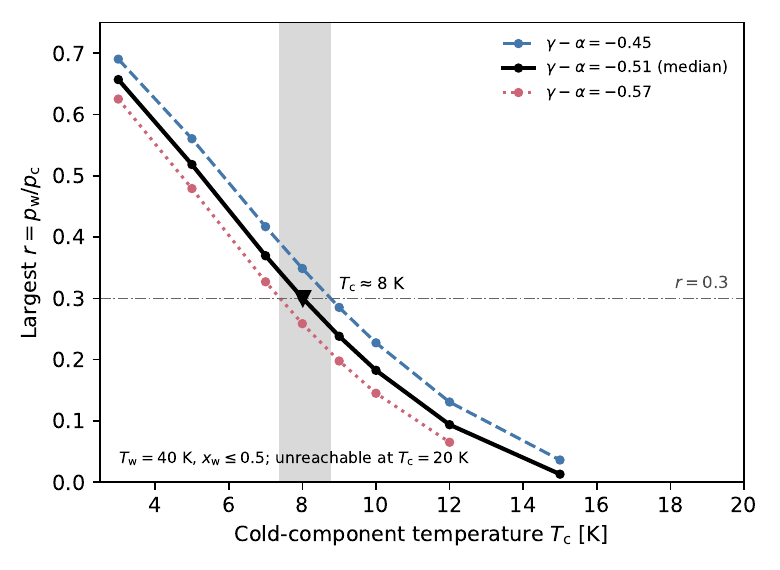}
\caption{Largest ratio $r=p_{\rm w}/p_{\rm c}$ of the effective polarization efficiencies of the warm and cold components for which the two-temperature model of appendix~\ref{app:Tmixing} reaches the observed median $\gamma-\alpha$, as a function of the cold temperature $T_{\rm c}$.
We maximize $r$ over warm mass fractions $x_{\rm w}\leq0.5$ and warm temperatures $T_{\rm w}=25$--40\,K; the maximum occurs at $T_{\rm w}=40$\,K at every cold temperature.
The solid curve is for the measured median, $-0.51$.
The dashed and dotted curves are for $-0.45$ and $-0.57$, respectively, the two ends of its range with the $3\sigma$ limits at their two extremes (equation~\ref{eq:gammaMinusAlphaGlobal}).
The horizontal line marks $r=0.3$.
The shaded band covers the thresholds of $T_{\rm c}$ below which the model with $r=0.3$ reaches each target, from 7.4\,K for $-0.57$ to 8.8\,K for $-0.45$.
The triangle marks 8.0\,K for the median.
No target is reached at $T_{\rm c}=20$\,K, and $-0.57$ is not reached at 15\,K.
The plotted values are those of table~\ref{tab:Tmix_required_r}.}
\label{fig:Tmix_rmax}
\end{figure}

Temperature mixing along the line of sight reproduces the observed median $\gamma-\alpha$ only if the warm dust is polarized less efficiently than the cold dust.
If $p_{\rm c}=p_{\rm w}$, then $PI_\nu$ is proportional to $I_\nu$, and $\gamma=\alpha$ identically for any temperatures and mass fractions (appendix~\ref{app:Tmixing_r1}).
A negative $\gamma-\alpha$ therefore requires a contrast $r=p_{\rm w}/p_{\rm c}<1$ between the effective polarization efficiencies of the warm and cold components.
Figure~\ref{fig:Tmix_rmax} shows the largest $r$ for which the model reaches the median as a function of the cold temperature $T_{\rm c}$, with the warm dust mass fraction $x_{\rm w}$ and temperature chosen to maximize $r$.
For $T_{\rm c}\geq10$\,K and a warm component at 40\,K or cooler, the median requires $r\lesssim0.18$ (appendix~\ref{app:Tmixing_contrast}).
A moderate contrast, $r=0.3$, suffices only if the cold component is colder than about 8\,K and carries most of the mass.
The limits on $r$ assume optically thin emission, a common polarization angle and dust opacity law, and equal modulation efficiency of POL-2 in the two bands (subsection~\ref{ss:Discussion_DiscriminatingPower}).

To test whether the mixtures are compatible with the observed thermal conditions, we compared them with the HGBS dust and \amm\ kinetic temperatures of four fields (appendix~\ref{app:Tmixing_data}).
In three of the four fields, no mixture with $\beta=2$ within the adopted model ranges both matches the two measured temperatures and reproduces the field's $\gamma-\alpha$ within $1\sigma$, even for unpolarized warm dust ($r=0$).
Within the adopted ranges, mixtures that match both measured temperatures reproduce the measured spectral-index difference in the fourth field, NGC~1333, within $1\sigma$ for $r\leq0.15$.
The test assumes equal gas and dust temperatures and \amm\ tracing the cold molecular gas.
We treat the result for OMC\,2 as tentative.
With the ranges widened from $T_{\rm w}=25$--40\,K and $x_{\rm w}\leq0.5$ to $T_{\rm w}=10$--100\,K and $x_{\rm w}\leq0.95$, mixtures reproduce the measured spectral-index differences in the three fields within $1\sigma$, but only with a very strong contrast, $r\leq0.041$.
We therefore find that, for these three fields, reducing $r$ alone is insufficient within the adopted ranges once the measured temperatures are imposed.

Within this model, mixing cold and warm dust can produce a polarization fraction that rises from 450 to 850\,$\mu$m when the cold dust is polarized more efficiently.
The warm dust then contributes a larger share of the total emission at 450 than at 850\,$\mu$m, lowering the polarization fraction at the shorter wavelength.

We speculate that the restrictions found in this test with the HGBS dust and \amm\ kinetic temperatures also apply to other LMSF and HMSF regions with comparable thermal conditions.
Our requirements for temperature coverage and agreement between the gas and dust temperatures limited this test to four fields; its applicability to other regions remains untested.
Whether this test can be extended to CMZ clouds is uncertain, particularly because the gas and dust temperatures can differ substantially there.

Neither the model calculation nor the test determines the origin of the required contrast in effective polarization efficiency (appendix~\ref{app:Tmixing_class3}).
We therefore need models that relate the thermal structure of a cloud to the polarized emission of its dust to assess how much temperature mixing contributes to the observed wavelength dependence.

\subsection{Discriminating Power and Remaining Degeneracies}
\label{ss:Discussion_DiscriminatingPower}
 
The observation $\gamma < \alpha$ in 69\,\%--77\,\% of the {\sc SpIdx} sample, supported by $P_{450}/P_{850} < 1$ in individual pixels, constrains but does not uniquely determine the underlying physics.
Our data establish the observed behavior.
The most straightforward interpretation, size-dependent grain alignment, requires confirmation from multiwavelength polarimetry extending to shorter (far-infrared emission) or longer (millimeter emission) wavelengths.
We group alternative explanations for $\gamma < \alpha$ into three classes: (Class~1) geometry and projection effects, (Class~2) instrumental and reduction systematics, and (Class~3) dust-emissivity and thermal degeneracies.
The analyses in subsection~\ref{ss:Discussion_GammaAlpha} and Paper~T disfavor or bound Classes~1 and~2, respectively, apart from the open terms named in sub-subsections~\ref{sss:class1_geometric} and~\ref{sss:class2_instrumental}.
For temperature mixing in Class~3, subsection~\ref{ss:Discussion_Tmixing} quantifies the required contrast in effective polarization efficiency and tests the mixtures against the dust and \amm\ temperatures of four fields.
Class~3 contains the remaining degeneracy that requires observations at additional wavelengths to resolve.
Table~\ref{tbl:degeneracy} summarizes the mechanisms in Class~3 and the observations that could distinguish them.

\subsubsection{Class~1: Geometric Origins}
\label{sss:class1_geometric}
Averaging within the beam, tangling of the $\vec{B}$-field along the line of sight, spatial mixing weighted by polarization, and depolarization induced by projection could in principle bias the wavelength dependence of polarization observables.
The present results constrain these geometric mechanisms.
Findings~1 and~3 show similar projected field orientations at the two wavelengths and no detected dependence of $P_{450}/P_{850}$ on the angle difference between the bands (subsection~\ref{ss:FourFindings}).
The $P$--$\mathcal{S}$ associations discussed in subsection~\ref{ss:Discussion_GammaAlpha} are consistent with depolarization in both bands, without demonstrating its cancellation in $\gamma-\alpha$ or $P_{450}/P_{850}$.
We therefore do not favor geometric effects as the origin of the $\gamma < \alpha$ trend across the survey, although they may contribute to the scatter among sources.

The effect of spatial averaging at different physical scales remains untested.
F1 and F3 compare the bands at a matched angular resolution and are insensitive to averaging at a \emph{physical} scale that varies with distance.
They therefore do not exclude a contribution from spatial mixing that depends on distance.
F1 also leaves a second possibility open.
A warm and a cold component with nearly orthogonal $\vec{B}$-field orientations can reproduce the observed median $\gamma-\alpha$ at equal polarization efficiency, over a limited range of warm fractions.
The orientations of the two components must differ by more than about $64\arcdeg$ for $T_{\rm c}=10$\,K and $T_{\rm w}=40$\,K, and by more than about $77\arcdeg$ for $T_{\rm c}=15$\,K and $T_{\rm w}=30$\,K.
In part of this range of orientations and warm fractions, the angles in the two bands differ by less than $10\arcdeg$ (appendix~\ref{app:Tmixing_r1}).

\subsubsection{Class~2: Instrumental and Reduction Systematics}
\label{sss:class2_instrumental}
The principal systematic uncertainties introduced by the reduction include differences in spatial-frequency filtering between the 450 and 850\,$\mu$m bands, absolute and relative flux calibration, matching the beam shapes, and residual additive zero-level offsets in the 450\,$\mu$m Stokes~$Q$ and~$U$ intensities.
These uncertainties affect polarization observables that compare the two bands (Paper~T~subsections~\ref{T-ss:beamval}, \ref{T-ss:flux}, and \ref{T-ss:noisemodel}).
Paper~T propagates the residual systematic uncertainties into those of the comparison between the bands.
The quantified contribution is much smaller than the observed $\gamma-\alpha\simeq-0.5$ and the amplitude of $P_{450}/P_{850}$ for individual sources.
The technical validation in Paper~T therefore bounds the multiplicative terms from flux calibration and the additive $Q/U$ zero-level offsets.
Two contributions remain to be quantified.
One is how residual errors in beam shape and filtering affect the ratios.
The other term, a difference in the modulation efficiency of POL-2 between the two bands, is not bounded.
Such a difference would scale $P_{450}/P_{850}$ by a common factor and shift $\gamma-\alpha$ by a common amount for all sources.
 
We interpret the measured levels of $\gamma-\alpha$ and $P_{450}/P_{850}$ on the assumption that the modulation efficiency of POL-2 is equal in the two bands.
The ranking of the sources and the correlations of $\gamma-\alpha$ and $P_{450}/P_{850}$ with source properties do not depend on this assumption.
The POL-2 attenuation factors describe only the loss of total power.
The modulation efficiency has not, to our knowledge, been calibrated in either band (Paper~T~subsection~\ref{T-ss:flux}).

\subsubsection{Class~3: Dust-Emissivity and Thermal Degeneracies}
\label{sss:class3_dust}
The degeneracies considered in this class (Class~3) are astrophysical in origin.
Temperature structure along the line of sight, effects of optical depth in the densest regions, and variations in grain composition can all modify the relation between total and polarized emission.
For a common polarization angle, temperature mixing reproduces the observed median $\gamma-\alpha$ only with a contrast in effective polarization efficiency between the warm and cold components (subsection~\ref{ss:Discussion_Tmixing}).
 
Differences in opacity slope among components along the line of sight do not generate $\gamma-\alpha\neq0$ if polarized intensity remains a wavelength-independent fraction of total intensity.
Composition becomes relevant only insofar as the grains dominating polarized emission have a different effective opacity law from those dominating total emission.
 
Beyond temperature mixing, differences in composition between the aligned grains and the bulk dust could also produce $\gamma < \alpha$.
We denote the opacity spectral indices of polarized and total emission by $\beta_{\rm P}$ and $\beta_{\rm I}$, respectively.
The indices $\beta_{\rm sil}$ and $\beta_{\rm C}$ refer to pure silicate and carbonaceous grain populations.
In the Rayleigh--Jeans limit valid for $h\nu \ll k_{\rm B}T_{\rm d}$ \citep[e.g.,][]{Hildebrand1983,Draine2011book}, $\gamma - \alpha \simeq \beta_{\rm P} - \beta_{\rm I}$.
A simple picture in which only silicate grains are aligned while carbonaceous grains remain unaligned therefore does not, by itself, produce $\gamma < \alpha$ for standard opacity indices ($\beta_{\rm sil} \gtrsim \beta_{\rm C}$).
Within this single-temperature picture, $\gamma < \alpha$ requires that the aligned grain population have a systematically smaller \emph{effective} opacity spectral index than the bulk dust.
Here, the effective index describes the population as a whole and can differ from the indices $\beta_{\rm sil}$ and $\beta_{\rm C}$ of pure silicate and carbonaceous grains.
Published models of submillimeter dust polarization differ in how they treat grain composition.
We first describe the wavelength dependence of the predicted polarization spectra and then relate their behavior between 450 and 850\,$\mu$m to the observed polarization-fraction ratio.
(i)~Models in which silicate and carbonaceous grains form separate populations and only the silicate grains are aligned \citep{Draine2009,Guillet2018,Lee2020,Tram2021} predict submillimeter polarization spectra whose shape depends on the adopted grain properties.
In models whose polarization spectrum peaks in the far-infrared or submillimeter, the polarization fraction falls toward longer wavelengths beyond the peak.
The peak moves to longer wavelengths as the dust becomes colder \citep{Guillet2018,Lee2020,Tram2021}.
The ratio is $P_{450}/P_{850} > 1$ when the peak lies shortward of 450\,$\mu$m, whereas cold dust can move the peak into the submillimeter (e.g., figures~13 and 15 of \citealt{Guillet2018}).
In those models in which the silicate grains contribute an increasing fraction of the emission as the wavelength increases, the polarization fraction instead increases steadily with wavelength through the submillimeter \citep{Draine2009}, so that $P_{450}/P_{850} < 1$.
(ii)~Models in which both the silicate and the carbonaceous grain populations are aligned predict a nearly flat submillimeter polarization spectrum \citep{Tram2021}.
The astrodust model of \citet{Hensley2023}, constructed for the diffuse ISM, treats large grains as a single composite material.
This model also predicts a nearly flat spectrum.
The astrodust model predicts that the polarization fraction changes only from 0.96 to 1.01 times its 850\,$\mu$m value between 250\,$\mu$m and 3\,mm.
(iii)~A model with a warm and a cold dust layer along the line of sight, in which the cold layer contains a silicate--carbon mixture, reproduces the polarization spectra of OMC\,1 that rise with wavelength beyond $\sim 300\,\mu$m, although it has difficulty with the steepest rises \citep{Tram2024}.
In the literature on radiative torques, ``composite'' grains often denote grains of low tensile strength \citep[e.g.,][]{Tram2021}.
Here we use the term only for composition.
The sample median of the ratio, $P_{450}/P_{850}=0.69$, corresponds to a polarization fraction that is typically higher at 850 than at 450\,$\mu$m.
This wavelength dependence is qualitatively consistent with the type~(i) models of \citet{Draine2009} and with model~(iii).
However, these models idealize the clouds: they do not include the internal structure, motions, or time evolution of the clouds.
On the observational side, the values for individual sources range from 0.46 to 1.28.
The uncertainty for an individual source due to residual $Q/U$ zero-level offsets is $\lesssim 0.05$ at the median polarized intensity of the sample (Paper~T subsection~\ref{T-sec:uncertainty}).
Paper~T tested the sensitivity to beam matching and spatial filtering for the polarization angles, but not for $P_{450}/P_{850}$ (Paper~T subsection~\ref{T-ss:beamval} and table~\ref{T-tbl:systematics}).
Given these uncertainties, models of each type may account for part of the present result, and none is decisively excluded by it.
Composition therefore offers a possible explanation for $\gamma < \alpha$ in addition to size-dependent alignment \citep{Andersson2015}.
However, the present data do not single out composition.
 
Non-radiative alignment mechanisms, such as mechanical alignment of irregular or helical grains by gas flows (often referred to as ``helical dust alignment''; \citealt{LazarianHoang2007L,Hoang2018,Das2016}), could in principle maintain grain alignment in environments where the radiation field is weak.
The positive correlation between $P_{850}$ and $T_{\rm d}$ in the full HGBS subsample, driven by OMC\,1, is consistent with enhanced RAT alignment.
It does not by itself establish RATs as the dominant alignment mechanism (appendix~\ref{app:tier2_env}).
This pooled correlation alone neither excludes mechanical alignment nor determines the importance of RATD, which can counteract the polarization increase associated with enhanced alignment \citep{Hoang2021}.
This correlation is carried entirely by OMC\,1: without it, $r_s = +0.03$ and $p = 0.55$ (Paper~S2 sub-subsection~\ref{S2-sss:S2_source_aware_correlation}).
Even this consistency with enhanced RAT alignment is therefore limited to conditions like those in OMC\,1.
Additionally, gas--grain drift \citep{Lazarian2008}, iron inclusions \citep{Draine1999,Lazarian2008}, and possibly cosmic-ray ionization may introduce secondary dependences of alignment efficiency on the environment.
These processes may be second-order modifiers rather than primary drivers of the systematic trends identified here.
To disentangle the relative contributions, we need polarimetric observations over a wider range of wavelengths than is accessible with POL-2 alone.
 
\begin{deluxetable*}{lll}
\tablecaption{Class~3 Dust-Emissivity and Thermal Degeneracies in $\gamma < \alpha$\label{tbl:degeneracy}}
\tablewidth{0pt}
\tablehead{
  \colhead{Mechanism} &
  \colhead{Predicted Signature} &
  \colhead{Discriminating Observation}
}
\startdata
Composition-dependent optical properties & Shallower polarized-emission spectrum & Far-IR emission polarimetry \\
Line-of-sight temperature mixing &
  \parbox[t]{0.30\textwidth}{\raggedright Warm and cold components contribute differently at 450 and 850\,$\mu$m.
  For a common polarization angle, a nonzero $\gamma-\alpha$ arises only if these components also have different effective polarization efficiencies.
  The observed median $\gamma-\alpha\simeq-0.5$ requires $p_{\rm w}/p_{\rm c}\lesssim0.18$ for $T_{\rm c}\geq10$\,K and $T_{\rm w}\leq40$\,K, or $T_{\rm c}\lesssim8$\,K for $p_{\rm w}/p_{\rm c}\approx0.3$ (appendix~\ref{app:Tmixing}).\strut} &
  \parbox[t]{0.34\textwidth}{\raggedright Dust and gas temperatures.
  Within the adopted model ranges ($T_{\rm c}=3$--20\,K, $T_{\rm w}=25$--40\,K, $x_{\rm w}\leq0.5$), in three of four fields (OMC\,2 tentative), no mixture matches the HGBS dust and \amm\ kinetic temperatures and reaches $\gamma-\alpha$ within $1\sigma$, for $\beta=2$.
  Mixtures with $T_{\rm w}=10$--100\,K and $x_{\rm w}\leq0.95$ reach these fields only for $p_{\rm w}/p_{\rm c}\leq0.041$.
  The test assumes equal gas and dust temperatures and \amm\ in the cold gas (appendix~\ref{app:Tmixing_data}).\strut} \\
Optical depth effects ($\tau \gtrsim 0.1$) & $\gamma - \alpha$ correlated with $N_{\rm H}$ & Optically thin tracers ($\lambda > 1$\,mm) \\
\enddata
\tablecomments{Three mechanisms involving dust emissivity and thermal structure that can produce $\gamma<\alpha$ (Class~3), considered alongside the geometric constraints of subsection~\ref{ss:Discussion_GammaAlpha} and the quantified instrumental contributions in Paper~T.
The observations do not uniquely distinguish these mechanisms.
Additional wavelengths would help constrain their relative contributions.}
\end{deluxetable*}

\section{Summary}
\label{sec:summary}
 
We have presented a polarimetric analysis of the BISTRO 450/850\,$\mu$m survey at two submillimeter wavelengths, focusing on the physical interpretation of size-dependent grain alignment.
Our analysis is based on a statistical subsample of 35 sources with well-constrained spectral indices, drawn from the full {\sc PolSrc} sample presented in Paper~S2.
We use the shared procedures for reduction and noise modeling in Paper~T.
Three principal observational findings constrain the physics of submillimeter polarization:
 
\begin{enumerate}
  \item \textit{Similar projected field orientations at the two wavelengths.}
The polarization angles at 450 and 850\,$\mu$m typically agree within $|\chi_{450} - \chi_{850}| \lesssim 10\arcdeg$, small compared to the $180\arcdeg$ range of polarization angles.
10 of 13 targets have a median $|\chi_{450} - \chi_{850}| < 10\arcdeg$, and the overall median is $6.7\arcdeg$.
The median $|\chi_{450} - \chi_{850}|$ is slightly above the $4\arcdeg$--$6\arcdeg$ processing floor of Paper~T, and the signed median is $+1.3\arcdeg$.
Larger differences arise where the polarized intensity is weak.
The median statistical uncertainty for individual pixels is $\sigma_{\Delta\chi} \approx 3.5\arcdeg$.
The Monoceros~R2 median is $12\arcdeg$ before and $4.0\arcdeg$ after the 450\,$\mu$m Stokes-$Q/U$ zero-level offset correction.
The net projected field direction thus changes little between the bands over most of the measured regional sample.
This constrains, but does not eliminate, geometric contributions to the wavelength dependence of the polarization fraction.
 
  \item \textit{Systematic $\gamma < \alpha$ across diverse environments.}
  The spectral index $\gamma$ of polarized intensity is smaller than the index $\alpha$ of total intensity in 24--27 of 35 sources (69\%--77\%).
  This range spans the two extreme placements of $3\sigma$ limits where the polarized intensity lies below its noise floor.
  The medians of the values for individual sources are $\langle\alpha\rangle = 3.33 \pm 0.11_{\rm sys} \pm 0.12_{\rm stat}$ and $\langle\gamma\rangle = 2.91 \pm 0.11_{\rm sys} \pm 0.13_{\rm stat}$.
  The statistical errors are SEs of the medians over the source values in the {\sc SpIdx} sample.
  The median of the differences for individual sources, not the difference of the two medians above, is $\gamma - \alpha = -0.51 \pm 0.04_{\rm sys} \pm 0.10_{\rm stat}$ ($-0.57$ to $-0.45$ with the $3\sigma$ limits at their two extremes, for which $\langle\gamma\rangle = 2.85$--$3.04$).
  The characteristic polarization ratio is $P_{450}/P_{850} = 0.69 \pm 0.02_{\rm sys} \pm 0.03_{\rm stat}$, with values for individual sources spanning $0.46$--$1.28$.
  Thus, the typical polarization spectrum in the two bands has a higher polarization fraction at 850 than at 450\,$\mu$m.
  This systematic difference is observed across low-mass cores (e.g., Ophiuchus~A), high-mass regions at intermediate distances (e.g., OMC~1), high-mass star-forming regions (e.g., NGC~6334), and Galactic Center clouds (e.g., Sagittarius~B2).
  We infer that polarized emission is dominated by well-aligned grains whose effective spectral dependence is shallower than that of the bulk dust emission.
 
  \item \textit{$P_{450}/P_{850} < 1$ prevails.}
  The polarization-fraction ratio does not show a systematic dependence on the angle difference $|\chi_{450} - \chi_{850}|$ between the bands.
  We detected statistically significant negative rank correlations between the polarization fraction and $\mathcal{S}$ at both wavelengths for the pixels above the detection floor in each band.
  They are $r_s = -0.45$ ($N = 308$, $p = 1.1 \times 10^{-16}$) at 450\,$\mu$m and $r_s = -0.73$ ($N = 358$, $p = 9.8 \times 10^{-61}$) at 850\,$\mu$m (subsection~\ref{ss:S_dispersion}).
  Both correlations remain significant without OMC\,1.
  These associations are consistent with depolarization accompanying variations in the projected field in both bands.
  However, these associations do not measure the relative depolarization at the two wavelengths or demonstrate its cancellation in the ratio.
  Together with Finding~1, the absence of a dependence on $|\chi_{450} - \chi_{850}|$ favors intrinsic grain physics---most naturally size-dependent alignment efficiency---over a geometric origin for the wavelength dependence of the polarization fraction.
\end{enumerate}
 
These three findings directly address the questions posed in section~\ref{sec:intro}.
Finding~1 shows similar projected field orientations at 450 and 850\,$\mu$m.
This disfavors geometric effects that rotate the angles as the dominant driver of the wavelength differences in polarization fraction.
Finding~2 demonstrates that the spectral index $\gamma$ of polarized intensity systematically falls below the index $\alpha$ of total intensity across diverse environments.
Finding~3 shows that the $P_{450}/P_{850}<1$ signature has no detected dependence on the angle difference between the bands, favoring intrinsic grain physics as its origin.
Together, these results favor size-dependent grain alignment in star-forming regions over a geometric origin for the wavelength dependence.
Our quantitative benchmarks are $\gamma - \alpha = -0.51 \pm 0.04_{\rm sys} \pm 0.10_{\rm stat}$, or $-0.57$ to $-0.45$ at the two extreme placements of the $3\sigma$ limits, and the sample median $P_{450}/P_{850} = 0.69 \pm 0.02_{\rm sys} \pm 0.03_{\rm stat}$, with a range among sources of $0.46$--$1.28$.
Theoretical dust models and magnetohydrodynamic simulations must reproduce these values within the combined systematic and statistical uncertainties.
 
Paper~S2 tests the {\em universality} of these findings (findings F4--F6).
F4 establishes the robustness of F1--F3 for the population represented by the BISTRO {\sc PolSrc} sample.
F5 tests whether the wavelength response depends on the regime.
The medians of $\gamma-\alpha$ and $R_p$ for the environment classes show no ordered LMSF--HMSF--CMZ sequence.
We do not detect a robust difference among the classes, although the high-mass class has the lowest median.
F6 describes how the polarization at each wavelength is associated with local $N_{\rm H}$ and $T_d$ within the Herschel Gould Belt Survey subsample of 9 fields.
The combined F1--F6 results link the detected signatures (F1--F3, this paper) to the tests across the population (F4--F6, Paper~S2).
Together, they form the observational benchmark that any candidate alignment mechanism must reproduce.
 
Temperature mixing along the line of sight reproduces the observed median $\gamma-\alpha\simeq-0.5$ only if the warm dust is polarized less efficiently than the cold dust, for a common polarization angle (subsection~\ref{ss:Discussion_Tmixing}).
If the cold and warm components have the same effective polarization efficiency and polarization angle, $\gamma=\alpha$ identically.
The observed value requires $p_{\rm w}/p_{\rm c}\lesssim0.18$ for a cold component at 10\,K or warmer and a warm component at 40\,K or cooler.
A moderate contrast, $p_{\rm w}/p_{\rm c}\approx0.3$, suffices only for a cold component colder than about 8\,K that contains most of the mass.

In three of the four fields tested, no mixture within the adopted model ranges both matches the HGBS dust and \amm\ kinetic temperatures and reproduces the field's $\gamma-\alpha$ within $1\sigma$, for $\beta=2$ (appendix~\ref{app:Tmixing_data}).
This test assumes equal gas and dust temperatures and \amm\ tracing the cold gas.
We treat its result for OMC\,2 as tentative.
With the ranges widened from $T_{\rm w}=25$--40\,K and $x_{\rm w}\leq0.5$ to $T_{\rm w}=10$--100\,K and $x_{\rm w}\leq0.95$, mixtures reproduce the measured spectral-index differences in the three fields within $1\sigma$, but only with a very strong contrast, $p_{\rm w}/p_{\rm c}\leq0.041$.
We therefore find that, within the adopted model ranges, reducing the polarization efficiency of the warm component alone does not account for these three fields once the measured temperatures are imposed.
The origin of the required contrast in effective polarization efficiency remains undetermined.
 
The spectral-index difference $\gamma - \alpha$, or equivalently $P_{450}/P_{850}$, provides a diagnostic of grain physics that does not require conversion using distance.
We also did not detect a trend in this quantity with the physical scale sampled by the beam.
Sources with $\gamma > \alpha$ may represent distinct physical conditions and warrant targeted follow-up.
This diagnostic has not yet been calibrated against independent evolutionary indicators, so we regard it as a hypothesis rather than an established metric.
 
We also note that the sample is heterogeneous in distance (139\,pc to 8.3\,kpc), in the physical scale sampled by the beam, and in evolutionary state.
The heterogeneity of the sample qualifies the physical interpretation.
The three findings themselves are stated in terms of directly measured quantities and do not assume a dust model.
 
We have demonstrated that the polarization ratio $P_{450}/P_{850}$, the spectral-index difference $\gamma - \alpha$, and the $P$--$\mathcal{S}$ relation used to test for geometric depolarization indirectly constrain size-dependent grain alignment without explicit normalization by dust column density or extinction.
Beyond the BISTRO {\sc SpIdx} subsample analyzed here, the three quantities comparing the two wavelengths, $R_{\rm p}$, $\gamma-\alpha$, and $|\chi_{450}-\chi_{850}|$, measure different aspects of the emission.
The first traces the net amplitude of the polarization spectrum.
The second traces the difference in emissivity behavior between polarized and total dust emission.
The third tests for differences in $\vec{B}$-field geometry between the two bands.
All three use polarimetry in the two bands alone, so any survey observing both bands can measure them.
They serve emission polarimetry as $p_V/A_V$ serves optical and infrared extinction studies.
 
In Paper~S2, we publish polarization maps, $\mathcal{S}$ maps, the proxy $\mathcal{S}\times\mathcal{P}^{\rm frac}$ for grain alignment efficiency, tables of spectral indices, and atlases of the $\vec{B}$-field at both wavelengths for 68 of the 69 sources, excluding W51.
Paper~S2 also compares low- and intermediate-mass star-forming regions, high-mass star-forming regions, and the Central Molecular Zone.
Paper~T presents the noise model, the 450\,$\mu$m Stokes-$Q/U$ zero-level offset correction, and the independent validation with SOFIA/HAWC+.

We distinguish four levels of claim strength according to how directly each result follows from the observations.
\textit{Directly measured} from the BISTRO data at matched resolution in the two bands are the polarization-angle differences $|\chi_{450}-\chi_{850}|$ for individual pixels, the spectral indices $\alpha$ and $\gamma$ for sources in the {\sc SpIdx} subsample, and the polarization-fraction ratios $R_p$ for individual pixels and sources.
\textit{Empirically inferred} from those measurements are the statements about the universality of F1--F3.
The field angles agree between the bands with a median $\widetilde{|\Delta\chi|} = 6.7\arcdeg$, slightly above the $4\arcdeg$--$6\arcdeg$ processing floor, across the sample of 21 sources with angles in both bands.
The result $\gamma<\alpha$ holds in 24--27 of 35 sources after accounting for the systematic floor, and $R_p<1$ in roughly four-fifths of the survey sources.

\textit{Physically interpreted} as a signature of size-dependent grain alignment is the combination of $\gamma<\alpha$, $R_p<1$, and no detected $R_p$--$|\Delta\chi|$ dependence.
This interpretation uses a model with two dust populations, consistent with RAT theory, in which the grains contributing most strongly to polarization have a shallower effective emissivity spectrum than the bulk.
\textit{Not yet uniquely determined} from these observations alone are the precise grain population beyond the interpretation in terms of size-dependent alignment, and the geometry of magnetic disorder in individual test cases such as Mon~R2.
The same holds for the contribution of temperature structure along the line of sight.
This contribution requires a contrast in effective polarization efficiency if the components share a polarization angle (subsection~\ref{ss:Discussion_Tmixing}).
We explicitly acknowledge these residual degeneracies in subsection~\ref{ss:Discussion_DiscriminatingPower}.
The observations also do not yet uniquely determine how much a difference in the modulation efficiency of POL-2 between the two bands contributes to the levels of $\gamma-\alpha$ and $R_p$.
We will make the data products described here for individual sources available electronically as supplementary material, including tables of angles in both bands, spectral-index statistics, and measurements of the polarization ratio summarized in table~\ref{tbl:S1_F1F3_master}.
 
\begin{acknowledgments}
RSF is deeply grateful to the medical staff who have continuously supported his recovery from cancer.
Their dedicated care made this work possible.
The authors thank Paul Ho, the former Director of the East Asian Observatory (EAO), for his full and unwavering support for this project, as well as the EAO staff for their continued assistance over a decade.
We also express our sincere gratitude to Dr.~Jessica Dempsey, former Deputy Director of the EAO, for her continuous and decisive leadership of the international instrument team throughout the commissioning and validation phase of POL-2.
Dr.~Dempsey sustained institutional support for the extensive test observations and performance assessments needed to establish POL-2 as a reliable scientific instrument, particularly to characterize instrumental polarization (IP).
She also enabled the participating teams to contribute their full capabilities.
In particular, she approved and supported dedicated IP validation experiments conducted with the JCMT dome cover removed, thereby enabling critical verification of the IP correction models developed independently by DB and PB.
We acknowledge Takeshi Inagaki, a UNIX systems engineer at the Subaru Telescope, who visited RSF for approximately two weeks each year from 2015 to 2020 and made substantial contributions to improving the efficiency and reliability of the data-analysis workflow.
We also thank the successive Directors of the Subaru Telescope, who not only approved Mr.~Inagaki's long-term annual visits, but also provided extensive in-kind support during JCMT observing runs from 2015 to 2019.
We gratefully acknowledge the Astronomy Data Center (ADC) of the National Astronomical Observatory of Japan (NAOJ) for long-term support and access to shared computing resources, including 32 UNIX machines and more than 100~TB of storage.
In 2022, Mr.~Inagaki also made \texttt{pol2map} run efficiently on all 32 of these machines, including the handling of jobs under the security settings of the ADC.
The authors are grateful to Drs.~Dan Bintley and Shaoliang Li (EAO), who generously shared their in-depth knowledge of SCUBA-2 instrumentation and the JCMT optics through close collaboration as members of the SCUBA-3 development team.
Their critical insights into instrumental characteristics and systematic effects informed the uncertainty analysis presented in this work.
We thank Prof.\ Doug Johnstone for helpful advice on this manuscript.
Discussions of observational uncertainties associated with submillimeter telescopes and bolometer cameras greatly benefited from interactions with Prof.~Ryohei~Kawabe and Dr.~Tai~Oshima (NAOJ).
We thank Prof.~Yoshito~Shimajiri (Kyushu Kyoritsu University, Japan) for discussions of bolometer imaging algorithms, Dr.~Hiroyuki~Hirashita (ASIAA, Taiwan) for discussions of dust properties, and Dr.~Riccardo~Cesaroni (INAF Arcetri, Italy) for advice on \emph{Herschel} data handling.
Discussions with Profs.~Hideki Takami and Yutaka Hayano (Subaru Telescope), drawing on their experience in adaptive optics development and operations, helped us understand atmospheric fluctuations at Maunakea relevant to submillimeter observations.
RSF expresses sincere gratitude to Prof.~Akira~Obuchi (Tokushima University, Japan) for extensive guidance and invaluable discussions on mathematical aspects relevant to this work.
Finally, we thank Mike Chen, Jungmi Kwon, Chi-Yan Law, Tetsuya Zenko, and Yapeng Zhang
for helpful discussions.
 
These observations were obtained by the James Clerk Maxwell Telescope, operated by the East Asian Observatory on behalf of The National Astronomical Observatory of Japan; Academia Sinica Institute of Astronomy and Astrophysics; the Korea Astronomy and Space Science Institute; the National Astronomical Research Institute of Thailand; Center for Astronomical Mega-Science (as well as the National Key R\&D Program of China with No.~2017YFA0402700).
Additional funding support is provided by the Science and Technology Facilities Council of the United Kingdom and participating universities and organizations in the United Kingdom and Canada.
The POL-2 data analyzed here were obtained under JCMT programs M16AL004, M17BL011, M20AL006 and M20AL018 (BISTRO Large Programs), together with E21AC002, E21BJ001, E21BJ002, M16AD003, M17AP019, M17AP067, M17AP073, M17AP074, M17BP037, M17BP041, M17BP046, M17BP050, M17BP070, M18AH27C, M18BD002, M18BP019, M18BP030, M18BP041, M18BP055, M19AP003, M19AP009, M19AP014, M19AP016, M19AP032, M19AP034, M19AP038, M19BP015, M19BP058, M19BP061, M20AP022, M20AP023, M20AP040, M20BP007, M20BP022, M20BP043, M20BP046, M21AH20A, M21AP016, M21AP021, M21BF004, M21BF005, M21BP004, M21BP045, M21BP083, M22AP018, M23BP011, R20XP002.
Programs M21BF004 and M21BF005, which observed fields in Orion~A, used observing time purchased through the EAO Individually Funded Program with JSPS KAKENHI Grant Numbers JP19H01938 and JP21H00045.
 
\textit{Use of AI tools.}
This study began in early 2016, and AI tools took part in it during roughly the last two years, from the autumn of 2024.
Their role changed as both the research and the tools developed.
We describe it here in some detail, because AI tools contributed substantially to the analysis code and to the text of all three papers of this series.
Before that period, until about 2023, RSF used the AI-based language tools Grammarly and DeepL under paid subscriptions.
He corrected the English of his drafts with Grammarly and checked their meaning by translating them back into Japanese with DeepL.
Some of the wording corrected with Grammarly probably remains in the present text, but it can no longer be identified.
 
Before AI tools were introduced, the observations and the basic data reduction had been completed by people.
All raw data were transferred from the JCMT Science Archive between 2017 March and 2024 February and reduced to Stokes $I$, $Q$, and $U$ maps (steps~1--3 of \texttt{pol2map}) by RSF on the computers of the NAOJ Astronomy Data Center.
No command or parameter of these steps was produced by an AI tool.
This reduction rests on the research of DB, who implemented \texttt{pol2map} in the Starlink SMURF package and advised on its use.
RSF produced the polarization-vector catalogs using his procedure; their scripts were later tidied with ChatGPT and refined with Claude Code.
The core analysis code was developed by RSF from 2015, together with Takeshi Inagaki until 2022, and RSF's notebooks of 2018--2024 underlie much of the present analysis.
From the autumn of 2024, AI tools took over part of the role that Mr.~Inagaki had played, helping RSF to debug this code and to run it in new computing environments.
Development of the analysis code proceeded iteratively between RSF and the AI tools, including substantial rewriting or re-implementation of parts of RSF's earlier code.
 
The principal tools were ChatGPT (OpenAI; GPT-4o from 2024 October, later GPT-5 and, from 2026 September, GPT-6 Astra) and Claude (Anthropic).
Claude was used through the AI assistant of the Warp terminal from the autumn of 2024 (Claude 3.5-series and other models; the model that answered cannot always be identified), through Claude Cowork, and through Claude Code.
The session logs of Claude Code, which survive from 2026 June onward, record the models Claude Opus 4.8, Claude Fable 5, Claude Opus 5, Claude Fable 5.1, and Claude Opus 5.5, with Claude Haiku 4.5 for auxiliary tasks.
The models used with Claude Code before 2026 June were not recorded.
The periods of use of these tools overlapped, and we do not assert the order in which they were adopted.
OpenAI Codex was used in 2026 September to review the work records and to draft working instructions.
In 2026 October, Codex also checked independently whether the cited papers support the statements for which we cite them (the model was not recorded in either period).
Gemini (Google) was used occasionally, including the summaries shown automatically with search results and Gemini Code Assist in a code editor.
Perplexity was used occasionally until 2026 May; it suggested points to check in the manuscripts, and text or arguments from it may have entered the papers.
 
The AI tools carried out work packages that could not have been completed by people in a reasonable time, and some work packages were led by them.
They shared the management of the large survey data set and the evaluation of calibration observations, for example of quasars, used as references when inspecting the science data.
They helped RSF to read and interpret DB's implementation in the Starlink software, the Starlink source code, and the accompanying documents, including e-mail correspondence.
They converted RSF's notebooks into stand-alone scripts and packages, wrote new analysis and plotting code to RSF's specifications, and modified and accelerated RSF's existing code.
They transcribed RSF's handwritten notes, articulated, organized, and consolidated RSF's ideas, and drafted and rewrote much of the English text.
In choosing the English language, RSF adopted many of their recommendations to match the strength of each claim to the statistical result.
The three papers were first planned as one, and their division was considered from late 2024 onward.
During this period the AI tools consolidated the many comments received in several rounds from more researchers than there are co-authors.
They were instructed to weigh each comment on its scientific merit alone, irrespective of who had made it, and this also eased communication in English among non-native speakers.
They kept the source counts and the numbers consistent among the text, tables, and figures of the three papers, a task that could not have been done by hand within realistic limits.
In the final stage they read the three manuscripts adversarially as simulated referees set up by RSF.
They also checked the cited statements against the cited papers, which RSF had obtained.
In 2026 October, they revised the English toward RSF's own style, as learned from his hand edits and his published papers.
The AI tools mainly handled short-term writing plans, while RSF set the medium- and long-term plans.
 
RSF set the astronomical motivation and the questions, specified the analyses, and verified the results from the astronomical point of view.
The verification from the data-science point of view was done mainly by the AI tools.
The statistical results were interpreted by RSF together with the AI tools.
RSF decided the outline of the astronomical conclusions, the AI tools then tested this outline adversarially, and RSF made the final decision on what to adopt.
RSF wrote the original versions of much of the analysis code, examined substantial parts of the code written or modified by the AI tools, and checked their outputs.
All visual checks, including separate displays of the original data, were made by RSF.
These checks found errors both in AI-written code and in RSF's earlier code, and the errors were corrected.
For example, RSF noticed that the statistics of one comparison of polarization angles were more tightly aligned than the maps suggested, which revealed a cut on the polarization-fraction ratio inherited from code written in 2021.
RSF also identified the cause of an error in which polarization-angle maps had been read in place of angle-dispersion maps.
 
The physical interpretation of the difference between the spectral indices of polarized and total intensity developed over several years in discussions among several of the authors, in the course of which RSF also used ChatGPT.
The contributions of the individuals and of the tool to these parts cannot now be separated, and we have not tried to reconstruct them from memory.
We assess the reliability of these parts from the analyses presented here and the authors' judgment, rather than their origin.
Whether, and how far, the co-authors used AI tools in preparing their own contributions and comments is not known to the first author.
The handling of unpublished data, manuscripts, and correspondence with these tools was judged comprehensively in light of the regulations of Tokushima University and the laws of Japan.
Under this judgment, RSF shared this material with the providers of these tools, mainly Anthropic, OpenAI, and Google.
The data were astronomical maps and catalogs that RSF had derived from raw data that are publicly available in the JCMT Science Archive.
RSF regards such sharing as a premise of using AI tools in research, and this study may have contributed modestly to the development of their models.
 
The figures in this paper were produced with Matplotlib or TikZ between 2026 January and 2026 October, that is, after AI tools had entered this work.
Figures~\ref{fig:S1_findings_overview}, \ref{fig:PA450vsPA850}, \ref{fig:IratioRp}--\ref{fig:Tmix_rmax}, and \ref{fig:Tmix_panels}--\ref{fig:F1_three_samples} were made with scripts written by Claude-based coding tools to RSF's specifications, in several cases re-implementing RSF's notebook code.
Figure~\ref{fig:S1_findings_overview} is a TikZ diagram drafted in the same way.
For submission, Claude Code typeset the TikZ diagrams and the figures with several panels as single figure files, without changing their content.
Figure~\ref{fig:Pratio_OphA_method3} was made with a notebook that RSF developed while using AI tools interactively.
The contributions of RSF and of the tools to it cannot be separated.
Figure~\ref{fig:CO32_map} was drawn with map-plotting code that RSF wrote before AI tools were used, from a contamination map computed with code developed with AI tools.
RSF inspected every figure against the underlying data.
 
No AI tool is an author of this paper.
The authors take full responsibility for its content, including every part prepared with the help of AI tools.
No output of these tools entered the paper without RSF's judgment, and RSF has read the whole text.
Without the commercially available models of OpenAI and Anthropic that took part in this work, these survey papers could not have been brought to submission before RSF's retirement.
\end{acknowledgments}
 
\section*{Funding}
RSF was supported by the Visiting Scholars Program provided by the NAOJ Research Coordination Committee, NINS (\mbox{NAOJ-RCC-23DS-050}).
RSF was supported by Grants-in-Aid for Scientific Research of Japan (JSPS KAKENHI Grant Numbers JP19H01938 and JP21H00045).
Team BISTRO-J gratefully acknowledges partial financial support from 260 individual contributors through a crowdfunding campaign.
 
\section*{Data Availability}
The raw SCUBA-2/POL-2 observations analyzed in this paper are available from the JCMT Science Archive hosted at the Canadian Astronomy Data Centre (CADC).
The \textit{Herschel} Gould Belt Survey column-density and dust-temperature maps used here are public.
The \amm\ kinetic-temperature maps of the Green Bank Ammonia Survey Data Release~2 \citep{Pineda2026} are available at \url{https://doi.org/10.11570/24.0091}.
 
\facilities{JCMT (SCUBA-2, POL-2), Herschel (SPIRE, PACS), GBT}
 
\software{Starlink \citep{Currie2014}, SMURF \citep{Chapin2013}, \texttt{pol2map} \citep{Parsons2021}, Astropy \citep{Astropy2013,Astropy2018,Astropy2022}, NumPy \citep{Harris2020}, SciPy \citep{Virtanen2020}, Matplotlib \citep{Hunter2007}, \texttt{reproject}}
 
\appendix

\section{Assessment of $^{12}$CO($J$=3--2) Contamination in the POL-2 850\,$\mu$m Band}
\label{app:12CO}
 
The SCUBA-2/POL-2 850~$\mu$m continuum band encompasses the $^{12}$CO($J$ = 3--2) transition at 345.796~GHz, and therefore strong $^{12}$CO ($J$ = 3--2) line emission can, in principle, contribute to the measured continuum intensity.
We evaluated the $^{12}$CO($J$=3--2) contamination in our Orion~A data set because it is expected to be greatest where dust emission is weak, such as molecular outflows and the edges of molecular clouds.

We used the map of the velocity-integrated intensity of $^{12}$CO($J$=3--2) in Orion~A obtained with JCMT/Heterodyne Array Receiver Program (HARP) by the JCMT Gould Belt Survey \citep{Graves2012,Coude2016}, expressed in units of K\,km\,s$^{-1}$.
We first corrected the map for the HARP main-beam efficiency ($\eta_{\rm MB}=0.61$; \citealt{Buckle2012}), converting the antenna to the main-beam temperatures.
We then reprojected the corrected $^{12}$CO($J$=3--2) map onto the POL-2 850~$\mu$m grid by bilinear interpolation.
 
To compare the line map with the POL-2 continuum data obtained from the ground, we applied a spatial high-pass filter similar to that used for SCUBA-2 to the $^{12}$CO($J$=3--2) map throughout the mosaicked region.
We removed structures larger than $\sim$300\arcsec\ by subtracting a map smoothed with a Gaussian.
This filtering prevents $^{12}$CO($J$ = 3--2) emission on large scales, which is not recovered in the POL-2 maps, from artificially increasing the estimated contamination.

To estimate the contribution of $^{12}$CO($J$=3--2) line emission to the 850~$\mu$m continuum band, we followed the method developed by \citet{Drabek2012}.
We converted the integrated intensity of the $^{12}$CO($J$=3--2)  emission to an equivalent continuum flux density with a bandpass-averaged factor that accounts for the dilution of the narrow line within the broad SCUBA-2 band.
 
We adopted a conversion coefficient of $C_{850} = 0.63$~mJy\,beam$^{-1}$ per K\,km\,s$^{-1}$, corresponding to Grade~1 weather conditions ($\tau_{225} < 0.05$) as tabulated in \citet[][table~1]{Drabek2012}.
This value is the most conservative estimate (the lowest conversion factor) and provides a lower bound on the expected $^{12}$CO($J$=3--2) contamination.
Under typical observing conditions (Grades~2--3), the conversion factor would be 8--11\% higher (0.68--0.70~mJy\,beam$^{-1}$ per K\,km\,s$^{-1}$).
 
This conversion factor was derived for the SCUBA-2 continuum bandpass without the POL-2 module.
The effective POL-2 bandpass, including the half-wave plate and wire-grid analyzer, has not been independently characterized for molecular line contamination.
We therefore use this conversion to estimate the contamination and treat the resulting values as approximate.
 
We first calculated the contribution of the $^{12}$CO($J$=3--2) emission in units of mJy\,beam$^{-1}$, then converted it to mJy\,arcsec$^{-2}$ using the POL-2 850~$\mu$m beam area $\Omega_{\rm beam}=1.133\,{\rm FWHM}^2$, with ${\rm FWHM}=12\farcs6$ \citep{Mairs2021}.
The coefficient of \citet{Drabek2012} refers to an 850~$\mu$m main-beam FWHM of 13\farcs8 (their section~2.1 and table~1).
Using that beam area would lower the contamination fractions given below by a factor $(12.6/13.8)^2 \simeq 0.83$, without changing our conclusions.
We verified that the POL-2 Stokes~$I$ map was in mJy\,arcsec$^{-2}$ for a consistent comparison.
The POL-2 Stokes~$I$ intensity near map edges is typically lower than the corresponding SCUBA-2 total intensity because of reduced integration time and coverage in the POL-2 scan pattern.
Consequently, the contamination ratio $R_{\rm CO}$ computed here may overestimate the true contamination level in these peripheral regions.

We define the contamination ratio $R_{\rm CO} \equiv I_{\rm CO}/I_{850}$ for each pixel as the estimated $^{12}$CO($J$=3--2) contribution divided by the observed POL-2 Stokes~$I$ intensity.
We excluded non-finite values and negative intensities.
For the statistical comparison relevant to the polarization analysis, we retained pixels with $I_{850}/\Delta I_{850} > 15$, using the intensity uncertainty $\Delta I_{850}$ from the variance extension of the POL-2 FITS files.
 
\begin{figure}[htbp]
\centering
\includegraphics[width=0.89\textwidth]{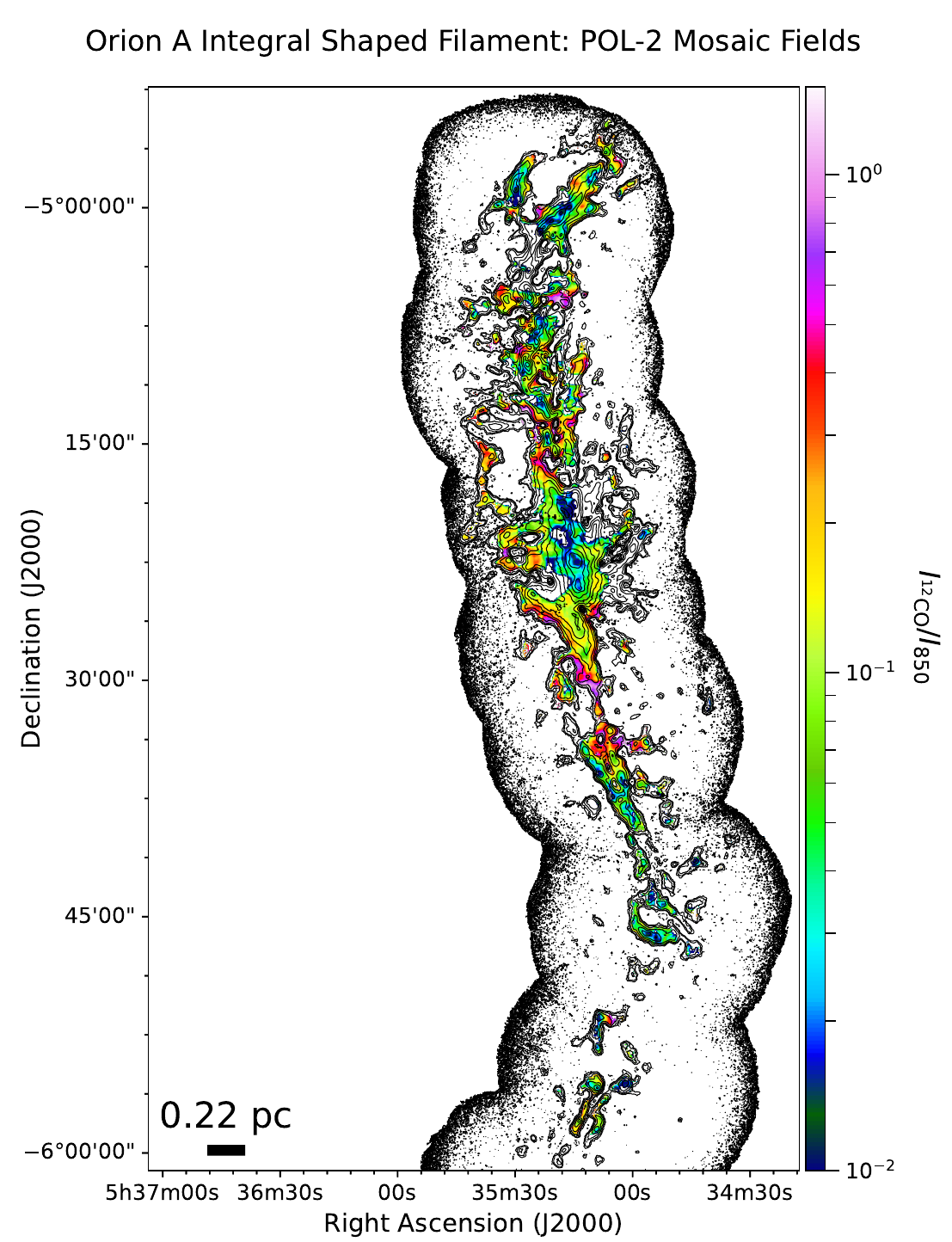}
\caption{
Map of the $^{12}$CO($J$=3--2) contamination ratio $R_{\rm CO} \equiv I_{\rm CO}/I_{850}$ at 850~$\mu$m for the mosaicked POL-2 fields in Orion~A.
Here $I_{\rm CO}$ is the estimated contribution of the line emission expressed as an equivalent continuum intensity in the same units as $I_{850}$.
The spatial coverage extends from OMC~1 through OMC~4.
Only pixels satisfying $I_{850}/\Delta I_{850} \geq 15$ are shown.
The color scale is logarithmic, and the overlaid contours trace the POL-2 Stokes~$I$ emission.
The highest contamination fractions occur preferentially in regions of low intensity surrounding dense filaments.
}
\label{fig:CO32_map}
\end{figure}

Figure~\ref{fig:CO32_map} shows the resulting map of the $^{12}$CO($J$=3--2) contamination ratio at 850~$\mu$m for the mosaicked POL-2 fields in Orion~A, covering the region from OMC~1 through OMC~4.
The spatial coverage corresponds to the maps of the individual fields presented in Paper~S2 (Orion~A atlas).
We mask pixels below the signal-to-noise threshold and display the map on a logarithmic scale to show the large range of contamination ratios.
 
We quantify how the inferred contamination is distributed across the selected Orion A lines of sight and how strongly the bright continuum regions contribute to it.
We characterize the distribution of $\log_{10}(R_{\rm CO})$ because the contamination ratio spans several orders of magnitude across the full mosaicked region.
For the pixels suitable for scientific analysis, which satisfy $I_{850}/\Delta I_{850} > 15$, the median contamination level is $\log_{10}(R_{\rm CO}) = -1.01$, corresponding to a contamination fraction of $\sim$10\%.
The central 90\% of the distribution (5th to 95th percentile) spans $\log_{10}(R_{\rm CO}) \simeq -2.00$ to $-0.30$, equivalent to contamination fractions of approximately 1\% to 50\%.
Approximately 5\% of the pixels have contamination below 1\%, 30\% fall below 5\%, and 51\% lie below 10\%.
 
When weighted by Stokes~$I$ intensity, the effective contamination level is lower, with a mean of $\sim$8\%.
We regard the mean weighted by intensity as more representative of contamination in polarimetric analyses because brighter pixels dominate the polarized flux and thus have greater influence on the derived $\vec{B}$-field statistics.
The weighted value is lower because the largest contamination fractions occur in faint regions.
This weighting describes the continuum budget.
The effect on a polarization-ratio statistic that gives equal weight to each pixel must be assessed on the pixels used for that statistic.

Polarized $^{12}$CO($J$=3--2) line emission via the Goldreich--Kylafis (GK) effect may in principle contribute to the observed polarization signal under specific physical conditions.
In particular, the GK effect is expected to be efficient only when the line optical depth is of the order of unity and the velocity field is sufficiently coherent, such that anisotropic radiative excitation dominates over collisional depolarization \citep[e.g.,][]{Goldreich1981, Goldreich1982, Cortes2005, Ching2016}.
In the present study, we expect any such contamination to be negligible for the global statistical results.
The polarized emission is overwhelmingly dominated by dust continuum emission, and our conclusions are based on ensemble properties rather than localized features.
 
Our findings are consistent with the comprehensive previous study of $^{12}$CO($J$=3--2) contamination in Orion~A by \citet{Coude2016}, who found that contamination levels exceeding 10\% are confined to regions with 850~$\mu$m flux densities below $\sim$300~mJy\,beam$^{-1}$.
Their analysis demonstrated that the optically thick $^{12}$CO($J$=3--2) line saturates more rapidly than the dust continuum, so that contamination becomes appreciable mainly in regions of low column density such as molecular outflows and cloud edges.
Our intensity-based pixel selection excludes many of those regions, but not all of them.
 
Even in regions where the $^{12}$CO($J$=3--2) contribution approaches the 10\% level, the inferred contamination remains comparable to the typical uncertainty in the absolute flux calibration of SCUBA-2 850~$\mu$m continuum observations \citep{Mairs2021}.
The contribution of $^{12}$CO($J$ = 3--2) should therefore have little effect on the polarization fractions or the derived quantities in the main text.
We do not apply an explicit correction for line contamination.

We did not attempt to evaluate potential contamination of the SCUBA-2 450~$\mu$m continuum band by the $^{12}$CO($J$=6--5) transition at 691.473~GHz.
No suitable wide-field $^{12}$CO($J$=6--5) data are available for Orion~A.
Inferring the contribution of this line from lower-$J$ transitions would require assumptions about excitation conditions that introduce large uncertainties.
Additionally, the $^{12}$CO($J$=6--5) line lies near the edge of the SCUBA-2 450~$\mu$m bandpass, where the instrumental transmission is low compared to the band center \citep{Drabek2012}.
The atmospheric transmission at these frequencies is highest under the driest conditions (JCMT Weather Band~1).
However, the $^{12}$CO($J$=6--5) conversion factor remains comparable to that of $^{12}$CO($J$=3--2) at 850~$\mu$m ($\sim$0.63~mJy\,beam$^{-1}$ per K\,km\,s$^{-1}$ for Grade~1; \citealt{Drabek2012}).
Given that the 450~$\mu$m dust continuum flux is typically a factor of 6--12 higher than at 850~$\mu$m for the dust temperatures in our targets, we expect any $^{12}$CO($J$=6--5) contamination to be proportionally smaller than the $^{12}$CO($J$=3--2) contamination at 850~$\mu$m.
For these reasons, we do not consider $^{12}$CO($J$=6--5) line contamination at 450~$\mu$m further in this study.
 
In summary, the 850~$\mu$m band is more susceptible to CO line contamination than the 450~$\mu$m band.
The $^{12}$CO($J$=3--2) line falls near the center of the 850~$\mu$m band, whereas the $^{12}$CO($J$=6--5) line lies at the edge of the 450~$\mu$m band, where the instrumental transmission is low \citep{Drabek2012}.
The 450~$\mu$m data in this study were taken exclusively under Weather Band~1 conditions ($\tau_{225}^{\rm zenith} < 0.05$), for which the Grade~1 conversion factor of \citet{Drabek2012} applies.
We did not measure the $^{12}$CO($J$=6--5) line-to-continuum ratio at 450~$\mu$m, but for the reasons given above, it should remain small.
 
\section{Per-Pixel Environmental Correlations: OMC\,1-Excluded vs Full HGBS Subsample}
\label{app:tier2_env}
The dust temperature and column density describe contrasting aspects of the cloud environment: the thermal state of the emitting material and the amount of material along the line of sight.
Comparing them with polarization at each wavelength tests whether the apparent environmental trends are shared among nearby clouds or depend strongly on the warm OMC\,1 field.
 
Subsection~\ref{ss:Discussion_theory} discusses Spearman rank correlations, calculated over individual pixels, between polarization fraction at each wavelength and local dust temperature $T_{\rm d}$ and column density $N_{\rm H_2}$ from HGBS.
Their magnitude and sign depend on whether OMC\,1, the warmest and most luminous single source in the HGBS subsample, is included.
This dependence limits the interpretation of these pooled correlations as a single environmental response across the HGBS fields.
Table~\ref{tbl:Tier2} gives the full numerical comparison and the $I_{450}/I_{850}$--$P_{450}/P_{850}$ anticorrelation used in the main text (subsection~\ref{ss:Iratio}, figure~\ref{fig:IratioRp}).
The latter expresses the $\gamma - \alpha$ structure for individual pixels and persists when OMC\,1 is excluded.
The persistence of this anticorrelation in the pooled sample does not establish the same relation within each cloud.
 
\begin{deluxetable*}{lcccccc}
\tablecaption{Per-Pixel Environmental Correlations: OMC\,1-Excluded vs Full HGBS Subsample\label{tbl:Tier2}}
\tablehead{
  \colhead{Pair} &
  \colhead{$r_s$ (excl OMC\,1)} & \colhead{$N$} & \colhead{$p$ (excl OMC\,1)} &
  \colhead{$r_s$ (full)} & \colhead{$N$} & \colhead{$p$ (full)}
}
\startdata
$T_{\rm d}, P_{450}$ & $-0.29$ & 281 & $5.9 \times 10^{-7}$ & $+0.13$ & 343 & $1.3 \times 10^{-2}$ \\
$T_{\rm d}, P_{850}$ & $+0.03$ & 321 & $5.5 \times 10^{-1}$ & $+0.36$ & 383 & $4.4 \times 10^{-13}$ \\
$N_{\rm H_2}, P_{450}$ & $-0.46$ & 281 & $1.8 \times 10^{-16}$ & $-0.07$ & 343 & $1.9 \times 10^{-1}$ \\
$N_{\rm H_2}, P_{850}$ & $-0.13$ & 321 & $1.7 \times 10^{-2}$ & $+0.21$ & 383 & $3.5 \times 10^{-5}$ \\
$I_{450}/I_{850}, P_{450}/P_{850}$ & $-0.35$ & 265 & $5.9 \times 10^{-9}$ & $-0.49$ & 327 & $1.9 \times 10^{-21}$ \\
\enddata
\tablecomments{Spearman rank correlations for the HGBS subsample of 9 sources (Method~3, detection mask $P_i \geq 3\,\delta P_i$ for individual pixels).
``Full'' refers to all qualifying pixels in the HGBS subsample; ``excl OMC\,1'' refers to the same subsample with the 62 pixels of OMC\,1 excluded.
The first four pairs ($T_{\rm d}$--$P$ and $N_{\rm H_2}$--$P$ in each band) change sign or change substantially in magnitude between the two samples.
They are sensitive to OMC\,1, which marks the high-radiation boundary of the physical conditions sampled here (subsection~\ref{ss:Discussion_theory}).
The relation between the ratios $I_{450}/I_{850}, P_{450}/P_{850}$ persists in both samples and is treated in subsection~\ref{ss:Iratio} as the counterpart of the $\gamma - \alpha$ spectral-index structure for individual pixels.
The detection limits for individual pixels at the adopted $\sigma=3$ threshold are propagated when constructing the masks based on detection floors in Paper~T~subsection~\ref{T-ss:noisemodel}.}
\end{deluxetable*}

\section{A Two-Temperature Line-of-Sight Mixing Test}
\label{app:Tmixing}
 
\subsection{Motivation, model, and assumptions}
\label{app:Tmixing_model}
To quantify the temperature-mixing interpretation discussed in subsection~\ref{ss:Discussion_Tmixing}, we calculate how much the polarization efficiencies of components along the line of sight must differ to reproduce the observed survey median $\gamma-\alpha\simeq-0.5$ (subsection~\ref{ss:SpIdx}; Paper~S2 table~\ref{S2-tbl:SpIdx_Global}).
We use a model with two emitting components along the line of sight: a cold component at temperature $T_{\rm c}$ with mass fraction $1-x_{\rm w}$, and a warm component at $T_{\rm w}$ with mass fraction $x_{\rm w}$.
We assume the same dust opacity per unit dust mass for both components, including its normalization and spectral index $\beta$.
The total intensity at frequency $\nu$ is
\begin{equation}
I_\nu \propto \nu^\beta \, [(1-x_{\rm w})\, B_\nu(T_{\rm c}) + x_{\rm w}\, B_\nu(T_{\rm w})],
\label{eq:Tmix_I}
\end{equation}
and the polarized intensity, with $p_{\rm c}$ and $p_{\rm w}$ representing the effective polarization efficiencies of the cold and warm components, is
\begin{equation}
PI_\nu \propto \nu^\beta \, [p_{\rm c}\,(1-x_{\rm w})\, B_\nu(T_{\rm c}) + p_{\rm w}\,x_{\rm w}\, B_\nu(T_{\rm w})].
\label{eq:Tmix_PI}
\end{equation}
We calculate $\alpha = \log(I_{450}/I_{850})/\log(\nu_{450}/\nu_{850})$ and $\gamma = \log(PI_{450}/PI_{850})/\log(\nu_{450}/\nu_{850})$ at $\nu_{450}=665.0$\,GHz and $\nu_{850}=349.5$\,GHz, the mid-frequencies of the SCUBA-2/POL-2 passbands, as used in Paper~T~subsection~\ref{T-ss:flux} and the main text.
We assume optically thin emission, a wavelength-independent polarization efficiency for each component, and a common polarization angle for the two components, so that their polarized intensities add as scalars.
For a single temperature equal to the HGBS dust temperature, the 450\,$\mu$m intensity at the brightest pixel of each of the four fields tested below corresponds to a dust optical depth $\tau_{450}\leq0.15$.
We took this pixel from the 450\,$\mu$m maps before their reconvolution to the 12\farcs6 beam (subsection~\ref{ss:data}).
The single-temperature estimate does not cover the cold component at the lowest temperatures considered below, because colder dust emits less per unit optical depth.
A line of sight through an internally heated protostellar core crosses cold dust in front of and behind the warm dust.
For optically thin emission, the order of the layers does not matter, and the two cold layers act as one cold component when they share $T_{\rm c}$ and $\beta$.
The two cold layers do not need to share a polarization angle.
Because the two layers have the same spectrum, their Stokes $Q$ and $U$ sum to cold emission with one angle in both bands and a lower effective $p_{\rm c}$, which raises $r$.
The common polarization angle that the model assumes is therefore shared by the warm component and the summed cold emission.
The same argument joins the warm layers on both sides of an externally heated cloud into one warm component.
Because the two components share $\beta$, the opacity factor $\nu^\beta$ cancels analytically from $\gamma-\alpha$.
The difference depends only on $T_{\rm c}$, $T_{\rm w}$, $x_{\rm w}$, and the contrast in polarization efficiency, $r \equiv p_{\rm w}/p_{\rm c}$.
Here $r$ is the ratio of the \emph{effective} polarization efficiencies assigned to the warm and cold components along the line of sight in the toy model.
It is not the observed polarization-fraction ratio between the bands, $R_{\rm p} \equiv P_{450}/P_{850}$ (subsection~\ref{ss:P_ratio}).

We let the cold component range from $T_{\rm c}=3$\,K, just above the temperature of the cosmic microwave background ($T_{\rm CMB}=2.725$\,K), to 20\,K, and the warm component from $T_{\rm w}=25$ to 40\,K.
We do not set the lower limit of $T_{\rm c}$ from measured dust temperatures.
Dust temperatures from single-temperature fits to the far-infrared spectrum, such as the HGBS temperatures adopted in Paper~S2 sub-subsection~\ref{S2-sss:SpIdxFramework}, lie between the temperatures of the components along the line of sight.
The fitted temperatures are pulled toward the warmer dust, which emits more per unit mass.
Such fitted temperatures therefore exceed the temperature of the coldest component.
The warm range extends above the representative $25\pm10$\,K that Paper~S2 adopts for high-mass star-forming and CMZ sources, because warmer components increase the 450/850 contrast and with it the effect of temperature mixing on $\gamma-\alpha$.
The upper limit of 40\,K is our choice, and we give the effect of a higher limit below.
Equations~(\ref{eq:Tmix_I}) and~(\ref{eq:Tmix_PI}) use $B_\nu(T)$.
Measured against the uniform cosmic microwave background, the emission of each component is proportional to $B_\nu(T)-B_\nu(T_{\rm CMB})$.
Using this difference changes the required contrast by less than 0.02 and the thresholds quoted below for the temperature of the cold component by less than 0.1\,K.

\subsection{Equal polarization efficiencies and differing angles}
\label{app:Tmixing_r1}
A useful limiting case is $p_{\rm w}=p_{\rm c}$ ($r=1$).
In this case $PI_\nu = p_{\rm c}\,I_\nu$ at both wavelengths, and therefore $\gamma=\alpha$ identically, independently of $T_{\rm c}$, $T_{\rm w}$, $x_{\rm w}$ and the adopted opacity index $\beta$.
We verified this cancellation numerically to machine precision over $\beta=1.5$--3.0.
The same cancellation holds even when the cold and warm components have different opacity slopes ($\beta_{\rm c}\neq\beta_{\rm w}$), provided that the polarization efficiency is common to both components.
Thus, neither mixing temperatures nor mixing opacity slopes alone produces a nonzero $\gamma-\alpha$.
The components must differ in effective polarization efficiency ($r\neq 1$).

The requirement $r\neq1$ holds only for a common polarization angle.
If the angles of the two components differ, their polarized intensities no longer add as scalars, because only their Stokes $Q$ and $U$ add.
Nearly orthogonal components can then reproduce the observed median even at $r=1$, over a limited range of warm fractions: the required angle difference is above $64\arcdeg$ at $T_{\rm c}=10$\,K and $T_{\rm w}=40$\,K, and above $77\arcdeg$ at (15\,K, 30\,K).
In part of this range, the angles in the two bands differ by less than $10\arcdeg$, so Finding~1 does not exclude this case.
We treat it as one of the geometric origins of Class~1 (sub-subsection~\ref{sss:class1_geometric}), where it remains untested.

\subsection{Required contrast in polarization efficiency}
\label{app:Tmixing_contrast}
\begin{figure*}[htbp]
\centering
\includegraphics[width=0.95\textwidth]{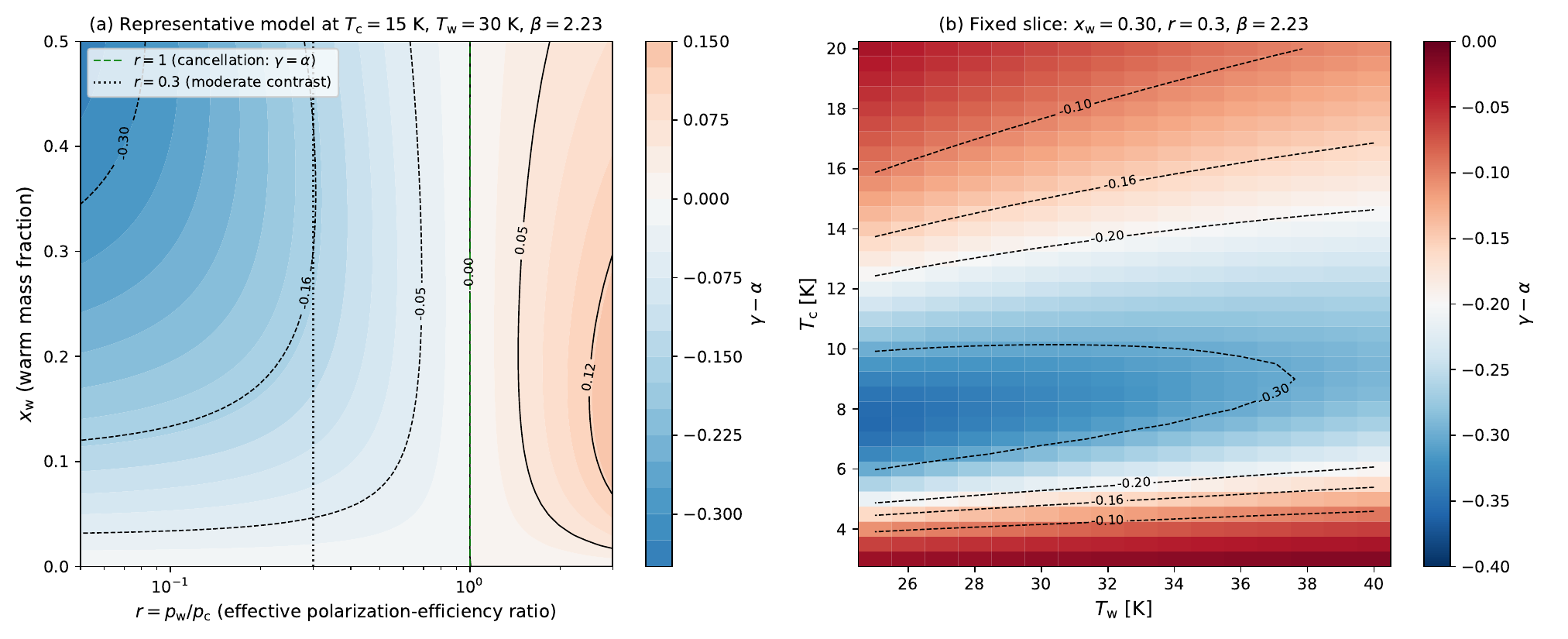}
\caption{
Test of two-temperature mixing along the line of sight for $\gamma-\alpha$.
The model parameters are the warm mass fraction $x_{\rm w}$ and the ratio $r=p_{\rm w}/p_{\rm c}$ of effective polarization efficiencies of the warm and cold components.
The ratio $r$ differs from the observed polarization-fraction ratio between the bands, $R_{\rm p}=P_{450}/P_{850}$ (subsection~\ref{ss:P_ratio}).
\textbf{(a)} $\gamma-\alpha$ versus $x_{\rm w}$ and $r$ for $T_{\rm c}=15$\,K and $T_{\rm w}=30$\,K.
The common emissivity index is $\beta=2.23$, but cancels from $\gamma-\alpha$.
The green dashed line at $r=1$ marks $PI_\nu\propto I_\nu$ and hence $\gamma=\alpha$.
Negative differences require $r<1$.
The black dotted line marks the moderate contrast $r=0.3$ discussed in the text.
The most negative plotted value is about $-0.34$ at $r=0.05$ and $x_{\rm w}=0.5$.
\textbf{(b)} Dependence on temperature at fixed mass fraction and ratio of polarization efficiencies, $x_{\rm w}=0.30$ and $r=0.3$, over $T_{\rm c}=3$--20\,K and $T_{\rm w}=25$--40\,K.
This slice does not reach the observed median.
Figure~\ref{fig:Tmix_rmax} and table~\ref{tab:Tmix_required_r} summarize solutions at other mass fractions.
\label{fig:Tmix_panels}
}
\end{figure*}

Warm dust contributes relatively more strongly at 450\,$\mu$m than at 850\,$\mu$m.
A mixture of warm and cold emission can therefore lower the polarization fraction at the shorter wavelength if the warm component contributes less polarized emission per unit total intensity.
Figure~\ref{fig:Tmix_panels}(a) shows $\gamma-\alpha$ as a function of $r$ and $x_{\rm w}$ at the representative temperature pair $T_{\rm c}=15$\,K, $T_{\rm w}=30$\,K, with the representative value $\beta=2.23$.
The choice of $\beta$ does not affect $\gamma-\alpha$, because the common $\nu^{\beta}$ factor cancels between $I_\nu$ and $PI_\nu$.
The vertical dashed line at $r=1$ marks the cancellation locus, and the dotted line at $r=0.3$ marks the moderate contrast discussed below.
Figure~\ref{fig:Tmix_panels}(b) shows the temperature dependence at fixed $x_{\rm w}=0.30$ and $r=0.3$.
Cooling the cold component does not make $\gamma-\alpha$ monotonically more negative.
The most negative value in this slice, $-0.36$, lies at $T_{\rm c}=7.5$\,K and $T_{\rm w}=25$\,K.
At $T_{\rm c}=3$\,K, $\gamma-\alpha$ increases to between $-0.03$ and $-0.02$.
At $x_{\rm w}=0.30$, a very cold component emits little at either wavelength, and the warm component then sets the polarization fraction in both bands.
With a moderate contrast, reaching the median at low $T_{\rm c}$ requires a much smaller warm mass fraction (table~\ref{tab:Tmix_required_r}).
 
Table~\ref{tab:Tmix_required_r} gives, for each cold temperature, the largest $r$ for which the model reaches the observed median.
We maximized $r$ over $x_{\rm w}\leq0.5$ and $T_{\rm w}=25$--40\,K, and the largest $r$ lies at $T_{\rm w}=40$\,K for every entry.
The required contrast depends steeply on $T_{\rm c}$.
For $T_{\rm c}\geq10$\,K, the median requires $r\lesssim0.18$, and for $T_{\rm c}\geq12$\,K it requires $r\lesssim0.094$.
For $T_{\rm c}\geq10$\,K and $r\geq0.3$, the most negative value is $\gamma-\alpha=-0.37$ (at $T_{\rm c}=10$\,K, $T_{\rm w}=40$\,K, $x_{\rm w}=0.13$, $r=0.30$).
With $r\geq0.3$, the median requires $T_{\rm c}\lesssim8$\,K and a cold component with at least five times the mass of the warm component.
At $T_{\rm c}=5$\,K, $r\approx0.5$ suffices.
Raising the upper limit of $T_{\rm w}$ from 40 to 100\,K relaxes the requirement at $T_{\rm c}=10$\,K from $r\lesssim0.18$ to $r\lesssim0.23$.
As in the main text, we compare the model with the observed median on the assumption that the modulation efficiency of POL-2 is equal in the two bands (sub-subsection~\ref{sss:class2_instrumental}).
 
At any temperatures, the polarization fraction of the model at each wavelength is the mean of the efficiencies of the two components, weighted by their emission.
For $r<1$, the model ratio $P_{450}/P_{850}$ therefore cannot fall below $r$, and $\gamma-\alpha\geq\log r/\log(\nu_{450}/\nu_{850})$.
Reaching the observed median thus requires $r<0.72$ at any $T_{\rm c}$, because the median implies $P_{450}/P_{850}=0.72$ through equation~(\ref{eq:Pratio_SpIdx}).
 
Without an independent value of $\beta$, the observed $\alpha$ does not constrain the mixture.
The opacity index cancels from $\gamma-\alpha$ but not from $\alpha=\beta+C$, where $C$ is the shape factor of the Planck function in equation~(\ref{eq:planck_correction}) with $B_\nu(T_{\rm d})$ replaced by the bracket of equation~(\ref{eq:Tmix_I}).
A mixture dominated by cold dust has a smaller $C$, so it matches the observed $\alpha$ only with a larger $\beta$.
The shape factor of a mixture lies between the shape factors of its two components.
Any mixture with $T_{\rm w}\leq40$\,K therefore requires $\beta\geq\alpha-1.68$, the bound that holds for a single temperature of 40\,K or less.
 
For $T_{\rm c}\geq10$\,K, temperature mixing along the line of sight reaches the observed median $\gamma-\alpha$ only if the components differ strongly in effective polarization efficiency.
A moderate difference ($r\approx0.3$) suffices only if a cold component below about 8\,K carries most of the mass.
The most negative value at $T_{\rm c}\geq10$\,K and $r\geq0.3$, $-0.37$, lies 1.4 times the statistical uncertainty above the median, $-0.51\pm0.04_{\rm sys}\pm0.10_{\rm stat}$ (equation~\ref{eq:gammaMinusAlphaGlobal}).
Over the range of the median with the $3\sigma$ limits at their two extremes, $-0.57$ to $-0.45$, the contrast required at $T_{\rm c}=10$\,K is $r\lesssim0.15$--0.23, and the cold limit for $r=0.3$ is 7.4--8.8\,K.
The calculation therefore maps the contrast that temperature mixing requires.
It does not exclude contributions from temperature mixing.

\subsection{Test with dust and ammonia temperatures}
\label{app:Tmixing_data}
\begin{figure}[htbp]
\centering
\includegraphics[width=\textwidth]{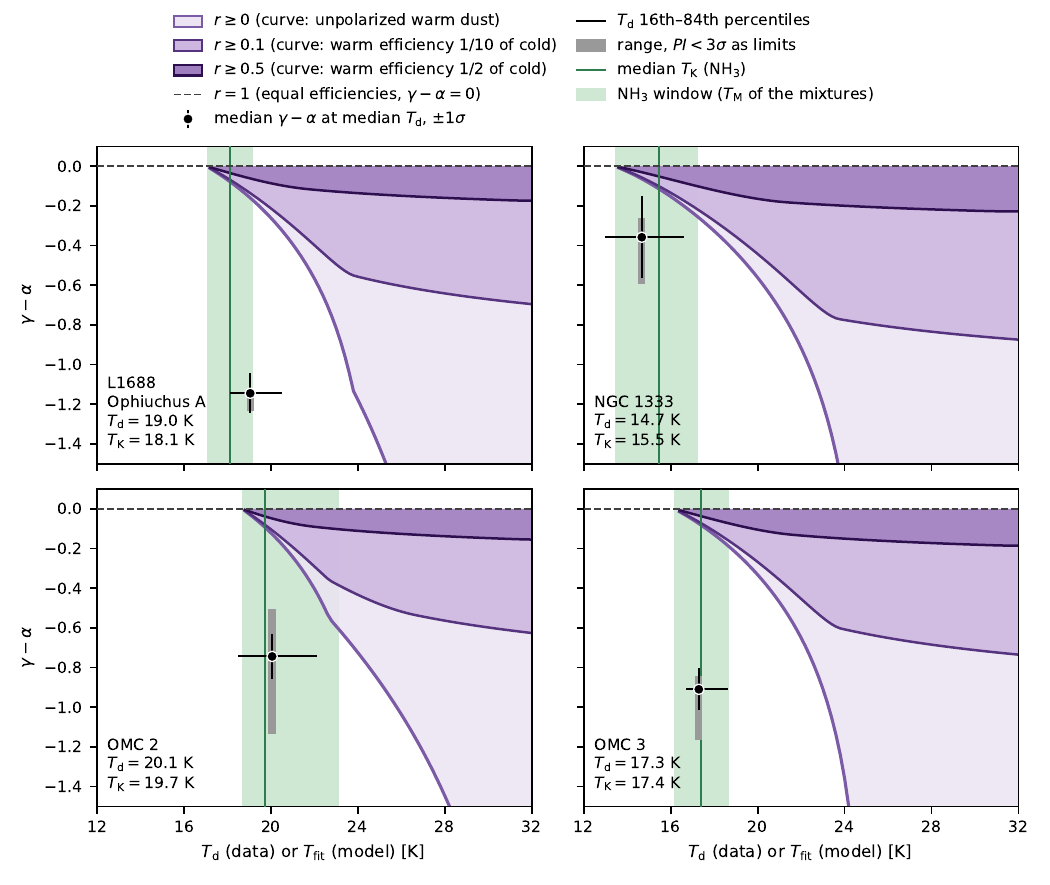}
\caption{Measured $\gamma-\alpha$ of the four fields that meet the criteria of appendix~\ref{app:Tmixing_data} and the most negative $\gamma-\alpha$ of two-temperature mixtures, against temperature.
The abscissa is the HGBS dust temperature $T_{\rm d}$ \citep{Andre2010} for the data and the temperature $T_{\rm fit}$ of a single-temperature fit for the mixtures.
We fit the 160--500\,$\mu$m intensities of each mixture with the same form of single-temperature spectrum used for the HGBS maps \citep{Konyves2015}.
The model uses $\beta=2$ for both components and for the fit, as the HGBS maps do, with $T_{\rm c}=3$--20\,K, $T_{\rm w}=25$--40\,K, and $x_{\rm w}\leq0.5$.
In each panel, we retain only mixtures whose mass-weighted dust temperature along the model line of sight, $T_{\rm M}=(1-x_{\rm w})\,T_{\rm c}+x_{\rm w}\,T_{\rm w}$, lies within the window of the \amm\ kinetic temperature $T_{\rm K}$ for that field (green band).
The window spans the 16th to 84th percentiles of $T_{\rm K}$ over the pixels used for $\gamma-\alpha$, widened at both ends by the median of its uncertainties for individual pixels.
The band limits $T_{\rm M}$ of the mixtures, not $T_{\rm fit}$, and the solid green line marks the median $T_{\rm K}$.
We take $T_{\rm K}$ from Data Release~2 \citep{Pineda2026} of the Green Bank Ammonia Survey \citep{Friesen2017}.
We use the window of $T_{\rm K}$ to constrain $T_{\rm M}$, assuming that gas and dust have equal temperatures and that $T_{\rm K}$ traces $T_{\rm M}$.
The ratio of the effective polarization efficiencies of the warm and cold components is $r=p_{\rm w}/p_{\rm c}$.
The three curves give the most negative $\gamma-\alpha$ attainable by the mixtures at each $T_{\rm fit}$, obtained by continuous minimization.
They are for unpolarized warm dust ($r=0$) and for warm dust whose effective polarization efficiency is one tenth ($r=0.1$) or one half ($r=0.5$) that of the cold dust.
Equal efficiencies ($r=1$) give the same polarization fraction in both bands and $\gamma-\alpha=0$ (dashed line), the limiting case discussed in the text.
Because $\gamma-\alpha$ rises with $r$ to 0 at $r=1$, the area shaded for a given $r$ contains the spectral-index differences attainable at that value of $r$ or larger ones, so the areas are nested.
The curve for $r=0$ continues below the plotted range.
In these mixtures, $T_{\rm fit}$ is never lower than $T_{\rm M}$, so the shaded areas begin near the lower end of the green band.
For the data, the circle shows the median $T_{\rm d}$ and the median $\gamma-\alpha$ of the field, with a vertical bar for its $1\sigma$ uncertainty (Paper~S2 table~\ref{S2-tbl:SpIdx_Global}).
The horizontal bar spans the 16th to 84th percentiles of $T_{\rm d}$ over the pixels used for $\gamma-\alpha$.
This range is the window of $T_{\rm fit}$ in the test described in the text.
Gray bars show the range of the median $\gamma-\alpha$ when measurements of polarized intensity below three times their uncertainty are treated as $3\sigma$ limits, as in figure~\ref{fig:alpha_vs_gamma}.
Each panel gives the median $T_{\rm d}$ and the median $T_{\rm K}$ of the field.
The shaded region associated with a given $r$ overlaps the field's allowed range when its lower boundary lies at or below the top of the vertical bar somewhere within the horizontal bar.
In these model ranges, NGC~1333 is reached within $1\sigma$ for $r\leq0.15$, whereas the other three fields are not reached even for $r=0$.
With wider ranges, $T_{\rm w}=10$--100\,K and $x_{\rm w}\leq0.95$, mixtures reach the other three fields within $1\sigma$, but only for $r\leq0.041$ (appendix~\ref{app:Tmixing_data}).}
\label{fig:Tmix_data}
\end{figure}

We ask whether mixtures that reproduce the polarization spectrum of each of the four fields also satisfy its dust and ammonia temperature constraints.
For $\beta=2$, no mixture in the ranges above both matches the HGBS dust and \amm\ kinetic temperatures of L1688 Ophiuchus~A, OMC\,2, and OMC\,3 and reproduces their $\gamma-\alpha$ within $1\sigma$ (figure~\ref{fig:Tmix_data}).
Within these ranges, mixtures that match both measured temperatures reproduce the measured spectral-index difference in the fourth field of figure~\ref{fig:Tmix_data}, NGC~1333, within $1\sigma$ for $r\leq0.15$.
We compared the HGBS dust temperature $T_{\rm d}$ with $T_{\rm fit}$, which we obtained by fitting a single-temperature spectrum with $\beta=2$ to each mixture, as in the derivation of the HGBS temperatures \citep{Konyves2015}.
Because $\nu^\beta$ cancels from $\gamma-\alpha$, the opacity index of the mixture enters the test only through $T_{\rm fit}$.
We took the \amm\ kinetic temperature $T_{\rm K}$ from Data Release~2 \citep{Pineda2026} of the Green Bank Ammonia Survey \citep[GAS;][]{Friesen2017}.
We compared $T_{\rm K}$ with the mass-weighted dust temperature of each mixture, $T_{\rm M}=(1-x_{\rm w})\,T_{\rm c}+x_{\rm w}\,T_{\rm w}$.
For the temperature comparison, we accept a mixture when $T_{\rm fit}$ and $T_{\rm M}$ lie within the 16th--84th percentile ranges of $T_{\rm d}$ and $T_{\rm K}$, respectively, over the pixels used for $\gamma-\alpha$.
We widened the window of $T_{\rm K}$ at both ends by the median of its uncertainties for individual pixels.

We used only fields in which $T_{\rm K}$ is valid at two thirds or more of these pixels and its median agrees with that of $T_{\rm d}$ within their combined uncertainties.
We took the uncertainty of $T_{\rm d}$ as the larger of the standard error of its median over independent beams and half its 16th--84th percentile range.
For $T_{\rm K}$, we added in quadrature the standard error of its median over independent beams and the median of its uncertainties for individual pixels.
We did not match the angular resolutions of the temperature maps for this comparison.
We used the beam sizes of the HGBS and GAS maps, $36\arcsec$ and $30.76\arcsec$, only to count the independent beams in these standard errors.
The validity and agreement criteria kept four of the 12 {\sc SpIdx} sources with HGBS dust temperatures.
These criteria do not use $\gamma-\alpha$.
The agreement criterion excluded L1689~North and Perseus~IC348, where $T_{\rm d}$ exceeds $T_{\rm K}$ as expected for mixtures with $\beta=2$, but their $\gamma-\alpha$ needs no $r<1$ within $1\sigma$.
The validity criterion excluded OMC\,1 and the overlapping Orion~A Field~4, where $T_{\rm K}$ is valid at only 30 of 77 and 27 of 79 pixels, respectively.
The same criterion excluded Serpens~Main, Serpens~South, PGCC~G203.21$-$11.20, and Orion~A Field~5, where $T_{\rm K}$ is valid at 6 or fewer pixels.
The test therefore covers only fields with median $T_{\rm d}$ of 14.7--20.1\,K and leaves out OMC\,1, the warmest source of the HGBS subsample (appendix~\ref{app:tier2_env}).

L1688 Ophiuchus~A and OMC\,3 are not reached even at $\gamma-\alpha+3\sigma$, also with their opacity indices from Paper~S2 and with $T_{\rm d}$ raised by 0.35--0.48\,K for the optical depth at 160\,$\mu$m.
For $\beta=2$ and uncorrected $T_{\rm d}$, the closest mixture has a spectral-index difference only 0.004 above $\gamma-\alpha+3\sigma$ for OMC\,2.
With $\beta\simeq1.9$ from Paper~S2, mixtures reach OMC\,2 even within $1\sigma$ for $r\leq0.003$.
Mixtures with $\beta=2$ also reach it within $1\sigma$ if its median is taken at the upper end of its range with the $3\sigma$ limits, $-0.50$.
We therefore regard the result for this field as tentative.
Raising the lower limit of $T_{\rm c}$ to 10\,K does not change the outcome for any of the four fields.
When we impose both temperature windows but widen the ranges to $T_{\rm w}=10$--100\,K ($\geq T_{\rm c}+0.5$\,K) and $x_{\rm w}\leq0.95$, mixtures reproduce the measured spectral-index difference in each of the three fields within $1\sigma$ only for $r\leq0.041$, with most of the mass at 17--23\,K.

The test assumes equal gas and dust temperatures and \amm\ tracing the cold gas, although GAS likely does not resolve a possible depletion of \amm\ \citep{Pineda2026}.
The conclusion for the three fields holds even at $\gamma-\alpha+3\sigma$ if $T_{\rm K}$ follows $T_{\rm c}$ instead of $T_{\rm M}$.
If \amm\ traced only the warm component, the test would not apply, because $T_{\rm w}\geq25$\,K exceeds the upper end of every \amm\ window.
Without the \amm\ temperature window, the constraint from $T_{\rm d}$ alone allows $r$ up to 0.37--0.58 while reproducing the measured spectral-index differences within $1\sigma$ in the three fields.
We therefore regard the result for the three fields as conditional on \amm\ tracing the cold gas.
 
\begin{table}[htbp]
\centering
\caption{Largest ratio $r=p_{\rm w}/p_{\rm c}$ of the effective polarization efficiencies of the warm and cold components for which the two-temperature model reaches the observed median $\gamma-\alpha$, as a function of the cold temperature $T_{\rm c}$.
The three middle columns correspond to the measured median, $-0.51$ (equation~\ref{eq:gammaMinusAlphaGlobal}), and to the two ends of its range with the $3\sigma$ limits at their two extremes, $-0.45$ and $-0.57$.
For each entry we maximized $r$ over $x_{\rm w}\leq0.5$ and $T_{\rm w}=25$--40\,K.
The largest $r$ lies at $T_{\rm w}=40$\,K for every entry, and for $-0.51$ at $T_{\rm c}=15$\,K it lies at the limit $x_{\rm w}=0.5$.
The last column gives the cold-to-warm mass ratio $(1-x_{\rm w})/x_{\rm w}$ at that maximum for $-0.51$.
Smaller $r$ reaches the target over a wider range of $x_{\rm w}$.
An entry ``unreachable'' means that the target is not reached for any $r\geq10^{-4}$ with $x_{\rm w}\leq0.5$ and $T_{\rm w}\leq40$\,K.
The values of $r$ use $B_\nu(T)$.
Subtracting $B_\nu(T_{\rm CMB})$ changes each value of $r$ by less than 0.02 and roughly doubles the mass ratio at $T_{\rm c}=3$\,K.
Numerical inversion uses \texttt{scipy.optimize.brentq} on $r\in[10^{-4}, 1]$.\label{tab:Tmix_required_r}}
\begin{tabular}{ccccc}
\hline
$T_{\rm c}$ [K] & \multicolumn{3}{c}{Largest $r$ for $\gamma-\alpha=$} & $(1-x_{\rm w})/x_{\rm w}$ \\
 & $-0.45$ & $-0.51$ & $-0.57$ & \\
\hline
3 & 0.69 & 0.66 & 0.63 & 3400 \\
5 & 0.56 & 0.52 & 0.48 & 120 \\
7 & 0.42 & 0.37 & 0.33 & 23 \\
8 & 0.35 & 0.30 & 0.26 & 13 \\
9 & 0.29 & 0.24 & 0.20 & 8.3 \\
10 & 0.23 & 0.18 & 0.15 & 5.4 \\
12 & 0.13 & 0.094 & 0.065 & 2.4 \\
15 & 0.036 & 0.013 & unreachable & 1.0 \\
20 & unreachable & unreachable & unreachable & -- \\
\hline
\end{tabular}
\end{table}

\subsection{Conclusions and the Class~3 degeneracy}
\label{app:Tmixing_class3}
Temperature mixing reproduces the observed median $\gamma-\alpha$ only if the warm component is polarized less efficiently than the cold component ($r=p_{\rm w}/p_{\rm c}<1$), for a common polarization angle (appendix~\ref{app:Tmixing_r1}).
For $T_{\rm w}\leq40$\,K, the model reaches the median only for $r\lesssim0.18$ if $T_{\rm c}\geq10$\,K and only for $r\lesssim0.094$ if $T_{\rm c}\geq12$\,K (appendix~\ref{app:Tmixing_contrast}).
At any temperatures, the median requires $r<0.72$.
A moderate contrast, $r=0.3$, suffices only if the cold component is colder than about 8\,K and carries at least five times the mass of the warm component.
Warm and cold components with nearly orthogonal polarization angles can reach the median even at $r=1$.
We treat this untested case as one of the geometric origins of Class~1 (sub-subsection~\ref{sss:class1_geometric}).

In three of the four fields tested, no mixture within the adopted model ranges both matches the HGBS dust and \amm\ kinetic temperatures and reproduces the field's $\gamma-\alpha$ within $1\sigma$, for $\beta=2$ (appendix~\ref{app:Tmixing_data}).
For two of these fields, L1688 Ophiuchus~A and OMC\,3, the mixtures do not reach even $\gamma-\alpha+3\sigma$.
Mixtures reach the fourth field, NGC~1333, within $1\sigma$ for $r\leq0.15$.
The test assumes identical gas and dust temperatures and \amm\ tracing the cold molecular gas, and it covers only fields with median $T_{\rm d}$ of 14.7--20.1\,K.
We treat its result for OMC\,2 as tentative.
With wider ranges, $T_{\rm w}=10$--100\,K and $x_{\rm w}\leq0.95$, mixtures reach the three fields within $1\sigma$, but only for $r\leq0.041$, with a warm component at 17--23\,K that carries most of the mass.
We therefore find that, within the adopted model ranges, lowering $r$ alone does not reproduce the measured spectral-index differences in these three fields when the mixtures must also match the measured temperatures.
Neither the model nor the test determines the origin of the required contrast in the effective polarization efficiency.

The model calculation refines the Class~3 degeneracy from temperature mixing in table~\ref{tbl:degeneracy}.
Differences in opacity slope or temperature among components along the line of sight do not generate $\gamma-\alpha\neq 0$ if all components have the same polarization efficiency and polarization angle.
Composition becomes relevant in $\gamma-\alpha$ only insofar as the grains dominating polarized emission have a \emph{different effective opacity law} from those dominating total emission.
Our model uses a common dust opacity per unit dust mass for the two components and does not describe this difference in grain populations.
This difference therefore remains a degeneracy in Class~3 (sub-subsection~\ref{sss:class3_dust}).
To distinguish the remaining possibilities---differences in emissivity or in grain populations among components---we need polarimetry at additional wavelengths or laboratory constraints on dust emissivity, beyond the scope of this analysis.

\section{Secondary Tests of the Statistical Methods Used in the Main Text}
\label{app:S1_secondary}
The statistical tests used for the astronomical conclusions make assumptions about the distribution of the data.
In this appendix, we repeat each primary test with an alternative appropriate to the same comparison to check these assumptions.

Subsection~\ref{ss:SpIdx} states Finding~2 as a count, $\gamma<\alpha$ in 24--27 of the 35 {\sc SpIdx} sources, together with the median difference $\gamma-\alpha$ between $-0.57$ and $-0.45$.
The two ends correspond to placing the $3\sigma$ limits at their extremes for pixels below the noise floor in polarized intensity.
The count addresses the prevalence of the negative differences, and the median addresses their magnitude.
For matched measurements, these negative differences correspond to $P_{450}<P_{850}$.
We test whether negative values occur in more than half of the measured sources and whether the distribution across sources is displaced below zero, using the values for individual sources in Paper~S2, table~\ref{S2-tbl:SpIdx_Global}.
These tests do not introduce new measurements.
The sign count supports a predominance of negative differences within the selected sample of sources, and the bootstrap interval places the sample median below zero.
These tests characterize the observed polarization spectrum; they do not distinguish among the dust and line-of-sight mechanisms that may produce it.
 
\begin{deluxetable}{lcc}
\tablecaption{Tests of $\gamma<\alpha$ on the 35-source {\sc SpIdx} sample\label{tbl:S1_secondary_F2}}
\tablewidth{0pt}
\tablehead{\colhead{Procedure} & \colhead{Quantity tested} & \colhead{$p$}}
\startdata
Wilcoxon signed-rank, one-sided   & median $\gamma-\alpha$ & $5.9\times10^{-5}$ -- $5.1\times10^{-3}$ \\
Student $t$, one-sided            & mean $\gamma-\alpha$   & \nodata\ -- $1.0\times10^{-2}$ \\
Sign test, one-sided              & incidence $27/35$ -- $24/35$ & $9.4\times10^{-4}$ -- $2.0\times10^{-2}$ \\
Binomial against $1/2$, two-sided & incidence $27/35$ -- $24/35$ & $1.9\times10^{-3}$ -- $4.1\times10^{-2}$ \\
\enddata
\tablecomments{The bootstrap $95\%$ confidence interval on the median ($2\times10^{5}$ resamples) is $[-0.78,-0.22]$ at the lower end and $[-0.62,-0.03]$ at the upper end of the limit range.
The mean is undefined at the lower end, where the values for two sources (Serpens~South and SDC~13 mm-2) have upper bounds but no lower bounds.
The Clopper--Pearson $95\%$ interval on the incidence is $[0.60,0.90]$ and $[0.51,0.83]$ at the two ends.
The two families test different statements: the first two address the magnitude of the offset and the last two only its incidence.
The magnitude is the better constrained of the two, and is the quantity the dust-model comparison of subsection~\ref{ss:Discussion_DiscriminatingPower} uses.}
\end{deluxetable}

\section{Inter-band Polarization-Angle Differences: Angles before the Stokes-$Q/U$ Zero-Level Offset Correction and the Effect of the Pixel Selection}
\label{app:F1_uncorrected}
 
\subsection{450\,$\mu$m angles before the Stokes-$Q/U$ zero-level offset correction}
\label{app:F1_uncorrected_angles}
 
\begin{figure*}
\centering
\includegraphics[width=0.95\textwidth]{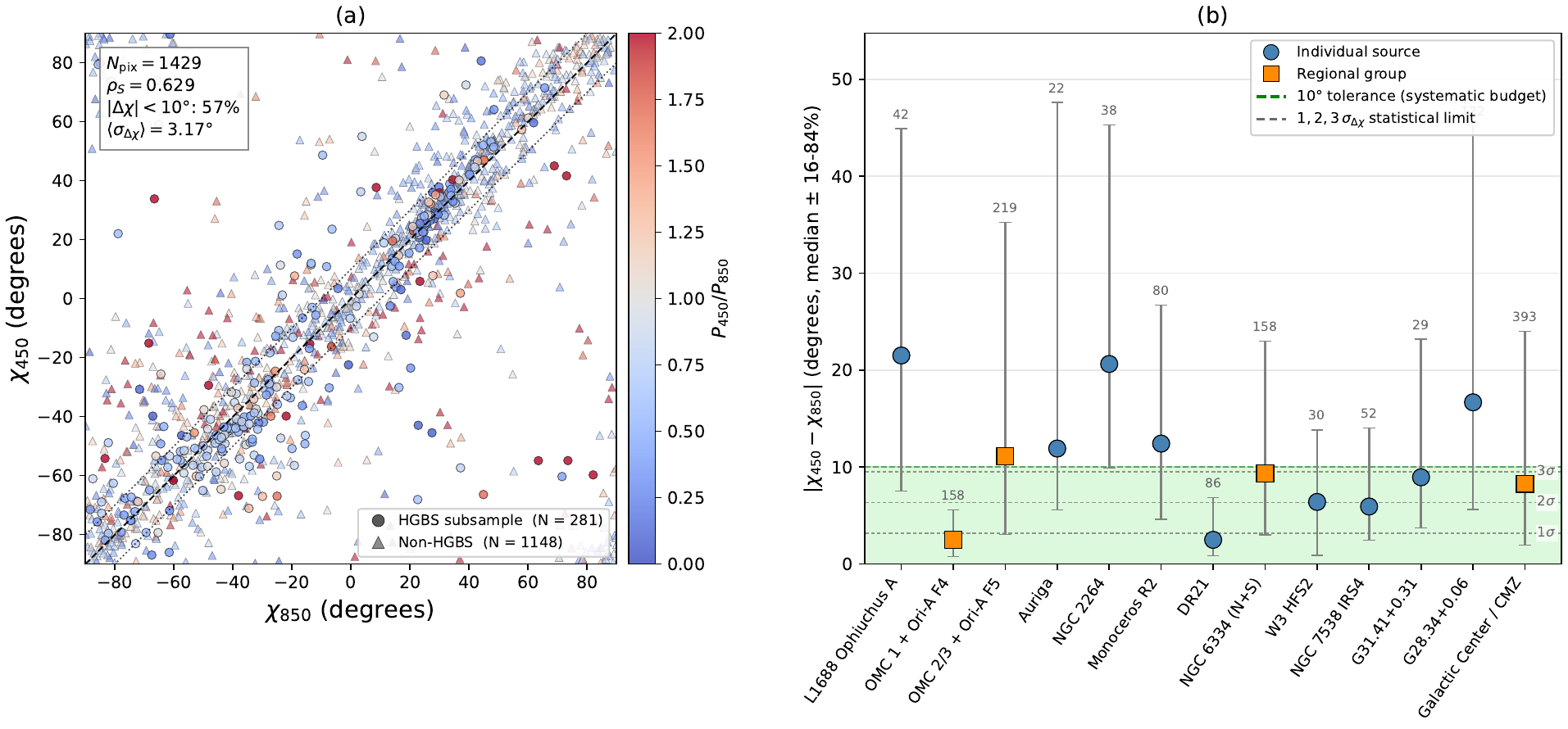}
\caption{Same as figure~\ref{fig:PA450vsPA850}, but with the 450\,$\mu$m polarization angles before the Stokes-$Q/U$ zero-level offset correction of Paper~T, for the 1429 matched pixels of the 21 sources.
The fraction with $|\chi_{450}-\chi_{850}|<10\arcdeg$ is 57\%, and seven of the thirteen targets have a median below $10\arcdeg$.
The median for Monoceros~R2 is $12\arcdeg$, against $4.0\arcdeg$ after the correction, illustrating the sensitivity of the angles to it (appendix~\ref{app:F1_uncorrected_angles}).}
\label{fig:PA450vsPA850_uncorrected}
\end{figure*}

Figure~\ref{fig:PA450vsPA850} uses 450\,$\mu$m angles after the Stokes-$Q/U$ zero-level offset correction of Paper~T.
The common-slope step of this correction (Step~2 of the three-step regression of Paper~T) assumes that the polarization angles do not depend on wavelength.
We therefore repeated the comparison using the 450\,$\mu$m angles before correction for all 1429 pixels in the parent set selected by intensity (figure~\ref{fig:PA450vsPA850_uncorrected}).
Figure~\ref{fig:PA450vsPA850} uses the 1420 pixels in that set with finite angles after correction.
Before correction, 57\% of the pixels have $|\chi_{450}-\chi_{850}|<10\arcdeg$, the median $|\chi_{450}-\chi_{850}|$ is $8.2\arcdeg$, and the median signed difference is $+1.4\arcdeg$.
Seven of the thirteen targets have a median below $10\arcdeg$.
For Monoceros~R2, the median $|\chi_{450}-\chi_{850}|$ is $12\arcdeg$ (median signed angle difference $+8.1\arcdeg$) before the correction and $4.0\arcdeg$ after it, which makes it a useful diagnostic of the angle sensitivity to this correction.
The uncorrected data already show similar projected field orientations for a majority of the sampled pixels, while the degree of agreement for individual targets remains sensitive to the correction.
 
\subsection{Effect of the pixel selection}
\label{app:F1_selection}
 
\begin{figure*}
\begin{center}
\includegraphics[width=\textwidth]{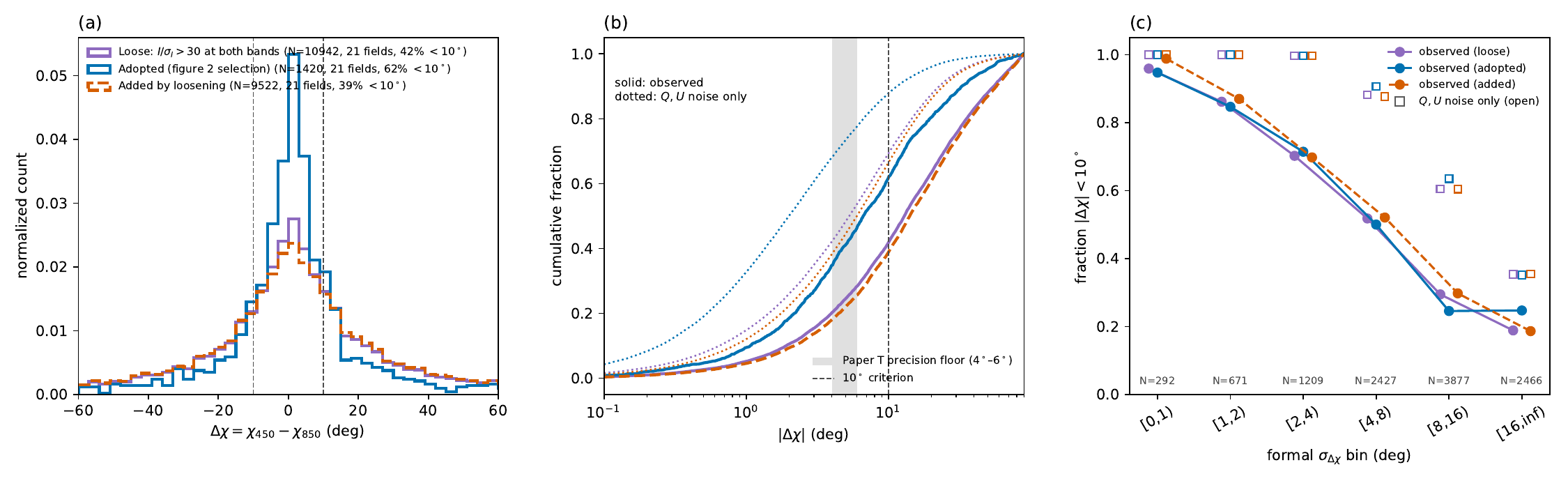}
\end{center}
\caption{Effect of the pixel selection on the difference in polarization angle between the bands, $\Delta\chi=\chi_{450}-\chi_{850}$, with $\chi_{450}$ after the 450\,$\mu$m Stokes-$Q/U$ zero-level offset correction of Paper~T (the convention of figure~\ref{fig:PA450vsPA850}).
Three samples from the same 21 fields are shown: loose ($I/\sigma_I>30$ in both bands only; purple, 10942 pixels), adopted (the selection of figure~\ref{fig:PA450vsPA850}; blue, 1420 pixels), and the pixels added by loosening (orange dashed, 9522 pixels).
(a) Normalized histograms of $\Delta\chi$.
The dashed lines mark the $10\arcdeg$ acceptance criterion.
(b) Cumulative distributions of $|\Delta\chi|$: observed (solid) and the prediction from the $Q,U$ uncertainties alone (dotted; 200 Gaussian realizations per pixel).
The light-gray band marks the Paper~T precision floor of $4\arcdeg$--$6\arcdeg$, below which differences are smaller than the measurement precision.
The dashed line marks the $10\arcdeg$ criterion.
(c) Fraction of pixels with $|\Delta\chi|<10\arcdeg$ in bins of the formal uncertainty $\sigma_{\Delta\chi}=(\sigma_{\chi,450}^2+\sigma_{\chi,850}^2)^{1/2}$: observed (filled circles) and $Q,U$ noise only (open squares).
The numbers give the count of pixels of the loose sample in each bin.}
\label{fig:F1_three_samples}
\end{figure*}

The sample adopted in figure~\ref{fig:PA450vsPA850} requires $I/\sigma_I>30$ in both bands and retains the 10\% of pixels with the highest $I$ in each field at both wavelengths.
We test the selection in figure~\ref{fig:F1_three_samples} using the same 21 fields and matched polarization measurements.
We compare a loose sample selected only by $I/\sigma_I>30$ in both bands (10942 pixels), the adopted sample (1420 pixels), and the pixels added by loosening the selection (9522 pixels).
The fraction with $|\Delta\chi|<10\arcdeg$ falls from 62\% (adopted) to 42\% (loose), and the median $|\Delta\chi|$ rises from $6.7\arcdeg$ to $13.0\arcdeg$.
The median signed difference stays at $+1.3\arcdeg$ in all three samples.
We attribute the change to the precision of individual pixels rather than to the selection itself.
In each bin of formal uncertainty $\sigma_{\Delta\chi}$, the three samples have nearly the same fraction within $10\arcdeg$ (figure~\ref{fig:F1_three_samples}c).
Among pixels with $\sigma_{\Delta\chi}<1\arcdeg$, 96\% of the loose sample lie within $10\arcdeg$.
In contrast, 85\% of the added pixels have $\sigma_{\Delta\chi}\geq 4\arcdeg$, where the $10\arcdeg$ criterion cannot separate agreement from noise.
 
The $Q,U$ uncertainties alone do not account for the observed spread.
For $\sigma_{\Delta\chi}<1\arcdeg$, a Monte Carlo realization of the $Q,U$ noise places every pixel within $10\arcdeg$ and gives a median absolute deviation of $0.4\arcdeg$.
The observed value is $2.3\arcdeg$.
Matching the observed spread requires an additional Gaussian term of $3\arcdeg$--$6\arcdeg$ for $\sigma_{\Delta\chi}<2\arcdeg$, comparable to the precision floor of $4\arcdeg$--$6\arcdeg$ derived in Paper~T (Paper~T equation~\ref{T-eq:precision_floors}).
A dedicated analysis is needed to separate the contributions to this additional spread from residual systematics, underestimated uncertainties in the maps, and genuine angle differences between the bands.
 
\bibliography{450um1G}{}

@ARTICLE{Andre2010,
       author = {{Andr{\'e}}, Ph. and {Men'shchikov}, A. and {Bontemps}, S. and {K{\"o}nyves}, V. and {Motte}, F. and {Schneider}, N. and {Didelon}, P. and {Minier}, V. and {Saraceno}, P. and {Ward-Thompson}, D. and {di Francesco}, J. and {White}, G. and {Molinari}, S. and {Testi}, L. and {Abergel}, A. and {Griffin}, M. and {Henning}, Th. and {Royer}, P. and {Mer{\'\i}n}, B. and {Vavrek}, R. and {Attard}, M. and {Arzoumanian}, D. and {Wilson}, C.~D. and {Ade}, P. and {Aussel}, H. and {Baluteau}, J. -P. and {Benedettini}, M. and {Bernard}, J. -Ph. and {Blommaert}, J.~A.~D.~L. and {Cambr{\'e}sy}, L. and {Cox}, P. and {di Giorgio}, A. and {Hargrave}, P. and {Hennemann}, M. and {Huang}, M. and {Kirk}, J. and {Krause}, O. and {Launhardt}, R. and {Leeks}, S. and {Le Pennec}, J. and {Li}, J.~Z. and {Martin}, P.~G. and {Maury}, A. and {Olofsson}, G. and {Omont}, A. and {Peretto}, N. and {Pezzuto}, S. and {Prusti}, T. and {Roussel}, H. and {Russeil}, D. and {Sauvage}, M. and {Sibthorpe}, B. and {Sicilia-Aguilar}, A. and {Spinoglio}, L. and {Waelkens}, C. and {Woodcraft}, A. and {Zavagno}, A.},
        title = "{From filamentary clouds to prestellar cores to the stellar IMF: Initial highlights from the Herschel Gould Belt Survey}",
      journal = {\aap},
         year = 2010,
        month = jul,
       volume = {518},
          eid = {L102},
        pages = {L102},
          doi = {10.1051/0004-6361/201014666},
archivePrefix = {arXiv},
       eprint = {1005.2618},
 primaryClass = {astro-ph.GA},
       adsurl = {https://ui.adsabs.harvard.edu/abs/2010A&A...518L.102A}
}

@ARTICLE{Andersson2015,
       author = {{Andersson}, B. -G. and {Lazarian}, A. and {Vaillancourt}, John E.},
        title = "{Interstellar Dust Grain Alignment}",
      journal = {Annual Review of Astronomy and Astrophysics},
         year = "2015",
        month = "Aug",
       volume = {53},
        pages = {501-539},
          doi = {10.1146/annurev-astro-082214-122414},
       adsurl = {https://ui.adsabs.harvard.edu/abs/2015ARA&A..53..501A}
}

@ARTICLE{Chapin2013,
       author = {{Chapin}, Edward L. and {Berry}, David S. and {Gibb}, Andrew G. and
         {Jenness}, Tim and {Scott}, Douglas and {Tilanus}, Remo P.~J. and
         {Economou}, Frossie and {Holland}, Wayne S.},
        title = "{SCUBA-2: iterative map-making with the Sub-Millimetre User Reduction Facility}",
      journal = {Monthly Notices of the Royal Astronomical Society},
         year = "2013",
        month = "Apr",
       volume = {430},
       number = {4},
        pages = {2545-2573},
          doi = {10.1093/mnras/stt052},
archivePrefix = {arXiv},
       eprint = {1301.3652},
 primaryClass = {astro-ph.IM},
       adsurl = {https://ui.adsabs.harvard.edu/abs/2013MNRAS.430.2545C}
}

@ARTICLE{Coude2019,
       author = {{Coud{\'e}}, Simon and {Bastien}, Pierre and {Houde}, Martin and
         {Sadavoy}, Sarah and {Friesen}, Rachel and {Di Francesco}, James and
         {Johnstone}, Doug and {Mairs}, Steve and {Hasegawa}, Tetsuo and
         {Kwon}, Woojin},
        title = "{The JCMT BISTRO Survey: The Magnetic Field of the Barnard 1 Star-forming Region}",
      journal = {The Astrophysical Journal},
         year = "2019",
        month = "Jun",
       volume = {877},
       number = {2},
          eid = {88},
        pages = {88},
          doi = {10.3847/1538-4357/ab1b23},
archivePrefix = {arXiv},
       eprint = {1904.07221},
 primaryClass = {astro-ph.GA},
       adsurl = {https://ui.adsabs.harvard.edu/abs/2019ApJ...877...88C}
}

@ARTICLE{Crutcher2012,
       author = {{Crutcher}, Richard M.},
        title = "{Magnetic Fields in Molecular Clouds}",
      journal = {\araa},
         year = "2012",
        month = "Sep",
       volume = {50},
        pages = {29-63},
          doi = {10.1146/annurev-astro-081811-125514},
       adsurl = {https://ui.adsabs.harvard.edu/abs/2012ARA&A..50...29C}
}

@ARTICLE{DavisGreenstein1951,
   author = {{Davis}, Jr., L. and {Greenstein}, J.~L.},
    title = "{The Polarization of Starlight by Aligned Dust Grains.}",
  journal = {The Astrophysical Journal},
     year = 1951,
    month = sep,
   volume = 114,
    pages = {206},
      doi = {10.1086/145464},
   adsurl = {https://ui.adsabs.harvard.edu/abs/1951ApJ...114..206D}
}

@ARTICLE{Drabek2012,
       author = {{Drabek}, E. and {Hatchell}, J. and {Friberg}, P. and {Richer}, J. and {Graves}, S. and {Buckle}, J.~V. and {Nutter}, D. and {Johnstone}, D. and {Di Francesco}, J.},
        title = "{Molecular line contamination in the SCUBA-2 450 and 850 {\ensuremath{\mu}}m continuum data}",
      journal = {\mnras},
         year = 2012,
        month = oct,
       volume = {426},
       number = {1},
        pages = {23-39},
          doi = {10.1111/j.1365-2966.2012.21140.x},
archivePrefix = {arXiv},
       eprint = {1204.6180},
 primaryClass = {astro-ph.GA},
       adsurl = {https://ui.adsabs.harvard.edu/abs/2012MNRAS.426...23D}
}

@ARTICLE{Draine2003,
       author = {{Draine}, B.~T.},
        title = "{Interstellar Dust Grains}",
      journal = {\araa},
         year = "2003",
        month = "Jan",
       volume = {41},
        pages = {241-289},
          doi = {10.1146/annurev.astro.41.011802.094840},
archivePrefix = {arXiv},
       eprint = {astro-ph/0304489},
 primaryClass = {astro-ph},
       adsurl = {https://ui.adsabs.harvard.edu/abs/2003ARA&A..41..241D}
}

@ARTICLE{Fissel2016,
   author = {{Fissel}, L.~M. and {Ade}, P.~A.~R. and {Angil{\`e}}, F.~E. and 
	{Ashton}, P. and {Benton}, S.~J. and {Devlin}, M.~J. and {Dober}, B. and 
	{Fukui}, Y. and {Galitzki}, N. and {Gandilo}, N.~N. and {Klein}, J. and 
	{Korotkov}, A.~L. and {Li}, Z.-Y. and {Martin}, P.~G. and {Matthews}, T.~G. and 
	{Moncelsi}, L. and {Nakamura}, F. and {Netterfield}, C.~B. and 
	{Novak}, G. and {Pascale}, E. and {Poidevin}, F. and {Santos}, F.~P. and 
	{Savini}, G. and {Scott}, D. and {Shariff}, J.~A. and {Diego Soler}, J. and 
	{Thomas}, N.~E. and {Tucker}, C.~E. and {Tucker}, G.~S. and 
	{Ward-Thompson}, D.},
    title = "{Balloon-Borne Submillimeter Polarimetry of the Vela C Molecular Cloud: Systematic Dependence of Polarization Fraction on Column Density and Local Polarization-Angle Dispersion}",
  journal = {\apj},
archivePrefix = "arXiv",
   eprint = {1509.05298},
     year = 2016,
    month = jun,
   volume = 824,
      eid = {134},
    pages = {134},
      doi = {10.3847/0004-637X/824/2/134},
   adsurl = {https://ui.adsabs.harvard.edu/abs/2016ApJ...824..134F}
}

@INPROCEEDINGS{Friberg2016,
       author = {{Friberg}, Per and {Bastien}, Pierre and {Berry}, David and
         {Savini}, Giorgio and {Graves}, Sarah F. and {Pattle}, Kate},
        title = "{POL-2: a polarimeter for the James-Clerk-Maxwell telescope}",
    booktitle = {Proceedings of the SPIE},
         year = "2016",
       series = {Society of Photo-Optical Instrumentation Engineers (SPIE) Conference Series},
       volume = {9914},
        month = "Jul",
          eid = {991403},
        pages = {991403},
          doi = {10.1117/12.2231943},
       adsurl = {https://ui.adsabs.harvard.edu/abs/2016SPIE.9914E..03F}
}

@ARTICLE{Goldreich1981,
       author = {{Goldreich}, P. and {Kylafis}, N.~D.},
        title = "{On mapping the magnetic field direction in molecular clouds by polarization measurements}",
      journal = {\apjl},
         year = "1981",
        month = "Jan",
       volume = {243},
        pages = {L75-L78},
          doi = {10.1086/183446},
       adsurl = {https://ui.adsabs.harvard.edu/abs/1981ApJ...243L..75G}
}

@ARTICLE{Gandilo2016,
       author = {{Gandilo}, Natalie N. and {Ade}, Peter A.~R. and {Angil{\`e}}, Francesco E. and {Ashton}, Peter and {Benton}, Steven J. and {Devlin}, Mark J. and {Dober}, Bradley and {Fissel}, Laura M. and {Fukui}, Yasuo and {Galitzki}, Nicholas and {Klein}, Jeffrey and {Korotkov}, Andrei L. and {Li}, Zhi-Yun and {Martin}, Peter G. and {Matthews}, Tristan G. and {Moncelsi}, Lorenzo and {Nakamura}, Fumitaka and {Netterfield}, Calvin B. and {Novak}, Giles and {Pascale}, Enzo and {Poidevin}, Fr{\'e}d{\'e}rick and {Santos}, Fabio P. and {Savini}, Giorgio and {Scott}, Douglas and {Shariff}, Jamil A. and {Diego Soler}, Juan and {Thomas}, Nicholas E. and {Tucker}, Carole E. and {Tucker}, Gregory S. and {Ward-Thompson}, Derek},
        title = "{Submillimeter Polarization Spectrum in the Vela C Molecular Cloud}",
      journal = {\apj},
         year = 2016,
        month = jun,
       volume = {824},
       number = {2},
          eid = {84},
        pages = {84},
          doi = {10.3847/0004-637X/824/2/84},
archivePrefix = {arXiv},
       eprint = {1512.06745},
 primaryClass = {astro-ph.IM},
       adsurl = {https://ui.adsabs.harvard.edu/abs/2016ApJ...824...84G}
}

@ARTICLE{Hildebrand1999,
       author = {{Hildebrand}, R.~H. and {Dotson}, J.~L. and {Dowell}, C.~D. and
         {Schleuning}, D.~A. and {Vaillancourt}, J.~E.},
        title = "{The Far-Infrared Polarization Spectrum: First Results and Analysis}",
      journal = {\apj},
         year = "1999",
        month = "May",
       volume = {516},
       number = {2},
        pages = {834-842},
          doi = {10.1086/307142},
       adsurl = {https://ui.adsabs.harvard.edu/abs/1999ApJ...516..834H}
}

@ARTICLE{Hildebrand1983,
       author = {{Hildebrand}, R.~H.},
        title = "{The determination of cloud masses and dust characteristics from submillimetre thermal emission.}",
      journal = {\qjras},
         year = "1983",
        month = "Sep",
       volume = {24},
        pages = {267-282},
       adsurl = {https://ui.adsabs.harvard.edu/abs/1983QJRAS..24..267H}
}

@ARTICLE{Hildebrand2009,
       author = {{Hildebrand}, Roger H. and {Kirby}, Larry and {Dotson}, Jessie L. and
         {Houde}, Martin and {Vaillancourt}, John E.},
        title = "{Dispersion of Magnetic Fields in Molecular Clouds. I}",
      journal = {\apj},
         year = "2009",
        month = "May",
       volume = {696},
       number = {1},
        pages = {567-573},
          doi = {10.1088/0004-637X/696/1/567},
archivePrefix = {arXiv},
       eprint = {0811.0813},
 primaryClass = {astro-ph},
       adsurl = {https://ui.adsabs.harvard.edu/abs/2009ApJ...696..567H}
}

@ARTICLE{Holland2013,
       author = {{Holland}, W.~S. and {Bintley}, D. and {Chapin}, E.~L. and
         {Chrysostomou}, A. and {Davis}, G.~R. and {Dempsey}, J.~T. and
         {Duncan}, W.~D. and {Fich}, M. and {Friberg}, P. and {Halpern}, M. and
         {Irwin}, K.~D. and {Jenness}, T. and {Kelly}, B.~D. and
         {MacIntosh}, M.~J. and {Robson}, E.~I. and {Scott}, D. and
         {Ade}, P.~A.~R. and {Atad-Ettedgui}, E. and {Berry}, D.~S. and
         {Craig}, S.~C. and {Gao}, X. and {Gibb}, A.~G. and {Hilton}, G.~C. and
         {Hollister}, M.~I. and {Kycia}, J.~B. and {Lunney}, D.~W. and
         {McGregor}, H. and {Montgomery}, D. and {Parkes}, W. and
         {Tilanus}, R.~P.~J. and {Ullom}, J.~N. and {Walther}, C.~A. and
         {Walton}, A.~J. and {Woodcraft}, A.~L. and {Amiri}, M. and
         {Atkinson}, D. and {Burger}, B. and {Chuter}, T. and {Coulson}, I.~M. and
         {Doriese}, W.~B. and {Dunare}, C. and {Economou}, F. and
         {Niemack}, M.~D. and {Parsons}, H.~A.~L. and {Reintsema}, C.~D. and
         {Sibthorpe}, B. and {Smail}, I. and {Sudiwala}, R. and {Thomas}, H.~S.},
        title = "{SCUBA-2: the 10 000 pixel bolometer camera on the James Clerk Maxwell Telescope}",
      journal = {Monthly Notices of the Royal Astronomical Society},
         year = "2013",
        month = "Apr",
       volume = {430},
       number = {4},
        pages = {2513-2533},
          doi = {10.1093/mnras/sts612},
archivePrefix = {arXiv},
       eprint = {1301.3650},
 primaryClass = {astro-ph.IM},
       adsurl = {https://ui.adsabs.harvard.edu/abs/2013MNRAS.430.2513H}
}

@ARTICLE{Hull2014,
       author = {{Hull}, Charles L.~H. and {Plambeck}, Richard L. and {Kwon}, Woojin and
         {Bower}, Geoffrey C. and {Carpenter}, John M. and
         {Crutcher}, Richard M. and {Fiege}, Jason D. and {Franzmann}, Erica and
         {Hakobian}, Nicholas S. and {Heiles}, Carl and {Houde}, Martin and
         {Hughes}, A. Meredith and {Lamb}, James W. and {Looney}, Leslie W. and
         {Marrone}, Daniel P. and {Matthews}, Brenda C. and {Pillai}, Thushara and
         {Pound}, Marc W. and {Rahman}, Nurur and {Sandell}, G{\"o}ran and
         {Stephens}, Ian W. and {Tobin}, John J. and {Vaillancourt}, John E. and
         {Volgenau}, N.~H. and {Wright}, Melvyn C.~H.},
        title = "{TADPOL: A 1.3 mm Survey of Dust Polarization in Star-forming Cores and Regions}",
      journal = {\apjs},
         year = "2014",
        month = "Jul",
       volume = {213},
       number = {1},
          eid = {13},
        pages = {13},
          doi = {10.1088/0067-0049/213/1/13},
archivePrefix = {arXiv},
       eprint = {1310.6653},
 primaryClass = {astro-ph.SR},
       adsurl = {https://ui.adsabs.harvard.edu/abs/2014ApJS..213...13H}
}

@ARTICLE{Hull2019,
       author = {{Hull}, Charles L.~H. and {Zhang}, Qizhou},
        title = "{Interferometric observations of magnetic fields in forming stars}",
      journal = {Frontiers in Astronomy and Space Sciences},
         year = "2019",
        month = "Mar",
       volume = {6},
          eid = {3},
        pages = {3},
          doi = {10.3389/fspas.2019.00003},
archivePrefix = {arXiv},
       eprint = {1903.03177},
 primaryClass = {astro-ph.SR},
       adsurl = {https://ui.adsabs.harvard.edu/abs/2019FrASS...6....3H}
}

@ARTICLE{Kwon2018,
       author = {{Kwon}, Jungmi and {Doi}, Yasuo and {Tamura}, Motohide and
         {Matsumura}, Masafumi and {Pattle}, Kate and {Berry}, David and
         {Sadavoy}, Sarah and {Matthews}, Brenda C. and {Ward-Thompson}, Derek and
         {Hasegawa}, Tetsuo},
        title = "{A First Look at BISTRO Observations of the {\ensuremath{\rho}} Oph-A core}",
      journal = {The Astrophysical Journal},
         year = "2018",
        month = "May",
       volume = {859},
       number = {1},
          eid = {4},
        pages = {4},
          doi = {10.3847/1538-4357/aabd82},
archivePrefix = {arXiv},
       eprint = {1804.09313},
 primaryClass = {astro-ph.GA},
       adsurl = {https://ui.adsabs.harvard.edu/abs/2018ApJ...859....4K}
}

@ARTICLE{Lazarian2007,
       author = {{Lazarian}, A.},
        title = "{Tracing magnetic fields with aligned grains}",
      journal = {\jqsrt},
         year = 2007,
        month = jul,
       volume = {106},
        pages = {225-256},
          doi = {10.1016/j.jqsrt.2007.01.038},
archivePrefix = {arXiv},
       eprint = {0707.0858},
 primaryClass = {astro-ph},
       adsurl = {https://ui.adsabs.harvard.edu/abs/2007JQSRT.106..225L}
}

@ARTICLE{LazarianHoang2007,
       author = {{Lazarian}, A. and {Hoang}, Thiem},
        title = "{Radiative torques: analytical model and basic properties}",
      journal = {\mnras},
         year = 2007,
        month = jul,
       volume = {378},
       number = {3},
        pages = {910-946},
          doi = {10.1111/j.1365-2966.2007.11817.x},
archivePrefix = {arXiv},
       eprint = {0707.0886},
 primaryClass = {astro-ph},
       adsurl = {https://ui.adsabs.harvard.edu/abs/2007MNRAS.378..910L}
}

@ARTICLE{Lyo2021,
       author = {{Lyo}, A. -Ran and {Kim}, Jongsoo and {Sadavoy}, Sarah and {Johnstone}, Doug and {Berry}, David and {Pattle}, Kate and {Kwon}, Woojin and {Bastien}, Pierre and {Onaka}, Takashi and {Di Francesco}, James and {Kang}, Ji-Hyun and {Furuya}, Ray and {Hull}, Charles L.~H. and {Tamura}, Motohide and {Koch}, Patrick M. and {Ward-Thompson}, Derek and {Hasegawa}, Tetsuo and {Hoang}, Thiem and {Arzoumanian}, Doris and {Won Lee}, Chang and {Lee}, Chin-Fei and {Byun}, Do-Young and {Kirchschlager}, Florian and {Doi}, Yasuo and {Kim}, Kee-Tae and {Hwang}, Jihye and {Diep}, Pham Ngoc and {Fanciullo}, Lapo and {Lee}, Sang-Sung and {Park}, Geumsook and {Yoo}, Hyunju and {Chung}, Eun Jung and {Whitworth}, Anthony and {Mairs}, Steve and {Soam}, Archana and {Liu}, Tie and {Tang}, Xindi and {Coud{\'e}}, Simon and {Andr{\'e}}, Philippe and {Bourke}, Tyler L. and {Vivien Chen}, Huei-Ru and {Chen}, Zhiwei and {Ping Chen}, Wen and {Chen}, Mike and {Ching}, Tao-Chung and {Cho}, Jungyeon and {Choi}, Minho and {Choi}, Yunhee and {Chrysostomou}, Antonio and {Dai}, Sophia and {Dowell}, C. Darren and {Duan}, Hao-Yuan and {Duan}, Yan and {Eden}, David and {Eswaraiah}, Chakali and {Eyres}, Stewart and {Fiege}, Jason and {Fissel}, Laura M. and {Franzmann}, Erica and {Friberg}, Per and {Friesen}, Rachel and {Fuller}, Gary and {Gledhill}, Tim and {Graves}, Sarah and {Greaves}, Jane and {Griffin}, Matt and {Gu}, Qilao and {Han}, Ilseung and {Hatchell}, Jannifer and {Hayashi}, Saeko and {Houde}, Martin and {Inoue}, Tsuyoshi and {Inutsuka}, Shu-ichiro and {Iwasaki}, Kazunari and {Jeong}, Il-Gyo and {Kang}, Miju and {Kataoka}, Akimasa and {Kawabata}, Koji and {Kemper}, Francisca and {Kim}, Gwanjeong and {Kim}, Mi-Ryang and {Kim}, Shinyoung and {Kim}, Kyoung Hee and {Kirk}, Jason and {Kobayashi}, Masato I.~N. and {K{\"o}nyves}, Vera and {Kusune}, Takayoshi and {Kwon}, Jungmi and {Lacaille}, Kevin and {Lai}, Shih-Ping and {Law}, Chi-Yan and {Lee}, Jeong-Eun and {Lee}, Yong-Hee and {Lee}, Hyeseung and {Li}, Dalei and {Li}, Di and {Li}, Hua-Bai and {Liu}, Hong-Li and {Liu}, Junhao and {Liu}, Sheng-Yuan and {Lu}, Xing and {Matsumura}, Masafumi and {Matthews}, Brenda and {Moriarty-Schieven}, Gerald and {Nagata}, Tetsuya and {Nakamura}, Fumitaka and {Nakanishi}, Hiroyuki and {Bich Ngoc}, Nguyen and {Ohashi}, Nagayoshi and {Parsons}, Harriet and {Peretto}, Nicolas and {Priestley}, Felix and {Pyo}, Tae-soo and {Qian}, Lei and {Qiu}, Keping and {Rao}, Ramprasad and {Rawlings}, Jonathan and {Rawlings}, Mark G. and {Retter}, Brendan and {Richer}, John and {Rigby}, Andrew and {Saito}, Hiro and {Savini}, Giorgio and {Scaife}, Anna and {Seta}, Masumichi and {Shimajiri}, Yoshito and {Shinnaga}, Hiroko and {Tahani}, Mehrnoosh and {Tang}, Ya-Wen and {Tomisaka}, Kohji and {Tram}, Le Ngoc and {Tsukamoto}, Yusuke and {Viti}, Serena and {Wang}, Jia-Wei and {Wang}, Hongchi and {Xie}, Jinjin and {Yen}, Hsi-Wei and {Yuan}, Jinghua and {Yun}, Hyeong-Sik and {Zenko}, Tetsuya and {Zhang}, Guoyin and {Zhang}, Chuan-Peng and {Zhang}, Yapeng and {Zhou}, Jianjun and {Zhu}, Lei and {de Looze}, Ilse and {Dowell}, C. Darren and {Falle}, Sam and {Robitaille}, Jean-Fran{\c{c}}ois and {van Loo}, Sven},
        title = "{The JCMT BISTRO Survey: An 850/450 {\ensuremath{\mu}}m Polarization Study of NGC 2071IR in Orion B}",
      journal = {\apj},
         year = 2021,
        month = sep,
       volume = {918},
       number = {2},
          eid = {85},
        pages = {85},
          doi = {10.3847/1538-4357/ac0ce9},
archivePrefix = {arXiv},
       eprint = {2109.13543},
 primaryClass = {astro-ph.GA},
       adsurl = {https://ui.adsabs.harvard.edu/abs/2021ApJ...918...85L}
}

@ARTICLE{Mairs2021,
       author = {{Mairs}, Steve and {Dempsey}, Jessica T. and {Bell}, Graham S. and {Parsons}, Harriet and {Currie}, Malcolm J. and {Friberg}, Per and {Jiang}, Xue-Jian and {Tetarenko}, Alexandra J. and {Bintley}, Dan and {Cookson}, Jamie and {Li}, Shaoliang and {Rawlings}, Mark G. and {Wouterloot}, Jan and {Berry}, David and {Graves}, Sarah and {Mizuno}, Izumi and {Acohido}, Alexis Ann and {Clark}, Alyssa and {Cox}, Jeff and {Fuchs}, Miriam and {Hoge}, James and {Kemp}, Johnathon and {Lee}, E'lisa and {Matulonis}, Callie and {Montgomerie}, William and {Silva}, Kevin and {Smith}, Patrice},
        title = "{A Decade of SCUBA-2: A Comprehensive Guide to Calibrating 450 {\ensuremath{\mu}}m and 850 {\ensuremath{\mu}}m Continuum Data at the JCMT}",
      journal = {\aj},
         year = 2021,
        month = nov,
       volume = {162},
       number = {5},
          eid = {191},
        pages = {191},
          doi = {10.3847/1538-3881/ac18bf},
archivePrefix = {arXiv},
       eprint = {2107.13558},
 primaryClass = {astro-ph.IM},
       adsurl = {https://ui.adsabs.harvard.edu/abs/2021AJ....162..191M}
}

@ARTICLE{Montier2015,
       author = {{Montier}, L. and {Plaszczynski}, S. and {Levrier}, F. and {Tristram}, M. and {Alina}, D. and {Ristorcelli}, I. and {Bernard}, J. -P. and {Guillet}, V.},
        title = "{Polarization measurement analysis. II. Best estimators of polarization fraction and angle}",
      journal = {\aap},
         year = 2015,
        month = feb,
       volume = {574},
          eid = {A136},
        pages = {A136},
          doi = {10.1051/0004-6361/201424451},
archivePrefix = {arXiv},
       eprint = {1407.0178},
 primaryClass = {astro-ph.IM},
       adsurl = {https://ui.adsabs.harvard.edu/abs/2015A&A...574A.136M}
}

@ARTICLE{Pattle2017,
       author = {{Pattle}, Kate and {Ward-Thompson}, Derek and {Berry}, David and
         {Hatchell}, Jennifer and {Chen}, Huei-Ru and {Pon}, Andy and
         {Koch}, Patrick M. and {Kwon}, Woojin and {Kim}, Jongsoo and
         {Bastien}, Pierre},
        title = "{The JCMT BISTRO Survey: The Magnetic Field Strength in the Orion A Filament}",
      journal = {The Astrophysical Journal},
         year = "2017",
        month = "Sep",
       volume = {846},
       number = {2},
          eid = {122},
        pages = {122},
          doi = {10.3847/1538-4357/aa80e5},
archivePrefix = {arXiv},
       eprint = {1707.05269},
 primaryClass = {astro-ph.GA},
       adsurl = {https://ui.adsabs.harvard.edu/abs/2017ApJ...846..122P}
}

@ARTICLE{Pattle2021,
       author = {{Pattle}, Kate and {Lai}, Shih-Ping and {Di Francesco}, James and {Sadavoy}, Sarah and {Ward-Thompson}, Derek and {Johnstone}, Doug and {Hoang}, Thiem and {Arzoumanian}, Doris and {Bastien}, Pierre and {Bourke}, Tyler L. and {Coud{\'e}}, Simon and {Doi}, Yasuo and {Eswaraiah}, Chakali and {Fanciullo}, Lapo and {Furuya}, Ray S. and {Hwang}, Jihye and {Hull}, Charles L.~H. and {Kang}, Jihyun and {Kim}, Kee-Tae and {Kirchschlager}, Florian and {Kwon}, Jungmi and {Kwon}, Woojin and {Lee}, Chang Won and {Liu}, Tie and {Redman}, Matt and {Soam}, Archana and {Tahani}, Mehrnoosh and {Tamura}, Motohide and {Tang}, Xindi},
        title = "{JCMT POL-2 and BISTRO Survey Observations of Magnetic Fields in the L1689 Molecular Cloud}",
      journal = {\apj},
         year = 2021,
        month = feb,
       volume = {907},
       number = {2},
          eid = {88},
        pages = {88},
          doi = {10.3847/1538-4357/abcc6c},
archivePrefix = {arXiv},
       eprint = {2011.09765},
 primaryClass = {astro-ph.GA},
       adsurl = {https://ui.adsabs.harvard.edu/abs/2021ApJ...907...88P}
}

@ARTICLE{PlanckIntXIX2015,
       author = {{Planck Collaboration} and {Ade}, P.~A.~R. and {Aghanim}, N. and
         {Alina}, D. and {Alves}, M.~I.~R. and {Armitage-Caplan}, C. and
         {Arnaud}, M. and {Arzoumanian}, D. and {Ashdown}, M. and {Atrio-Barand
        ela}, F.},
        title = "{Planck intermediate results. XIX. An overview of the polarized thermal emission from Galactic dust}",
      journal = {Astronomy \& Astrophysics},
         year = "2015",
        month = "Apr",
       volume = {576},
          eid = {A104},
        pages = {A104},
          doi = {10.1051/0004-6361/201424082},
archivePrefix = {arXiv},
       eprint = {1405.0871},
 primaryClass = {astro-ph.GA},
       adsurl = {https://ui.adsabs.harvard.edu/abs/2015A&A...576A.104P}
}

@ARTICLE{Plaszczynski2014,
       author = {{Plaszczynski}, S. and {Montier}, L. and {Levrier}, F. and {Tristram}, M.},
        title = "{A novel estimator of the polarization amplitude from normally distributed Stokes parameters}",
      journal = {\mnras},
         year = 2014,
        month = apr,
       volume = {439},
       number = {4},
        pages = {4048-4056},
          doi = {10.1093/mnras/stu270},
archivePrefix = {arXiv},
       eprint = {1312.0437},
 primaryClass = {astro-ph.CO},
       adsurl = {https://ui.adsabs.harvard.edu/abs/2014MNRAS.439.4048P}
}

@ARTICLE{Shariff2019,
       author = {{Shariff}, Jamil A. and {Ade}, Peter A.~R. and
         {Angil{\`e}}, Francesco E. and {Ashton}, Peter and {Benton}, Steven J. and
         {Devlin}, Mark J. and {Dober}, Bradley and {Fissel}, Laura M. and
         {Fukui}, Yasuo and {Galitzki}, Nicholas},
        title = "{Submillimeter Polarization Spectrum of the Carina Nebula}",
      journal = {The Astrophysical Journal},
         year = "2019",
        month = "Feb",
       volume = {872},
       number = {2},
          eid = {197},
        pages = {197},
          doi = {10.3847/1538-4357/aaff5f},
archivePrefix = {arXiv},
       eprint = {1809.06375},
 primaryClass = {astro-ph.GA},
       adsurl = {https://ui.adsabs.harvard.edu/abs/2019ApJ...872..197S}
}

@ARTICLE{Soam2018,
       author = {{Soam}, Archana and {Pattle}, Kate and {Ward-Thompson}, Derek and
         {Lee}, Chang Won and {Sadavoy}, Sarah and {Koch}, Patrick M. and
         {Kim}, Gwanjeong and {Kwon}, Jungmi and {Kwon}, Woojin and
         {Arzoumanian}, Doris},
        title = "{Magnetic Fields toward Ophiuchus-B Derived from SCUBA-2 Polarization Measurements}",
      journal = {The Astrophysical Journal},
         year = "2018",
        month = "Jul",
       volume = {861},
       number = {1},
          eid = {65},
        pages = {65},
          doi = {10.3847/1538-4357/aac4a6},
archivePrefix = {arXiv},
       eprint = {1805.06131},
 primaryClass = {astro-ph.GA},
       adsurl = {https://ui.adsabs.harvard.edu/abs/2018ApJ...861...65S}
}

@ARTICLE{Vaillancourt2008,
       author = {{Vaillancourt}, John E. and {Dowell}, C. Darren and {Hildebrand
        }, Roger H. and {Kirby}, Larry and {Krejny}, Megan M. and {Li}, Hua-bai and
         {Novak}, Giles and {Houde}, Martin and {Shinnaga}, Hiroko and
         {Attard}, Michael},
        title = "{New Results on the Submillimeter Polarization Spectrum of the Orion Molecular Cloud}",
      journal = {\apjl},
         year = "2008",
        month = "May",
       volume = {679},
       number = {1},
        pages = {L25},
          doi = {10.1086/589152},
archivePrefix = {arXiv},
       eprint = {0803.4185},
 primaryClass = {astro-ph},
       adsurl = {https://ui.adsabs.harvard.edu/abs/2008ApJ...679L..25V}
}

@ARTICLE{VaillancourtMatthews2012,
       author = {{Vaillancourt}, John E. and {Matthews}, Brenda C.},
        title = "{Submillimeter Polarization of Galactic Clouds: A Comparison of 350 {\ensuremath{\mu}}m and 850 {\ensuremath{\mu}}m Data}",
      journal = {\apjs},
         year = 2012,
        month = aug,
       volume = {201},
       number = {2},
          eid = {13},
        pages = {13},
          doi = {10.1088/0067-0049/201/2/13},
archivePrefix = {arXiv},
       eprint = {1204.1378},
 primaryClass = {astro-ph.GA},
       adsurl = {https://ui.adsabs.harvard.edu/abs/2012ApJS..201...13V}
}

@ARTICLE{Ward-Thompson2017,
       author = {{Ward-Thompson}, Derek and {Pattle}, Kate and {Bastien}, Pierre and
         {Furuya}, Ray S. and {Kwon}, Woojin and {Lai}, Shih-Ping and
         {Qiu}, Keping and {Berry}, David and {Choi}, Minho and
         {Coud{\'e}}, Simon},
        title = "{First Results from BISTRO: A SCUBA-2 Polarimeter Survey of the Gould Belt}",
      journal = {The Astrophysical Journal},
         year = "2017",
        month = "Jun",
       volume = {842},
       number = {1},
          eid = {66},
        pages = {66},
          doi = {10.3847/1538-4357/aa70a0},
archivePrefix = {arXiv},
       eprint = {1704.08552},
 primaryClass = {astro-ph.GA},
       adsurl = {https://ui.adsabs.harvard.edu/abs/2017ApJ...842...66W}
}

@ARTICLE{Alves2014,
       author = {{Alves}, Felipe O. and {Frau}, Pau and {Girart}, Josep M. and {Franco}, Gabriel A.~P. and {Santos}, Fabio P. and {Wiesemeyer}, Helmut},
        title = "{On the radiation driven alignment of dust grains: Detection of the polarization hole in a starless core}",
      journal = {\aap},
         year = 2014,
        month = sep,
       volume = {569},
          eid = {L1},
        pages = {L1},
          doi = {10.1051/0004-6361/201424678},
archivePrefix = {arXiv},
       eprint = {1408.4126},
 primaryClass = {astro-ph.SR},
       adsurl = {https://ui.adsabs.harvard.edu/abs/2014A&A...569L...1A}
}

@ARTICLE{Planck2020-XII,
       author = {{Planck Collaboration} and {Aghanim}, N. and {Akrami}, Y. and {Alves}, M.~I.~R. and {Ashdown}, M. and {Aumont}, J. and {Baccigalupi}, C. and {Ballardini}, M. and {Banday}, A.~J. and {Barreiro}, R.~B. and {Bartolo}, N. and {Basak}, S. and {Benabed}, K. and {Bernard}, J. -P. and {Bersanelli}, M. and {Bielewicz}, P. and {Bock}, J.~J. and {Bond}, J.~R. and {Borrill}, J. and {Bouchet}, F.~R. and {Boulanger}, F. and {Bracco}, A. and {Bucher}, M. and {Burigana}, C. and {Calabrese}, E. and {Cardoso}, J. -F. and {Carron}, J. and {Chary}, R. -R. and {Chiang}, H.~C. and {Colombo}, L.~P.~L. and {Combet}, C. and {Crill}, B.~P. and {Cuttaia}, F. and {de Bernardis}, P. and {de Zotti}, G. and {Delabrouille}, J. and {Delouis}, J. -M. and {Di Valentino}, E. and {Dickinson}, C. and {Diego}, J.~M. and {Dor{\'e}}, O. and {Douspis}, M. and {Ducout}, A. and {Dupac}, X. and {Efstathiou}, G. and {Elsner}, F. and {En{\ss}lin}, T.~A. and {Eriksen}, H.~K. and {Falgarone}, E. and {Fantaye}, Y. and {Fernandez-Cobos}, R. and {Ferri{\`e}re}, K. and {Finelli}, F. and {Forastieri}, F. and {Frailis}, M. and {Fraisse}, A.~A. and {Franceschi}, E. and {Frolov}, A. and {Galeotta}, S. and {Galli}, S. and {Ganga}, K. and {G{\'e}nova-Santos}, R.~T. and {Gerbino}, M. and {Ghosh}, T. and {Gonz{\'a}lez-Nuevo}, J. and {G{\'o}rski}, K.~M. and {Gratton}, S. and {Green}, G. and {Gruppuso}, A. and {Gudmundsson}, J.~E. and {Guillet}, V. and {Handley}, W. and {Hansen}, F.~K. and {Helou}, G. and {Herranz}, D. and {Hivon}, E. and {Huang}, Z. and {Jaffe}, A.~H. and {Jones}, W.~C. and {Keih{\"a}nen}, E. and {Keskitalo}, R. and {Kiiveri}, K. and {Kim}, J. and {Krachmalnicoff}, N. and {Kunz}, M. and {Kurki-Suonio}, H. and {Lagache}, G. and {Lamarre}, J. -M. and {Lasenby}, A. and {Lattanzi}, M. and {Lawrence}, C.~R. and {Le Jeune}, M. and {Levrier}, F. and {Liguori}, M. and {Lilje}, P.~B. and {Lindholm}, V. and {L{\'o}pez-Caniego}, M. and {Lubin}, P.~M. and {Ma}, Y. -Z. and {Mac{\'\i}as-P{\'e}rez}, J.~F. and {Maggio}, G. and {Maino}, D. and {Mandolesi}, N. and {Mangilli}, A. and {Marcos-Caballero}, A. and {Maris}, M. and {Martin}, P.~G. and {Mart{\'\i}nez-Gonz{\'a}lez}, E. and {Matarrese}, S. and {Mauri}, N. and {McEwen}, J.~D. and {Melchiorri}, A. and {Mennella}, A. and {Migliaccio}, M. and {Miville-Desch{\^e}nes}, M. -A. and {Molinari}, D. and {Moneti}, A. and {Montier}, L. and {Morgante}, G. and {Moss}, A. and {Natoli}, P. and {Pagano}, L. and {Paoletti}, D. and {Patanchon}, G. and {Perrotta}, F. and {Pettorino}, V. and {Piacentini}, F. and {Polastri}, L. and {Polenta}, G. and {Puget}, J. -L. and {Rachen}, J.~P. and {Reinecke}, M. and {Remazeilles}, M. and {Renzi}, A. and {Ristorcelli}, I. and {Rocha}, G. and {Rosset}, C. and {Roudier}, G. and {Rubi{\~n}o-Mart{\'\i}n}, J.~A. and {Ruiz-Granados}, B. and {Salvati}, L. and {Sandri}, M. and {Savelainen}, M. and {Scott}, D. and {Sirignano}, C. and {Sunyaev}, R. and {Suur-Uski}, A. -S. and {Tauber}, J.~A. and {Tavagnacco}, D. and {Tenti}, M. and {Toffolatti}, L. and {Tomasi}, M. and {Trombetti}, T. and {Valiviita}, J. and {Vansyngel}, F. and {Van Tent}, B. and {Vielva}, P. and {Villa}, F. and {Vittorio}, N. and {Wandelt}, B.~D. and {Wehus}, I.~K. and {Zacchei}, A. and {Zonca}, A.},
        title = "{Planck 2018 results. XII. Galactic astrophysics using polarized dust emission}",
      journal = {\aap},
         year = 2020,
        month = sep,
       volume = {641},
          eid = {A12},
        pages = {A12},
          doi = {10.1051/0004-6361/201833885},
archivePrefix = {arXiv},
       eprint = {1807.06212},
 primaryClass = {astro-ph.GA},
       adsurl = {https://ui.adsabs.harvard.edu/abs/2020A&A...641A..12P}
}

@ARTICLE{Serkowski1975,
    author = {{Serkowski}, K. and {Mathewson}, D.~S. and {Ford}, V.~L.},
    title = "{Wavelength dependence of interstellar polarization and ratio of total to selective extinction}",
    journal = {\apj},
    year = 1975,
    volume = {196},
    pages = {261-290},
    doi = {10.1086/153410}
}

@ARTICLE{Guillet2018,
    author = {{Guillet}, V. and {Fanciullo}, L. and {Verstraete}, L. and 
              {Boulanger}, F. and {Jones}, A.~P. and {others}},
    title = "{Dust models compatible with Planck intensity and polarization data in translucent lines of sight}",
    journal = {\aap},
    year = 2018,
    volume = {610},
    pages = {A16},
    doi = {10.1051/0004-6361/201630271},
    archivePrefix = {arXiv},
    eprint = {1710.04598}
}

@ARTICLE{Hensley2023,
       author = {{Hensley}, Brandon S. and {Draine}, B.~T.},
        title = "{The Astrodust+PAH Model: A Unified Description of the Extinction, Emission, and Polarization from Dust in the Diffuse Interstellar Medium}",
      journal = {\apj},
         year = 2023,
        month = may,
       volume = {948},
       number = {1},
          eid = {55},
        pages = {55},
          doi = {10.3847/1538-4357/acc4c2},
archivePrefix = {arXiv},
       eprint = {2208.12365},
 primaryClass = {astro-ph.GA},
       adsurl = {https://ui.adsabs.harvard.edu/abs/2023ApJ...948...55H}
}

@ARTICLE{Hoang2021,
       author = {{Hoang}, Thiem and {Tram}, Le Ngoc and {Lee}, Hyeseung and {Diep}, Pham Ngoc and {Ngoc}, Nguyen Bich},
        title = "{Grain Alignment and Disruption by Radiative Torques in Dense Molecular Clouds and Implication for Polarization Holes}",
      journal = {\apj},
         year = 2021,
        month = feb,
       volume = {908},
       number = {2},
          eid = {218},
        pages = {218},
          doi = {10.3847/1538-4357/abd54f},
archivePrefix = {arXiv},
       eprint = {2010.07742},
 primaryClass = {astro-ph.GA},
       adsurl = {https://ui.adsabs.harvard.edu/abs/2021ApJ...908..218H}
}

@ARTICLE{Konyves2015,
    author = {{K{\"o}nyves}, V. and {Andr{\'e}}, P. and {Men'shchikov}, A. and others},
    title = "{A census of dense cores in the Aquila cloud complex: SPIRE/PACS observations from the Herschel Gould Belt survey}",
    journal = {\aap},
    year = 2015,
    volume = {584},
    pages = {A91},
    doi = {10.1051/0004-6361/201525861}
}

@ARTICLE{Tram2024,
       author = {{Tram}, Le Ngoc and {Hoang}, Thiem and {Wiesemeyer}, Helmut and {Ristorcelli}, Isabelle and {Menten}, Karl M. and {Ngoc}, Nguyen Bich and {Diep}, Pham Ngoc},
        title = "{Understanding the multi-wavelength thermal dust polarisation from the Orion molecular cloud in light of the radiative torque paradigm}",
      journal = {\aap},
         year = 2024,
        month = sep,
       volume = {689},
          eid = {A290},
        pages = {A290},
          doi = {10.1051/0004-6361/202450127},
archivePrefix = {arXiv},
       eprint = {2403.17088},
 primaryClass = {astro-ph.GA},
       adsurl = {https://ui.adsabs.harvard.edu/abs/2024A&A...689A.290T}
}

@ARTICLE{Draine2009,
       author = {{Draine}, Bruce T. and {Fraisse}, Aur{\'e}lien A.},
        title = "{Polarized Far-Infrared and Submillimeter Emission from Interstellar Dust}",
      journal = {\apj},
         year = 2009,
        month = may,
       volume = {696},
       number = {1},
        pages = {1-11},
          doi = {10.1088/0004-637X/696/1/1},
archivePrefix = {arXiv},
       eprint = {0809.2094},
 primaryClass = {astro-ph},
       adsurl = {https://ui.adsabs.harvard.edu/abs/2009ApJ...696....1D}
}

@ARTICLE{Ysard2019,
       author = {{Ysard}, N. and {Koehler}, M. and {Jimenez-Serra}, I. and {Jones}, A.~P. and {Verstraete}, L.},
        title = "{From grains to pebbles: the influence of size distribution and chemical composition on dust emission properties}",
      journal = {\aap},
         year = 2019,
        month = nov,
       volume = {631},
          eid = {A88},
        pages = {A88},
          doi = {10.1051/0004-6361/201936089},
archivePrefix = {arXiv},
       eprint = {1909.05015},
 primaryClass = {astro-ph.GA},
       adsurl = {https://ui.adsabs.harvard.edu/abs/2019A&A...631A..88Y}
}

@ARTICLE{Kohler2015,
       author = {{K{\"o}hler}, M. and {Ysard}, N. and {Jones}, A.~P.},
        title = "{Dust evolution in the transition towards the denser ISM: impact on dust temperature, opacity, and spectral index}",
      journal = {\aap},
         year = 2015,
        month = jul,
       volume = {579},
          eid = {A15},
        pages = {A15},
          doi = {10.1051/0004-6361/201525646},
archivePrefix = {arXiv},
       eprint = {1506.01533},
 primaryClass = {astro-ph.GA},
       adsurl = {https://ui.adsabs.harvard.edu/abs/2015A&A...579A..15K}
}

@ARTICLE{HoangLazarian2016,
       author = {{Hoang}, Thiem and {Lazarian}, A.},
        title = "{A Unified Model of Grain Alignment: Radiative Alignment of Interstellar Grains with Magnetic Inclusions}",
      journal = {\apj},
         year = 2016,
        month = nov,
       volume = {831},
       number = {2},
          eid = {159},
        pages = {159},
          doi = {10.3847/0004-637X/831/2/159},
archivePrefix = {arXiv},
       eprint = {1605.02828},
 primaryClass = {astro-ph.GA},
       adsurl = {https://ui.adsabs.harvard.edu/abs/2016ApJ...831..159H}
}

@ARTICLE{Fanciullo2022b,
       author = {{Fanciullo}, Lapo and {Kemper}, Francisca and {Pattle}, Kate and {Koch}, Patrick M. and {Sadavoy}, Sarah and {Coud{\'e}}, Simon and {Soam}, Archana and {Hoang}, Thiem and {Onaka}, Takashi and {Le Gouellec}, Valentin J.~M. and {Arzoumanian}, Doris and {Berry}, David and {Eswaraiah}, Chakali and {Chung}, Eun Jung and {Furuya}, Ray and {Hull}, Charles L.~H. and {Hwang}, Jihye and {Johnstone}, Douglas and {Kang}, Ji-hyun and {Kim}, Kyoung Hee and {Kirchschlager}, Florian and {K{\"o}nyves}, Vera and {Kwon}, Jungmi and {Kwon}, Woojin and {Lai}, Shih-Ping and {Lee}, Chang Won and {Liu}, Tie and {Lyo}, A.-Ran and {Stephens}, Ian and {Tamura}, Motohide and {Tang}, Xindi and {Ward-Thompson}, Derek and {Whitworth}, Anthony and {Shinnaga}, Hiroko},
        title = "{The JCMT BISTRO Survey: multiwavelength polarimetry of bright regions in NGC 2071 in the far-infrared/submillimetre range, with POL-2 and HAWC+}",
      journal = {\mnras},
         year = 2022,
        month = may,
       volume = {512},
       number = {2},
        pages = {1985-2002},
          doi = {10.1093/mnras/stac528},
archivePrefix = {arXiv},
       eprint = {2209.09604},
 primaryClass = {astro-ph.GA},
       adsurl = {https://ui.adsabs.harvard.edu/abs/2022MNRAS.512.1985F}
}

@ARTICLE{Goodman1995,
       author = {{Goodman}, Alyssa A. and {Jones}, Terry J. and {Lada}, Elizabeth A. and {Myers}, Philip C.},
        title = "{Does Near-Infrared Polarimetry Reveal the Magnetic Field in Cold Dark Clouds?}",
      journal = {\apj},
         year = 1995,
        month = aug,
       volume = {448},
        pages = {748},
          doi = {10.1086/176003},
       adsurl = {https://ui.adsabs.harvard.edu/abs/1995ApJ...448..748G}
}

@ARTICLE{Houde2009,
       author = {{Houde}, Martin and {Vaillancourt}, John E. and {Hildebrand}, Roger H. and {Chitsazzadeh}, Shadi and {Kirby}, Larry},
        title = "{Dispersion of Magnetic Fields in Molecular Clouds. II.}",
      journal = {\apj},
         year = 2009,
        month = dec,
       volume = {706},
       number = {2},
        pages = {1504-1516},
          doi = {10.1088/0004-637X/706/2/1504},
archivePrefix = {arXiv},
       eprint = {0909.5227},
 primaryClass = {astro-ph.GA},
       adsurl = {https://ui.adsabs.harvard.edu/abs/2009ApJ...706.1504H}
}

@ARTICLE{Coude2016,
       author = {{Coud{\'e}}, S. and {Bastien}, P. and {Kirk}, H. and {Johnstone}, D. and {Drabek-Maunder}, E. and {Graves}, S. and {Hatchell}, J. and {Chapin}, E.~L. and {Gibb}, A.~G. and {Matthews}, B.},
        title = "{The JCMT Gould Belt Survey: the effect of molecular contamination in SCUBA-2 observations of Orion A}",
      journal = {\mnras},
         year = 2016,
       volume = {457},
        pages = {2139-2150},
          doi = {10.1093/mnras/stv3009}
}

@ARTICLE{Draine2006,
       author = {{Draine}, B.~T.},
        title = "{On the Submillimeter Opacity of Protoplanetary Disks}",
      journal = {\apj},
         year = 2006,
        month = jan,
       volume = {636},
       number = {2},
        pages = {1114-1120},
          doi = {10.1086/498130},
archivePrefix = {arXiv},
       eprint = {astro-ph/0507292},
 primaryClass = {astro-ph},
       adsurl = {https://ui.adsabs.harvard.edu/abs/2006ApJ...636.1114D}
}

@ARTICLE{Hennebelle2019,
       author = {{Hennebelle}, Patrick and {Inutsuka}, Shu-ichiro},
        title = "{The role of magnetic field in molecular cloud formation and evolution}",
      journal = {Frontiers in Astronomy and Space Sciences},
         year = 2019,
        month = mar,
       volume = {6},
          eid = {5},
        pages = {5},
          doi = {10.3389/fspas.2019.00005},
archivePrefix = {arXiv},
       eprint = {1902.00798},
 primaryClass = {astro-ph.GA},
       adsurl = {https://ui.adsabs.harvard.edu/abs/2019FrASS...6....5H}
}

@INPROCEEDINGS{Pattle2023PPVII,
       author = {{Pattle}, K. and {Fissel}, L. and {Tahani}, M. and {Liu}, T. and {Ntormousi}, E.},
        title = "{Magnetic Fields in Star Formation: from Clouds to Cores}",
    booktitle = {Protostars and Planets VII},
         year = 2023,
       editor = {{Inutsuka}, S. and {Aikawa}, Y. and {Muto}, T. and {Tomida}, K. and {Tamura}, M.},
       series = {Astronomical Society of the Pacific Conference Series},
       volume = {534},
        month = jul,
    publisher = {ASP},
        pages = {193},
          doi = {10.48550/arXiv.2203.11179},
archivePrefix = {arXiv},
       eprint = {2203.11179},
 primaryClass = {astro-ph.GA},
       adsurl = {https://ui.adsabs.harvard.edu/abs/2023ASPC..534..193P}
}

@ARTICLE{Buckle2012,
       author = {{Buckle}, J.~V. and {Davis}, C.~J. and {di Francesco}, J. and {Graves}, S.~F. and {Nutter}, D. and {Richer}, J.~S. and {Roberts}, J.~F. and {Ward-Thompson}, D. and {White}, G.~J. and {Brunt}, C. and {Butner}, H.~M. and {Cavanagh}, B. and {Chrysostomou}, A. and {Curtis}, E.~I. and {Duarte-Cabral}, A. and {Etxaluze}, M. and {Fich}, M. and {Friberg}, P. and {Friesen}, R. and {Fuller}, G.~A. and {Greaves}, J.~S. and {Hatchell}, J. and {Hogerheijde}, M.~R. and {Johnstone}, D. and {Matthews}, B. and {Matthews}, H. and {Rawlings}, J.~M.~C. and {Sadavoy}, S. and {Simpson}, R.~J. and {Tothill}, N.~F.~H. and {Tsamis}, Y.~G. and {Viti}, S. and {Wouterloot}, J.~G.~A. and {Yates}, J.},
        title = "{The JCMT Legacy Survey of the Gould Belt: mapping $^{13}$CO and C$^{18}$O in Orion A}",
      journal = {\mnras},
         year = 2012,
        month = may,
       volume = {422},
       number = {1},
        pages = {521-541},
          doi = {10.1111/j.1365-2966.2012.20628.x},
archivePrefix = {arXiv},
       eprint = {1201.5483},
 primaryClass = {astro-ph.GA},
       adsurl = {https://ui.adsabs.harvard.edu/abs/2012MNRAS.422..521B}
}

@ARTICLE{Goldreich1982,
       author = {{Goldreich}, P. and {Kylafis}, N.~D.},
        title = "{Linear polarization of radio frequency lines in molecular clouds and circumstellar envelopes}",
      journal = {\apj},
         year = 1982,
        month = feb,
       volume = {253},
        pages = {606-621},
          doi = {10.1086/159663},
       adsurl = {https://ui.adsabs.harvard.edu/abs/1982ApJ...253..606G}
}

@ARTICLE{Cortes2005,
       author = {{Cortes}, P.~C. and {Crutcher}, R.~M. and {Watson}, W.~D.},
        title = "{Line Polarization of Molecular Lines at Radio Frequencies: The Case of DR 21(OH)}",
      journal = {\apj},
         year = 2005,
        month = aug,
       volume = {628},
       number = {2},
        pages = {780-788},
          doi = {10.1086/430815},
archivePrefix = {arXiv},
       eprint = {astro-ph/0504258},
 primaryClass = {astro-ph},
       adsurl = {https://ui.adsabs.harvard.edu/abs/2005ApJ...628..780C}
}

@ARTICLE{Ching2016,
       author = {{Ching}, Tao-Chung and {Lai}, Shih-Ping and {Zhang}, Qizhou and {Yang}, Louis and {Girart}, Josep M. and {Rao}, Ramprasad},
        title = "{Helical Magnetic Fields in the NGC 1333 IRAS 4A Protostellar Outflows}",
      journal = {\apj},
         year = 2016,
        month = mar,
       volume = {819},
       number = {2},
          eid = {159},
        pages = {159},
          doi = {10.3847/0004-637X/819/2/159},
archivePrefix = {arXiv},
       eprint = {1601.05229},
 primaryClass = {astro-ph.SR},
       adsurl = {https://ui.adsabs.harvard.edu/abs/2016ApJ...819..159C}
}

@ARTICLE{Michail2021,
       author = {{Michail}, Joseph M. and {Ashton}, Peter C. and {Berthoud}, Marc G. and {Chuss}, David T. and {Dowell}, C. Darren and {Guerra}, Jordan A. and {Harper}, Doyal A. and {Novak}, Giles and {Santos}, Fabio P. and {Siah}, Javad and et al.},
        title = "{Far-infrared Polarization Spectrum of the OMC-1 Star-forming Region}",
      journal = {\apj},
         year = 2021,
        month = jan,
       volume = {907},
       number = {1},
          eid = {46},
        pages = {46},
          doi = {10.3847/1538-4357/abd090},
archivePrefix = {arXiv},
       eprint = {2008.00310},
 primaryClass = {astro-ph.GA},
       adsurl = {https://ui.adsabs.harvard.edu/abs/2021ApJ...907...46M}
}

@ARTICLE{Ysard2024,
       author = {{Ysard}, N. and {Jones}, A.~P. and {Guillet}, V. and {Demyk}, K. and {Decleir}, M. and {Verstraete}, L. and {Choubani}, I. and {Miville-Desch{\^e}nes}, M.-A. and {Fanciullo}, L.},
        title = "{THEMIS 2.0: A self-consistent model for dust extinction, emission, and polarisation}",
      journal = {\aap},
         year = 2024,
        month = apr,
       volume = {684},
          eid = {A34},
        pages = {A34},
          doi = {10.1051/0004-6361/202348391},
archivePrefix = {arXiv},
       eprint = {2401.07739},
 primaryClass = {astro-ph.GA},
       adsurl = {https://ui.adsabs.harvard.edu/abs/2024A&A...684A..34Y}
}

@ARTICLE{LazarianHoang2007L,
       author = {{Lazarian}, A. and {Hoang}, Thiem},
        title = "{Subsonic Mechanical Alignment of Irregular Grains}",
      journal = {\apjl},
         year = 2007,
        month = nov,
       volume = {669},
       number = {2},
        pages = {L77-L80},
          doi = {10.1086/523849},
archivePrefix = {arXiv},
       eprint = {0707.3805},
 primaryClass = {astro-ph},
       adsurl = {https://ui.adsabs.harvard.edu/abs/2007ApJ...669L..77L}
}

@ARTICLE{Hoang2018,
       author = {{Hoang}, Thiem and {Cho}, Jungyeon and {Lazarian}, A.},
        title = "{Alignment of Irregular Grains by Mechanical Torques}",
      journal = {\apj},
         year = 2018,
        month = jan,
       volume = {852},
       number = {2},
          eid = {129},
        pages = {129},
          doi = {10.3847/1538-4357/aa9edc},
archivePrefix = {arXiv},
       eprint = {1704.02256},
 primaryClass = {astro-ph.GA},
       adsurl = {https://ui.adsabs.harvard.edu/abs/2018ApJ...852..129H}
}

@ARTICLE{Das2016,
       author = {{Das}, Indrajit and {Weingartner}, Joseph C.},
        title = "{Alignment of interstellar grains by mechanical torques: suprathermally rotating Gaussian random spheres}",
      journal = {\mnras},
         year = 2016,
        month = apr,
       volume = {457},
       number = {2},
        pages = {1958-1987},
          doi = {10.1093/mnras/stw146},
archivePrefix = {arXiv},
       eprint = {1602.04775},
 primaryClass = {astro-ph.GA},
       adsurl = {https://ui.adsabs.harvard.edu/abs/2016MNRAS.457.1958D}
}

@ARTICLE{Draine1999,
       author = {{Draine}, B.~T. and {Lazarian}, A.},
        title = "{Magnetic Dipole Microwave Emission from Dust Grains}",
      journal = {\apj},
         year = 1999,
        month = feb,
       volume = {512},
       number = {2},
        pages = {740-754},
          doi = {10.1086/306809},
archivePrefix = {arXiv},
       eprint = {astro-ph/9807009},
 primaryClass = {astro-ph},
       adsurl = {https://ui.adsabs.harvard.edu/abs/1999ApJ...512..740D}
}

@ARTICLE{Lazarian2008,
       author = {{Lazarian}, A. and {Hoang}, Thiem},
        title = "{Alignment of Dust with Magnetic Inclusions: Radiative Torques and Superparamagnetic Barnett and Nuclear Relaxation}",
      journal = {\apjl},
         year = 2008,
        month = mar,
       volume = {676},
       number = {1},
        pages = {L25},
          doi = {10.1086/586706},
archivePrefix = {arXiv},
       eprint = {0801.0265},
 primaryClass = {astro-ph},
       adsurl = {https://ui.adsabs.harvard.edu/abs/2008ApJ...676L..25L}
}

@ARTICLE{Whittet2008,
       author = {{Whittet}, D.~C.~B. and {Hough}, J.~H. and {Lazarian}, A. and {Hoang}, Thiem},
        title = "{The Efficiency of Grain Alignment in Dense Interstellar Clouds: a Reassessment of Constraints from Near-Infrared Polarization}",
      journal = {\apj},
         year = 2008,
        month = feb,
       volume = {674},
       number = {1},
        pages = {304-315},
          doi = {10.1086/525040},
archivePrefix = {arXiv},
       eprint = {0711.2536},
 primaryClass = {astro-ph},
       adsurl = {https://ui.adsabs.harvard.edu/abs/2008ApJ...674..304W}
}

@BOOK{Draine2011book,
       author = {{Draine}, B.~T.},
        title = "{Physics of the Interstellar and Intergalactic Medium}",
    publisher = {Princeton University Press},
         year = "2011",
       series = {Princeton Series in Astrophysics},
      address = {Princeton, NJ},
         isbn = {9780691122144},
       adsurl = {https://ui.adsabs.harvard.edu/abs/2011piim.book.....D}
}

@ARTICLE{Astropy2013,
       author = {{Astropy Collaboration} and {Robitaille}, Thomas P. and {Tollerud}, Erik J. and others},
        title = "{Astropy: A community Python package for astronomy}",
      journal = {\aap},
         year = 2013,
        month = oct,
       volume = {558},
          eid = {A33},
        pages = {A33},
          doi = {10.1051/0004-6361/201322068}
}

@ARTICLE{Astropy2018,
       author = {{Astropy Collaboration} and {Price-Whelan}, A.~M. and {Sip{\H{o}}cz}, B.~M. and others},
        title = "{The Astropy Project: Building an Open-science Project and Status of the v2.0 Core Package}",
      journal = {\aj},
         year = 2018,
        month = sep,
       volume = {156},
       number = {3},
          eid = {123},
        pages = {123},
          doi = {10.3847/1538-3881/aabc4f}
}

@ARTICLE{Astropy2022,
       author = {{Astropy Collaboration} and {Price-Whelan}, Adrian M. and {Lim}, Pey Lian and others},
        title = "{The Astropy Project: Sustaining and Growing a Community-oriented Open-source Project and the Latest Major Release (v5.0) of the Core Package}",
      journal = {\apj},
         year = 2022,
        month = aug,
       volume = {935},
       number = {2},
          eid = {167},
        pages = {167},
          doi = {10.3847/1538-4357/ac7c74}
}

@ARTICLE{Harris2020,
       author = {{Harris}, Charles R. and {Millman}, K. Jarrod and {van der Walt}, St{\'e}fan J. and others},
        title = "{Array programming with NumPy}",
      journal = {Nature},
         year = 2020,
        month = sep,
       volume = {585},
       number = {7825},
        pages = {357-362},
          doi = {10.1038/s41586-020-2649-2}
}

@ARTICLE{Virtanen2020,
       author = {{Virtanen}, Pauli and {Gommers}, Ralf and {Oliphant}, Travis E. and others},
        title = "{SciPy 1.0: fundamental algorithms for scientific computing in Python}",
      journal = {Nature Methods},
         year = 2020,
        month = mar,
       volume = {17},
        pages = {261-272},
          doi = {10.1038/s41592-019-0686-2}
}

@ARTICLE{Hunter2007,
       author = {{Hunter}, John D.},
        title = "{Matplotlib: A 2D Graphics Environment}",
      journal = {Computing in Science and Engineering},
         year = 2007,
        month = jan,
       volume = {9},
       number = {3},
        pages = {90-95},
          doi = {10.1109/MCSE.2007.55}
}

@INPROCEEDINGS{Currie2014,
       author = {{Currie}, M.~J. and {Berry}, D.~S. and {Jenness}, T. and {Gibb}, A.~G. and {Bell}, G.~S. and {Draper}, P.~W.},
        title = "{Starlink Software in 2013}",
    booktitle = {Astronomical Data Analysis Software and Systems XXIII},
         year = 2014,
       editor = {{Manset}, N. and {Forshay}, P.},
       series = {Astronomical Society of the Pacific Conference Series},
       volume = {485},
        month = may,
        pages = {391},
       adsurl = {https://ui.adsabs.harvard.edu/abs/2014ASPC..485..391C}
}

@MISC{Parsons2021,
       author = {{Parsons}, H.~A.~L. and {Berry}, D.~S. and {Rawlings}, M.~G. and {Graves}, S.~F.},
        title = "{The POL-2 Data Reduction Cookbook}",
         year = 2021,
        month = jul,
 howpublished = {Starlink Cookbook SC/22, version 1.1 (2021 July 15), East Asian Observatory; source in the Starlink software collection, \url{https://github.com/Starlink/starlink/tree/3a7e54c75281daedf99c1f3be6e87030115d7bba/docs/sc/022}},
}

@ARTICLE{Scheiter2026,
       author = {{Scheiter}, M.~L. and {Wolf}, S.},
        title = "{The role of radiative torques in the molecular cloud core L43}",
      journal = {\aap},
         year = 2026,
        month = jan,
       volume = {705},
          eid = {A207},
        pages = {A207},
          doi = {10.1051/0004-6361/202557363},
archivePrefix = {arXiv},
       eprint = {2602.16519},
 primaryClass = {astro-ph.SR},
       adsurl = {https://ui.adsabs.harvard.edu/abs/2026A&A...705A.207S}
}

@ARTICLE{Tram2021,
       author = {{Tram}, Le Ngoc and {Hoang}, Thiem and {Lee}, Hyeseung and {Santos}, Fabio P. and {Soam}, Archana and {Lesaffre}, Pierre and {Gusdorf}, Antoine and {Reach}, William T.},
        title = "{Understanding Polarized Dust Emission from {\ensuremath{\rho}} Ophiuchi A in Light of Grain Alignment and Disruption by Radiative Torques}",
      journal = {\apj},
         year = 2021,
        month = jan,
       volume = {906},
       number = {2},
          eid = {115},
        pages = {115},
          doi = {10.3847/1538-4357/abc6fe},
archivePrefix = {arXiv},
       eprint = {2007.10621},
 primaryClass = {astro-ph.GA},
       adsurl = {https://ui.adsabs.harvard.edu/abs/2021ApJ...906..115T}
}

@ARTICLE{Lee2020,
       author = {{Lee}, Hyeseung and {Hoang}, Thiem and {Le}, Ngan and {Cho}, Jungyeon},
        title = "{Physical Model of Dust Polarization by Radiative Torque Alignment and Disruption and Implications for Grain Internal Structures}",
      journal = {\apj},
         year = 2020,
        month = jun,
       volume = {896},
       number = {1},
          eid = {44},
        pages = {44},
          doi = {10.3847/1538-4357/ab8e33},
archivePrefix = {arXiv},
       eprint = {1911.00654},
 primaryClass = {astro-ph.GA},
       adsurl = {https://ui.adsabs.harvard.edu/abs/2020ApJ...896...44L}
}

@PHDTHESIS{Graves2012,
       author = {{Graves}, Sarah Frances},
        title = "{Submillimetre spectral imaging of clustered star formation}",
       school = {University of Cambridge, UK},
         year = 2012,
        month = jan,
       adsurl = {https://ui.adsabs.harvard.edu/abs/2012PhDT.......295G}
}

@ARTICLE{Friesen2017,
       author = {{Friesen}, Rachel K. and {Pineda}, Jaime E. and {Rosolowsky}, Erik and {Alves}, Felipe and {Chac{\'o}n-Tanarro}, Ana and {How-Huan Chen}, Hope and {Chun-Yuan Chen}, Michael and {Di Francesco}, James and {Keown}, Jared and {Kirk}, Helen and {Punanova}, Anna and {Seo}, Youngmin and {Shirley}, Yancy and {Ginsburg}, Adam and {Hall}, Christine and {Offner}, Stella S.~R. and {Singh}, Ayushi and {Arce}, H{\'e}ctor G. and {Caselli}, Paola and {Goodman}, Alyssa A. and {Martin}, Peter G. and {Matzner}, Christopher and {Myers}, Philip C. and {Redaelli}, Elena and {GAS Collaboration}},
        title = "{The Green Bank Ammonia Survey: First Results of NH$_{3}$ Mapping of the Gould Belt}",
      journal = {\apj},
         year = 2017,
        month = jul,
       volume = {843},
       number = {1},
          eid = {63},
        pages = {63},
          doi = {10.3847/1538-4357/aa6d58},
archivePrefix = {arXiv},
       eprint = {1704.06318},
 primaryClass = {astro-ph.GA},
       adsurl = {https://ui.adsabs.harvard.edu/abs/2017ApJ...843...63F}
}

@ARTICLE{Pineda2026,
       author = {{Pineda}, Jaime E. and {Friesen}, Rachel K. and {Rosolowsky}, Erik and {Chac{\'o}n-Tanarro}, Ana and {Chen}, Michael Chun-Yuan and {Di Francesco}, James and {Kirk}, Helen and {Punanova}, Anna and {Seo}, Youngmin and {Shirley}, Yancy and {Ginsburg}, Adam and {Offner}, Stella S.~R. and {Pandhi}, Ayush and {Singh}, Ayushi and {Quan}, Feiyu and {Arce}, H{\'e}ctor G. and {Caselli}, Paola and {Choudhury}, Spandan and {Goodman}, Alyssa A. and {Heitsch}, Fabian and {Martin}, Peter G. and {Matzner}, Christopher D. and {Myers}, Philip C. and {Redaelli}, Elena and {Scibelli}, Samantha and {GAS Collaboration}},
        title = "{The Green Bank Ammonia Survey: Data Release 2}",
      journal = {\apjs},
         year = 2026,
        month = jan,
       volume = {282},
       number = {1},
          eid = {18},
        pages = {18},
          doi = {10.3847/1538-4365/ae11b1},
archivePrefix = {arXiv},
       eprint = {2510.10607},
 primaryClass = {astro-ph.GA},
       adsurl = {https://ui.adsabs.harvard.edu/abs/2026ApJS..282...18P}
}
\bibliographystyle{aasjournal}
 
\end{document}